\documentclass[preprintnumbers, prd,  showpacs, floatfix,
preprintnumbers, letterpaper,nofootinbib, superscriptaddress]{revtex4}
\usepackage[dvips]{graphicx}
\usepackage{epsf}
\usepackage{amsmath}
\usepackage{amssymb}
\usepackage{graphicx}
\usepackage{dcolumn}
\usepackage{bm}
\usepackage{color}
\usepackage{subfigure}
\usepackage{threeparttable}
\usepackage{pdflscape}
\usepackage{rotating}
\usepackage[latin1]{inputenc}
\newcommand{\be}{\begin{equation}}
\newcommand{\ee}{\end{equation}}
\newcommand{\bea}{\begin{eqnarray}}
\newcommand{\eea}{\end{eqnarray}}

\begin{document}

\title[]{Asymptotic behavior of a scalar field with an arbitrary potential trapped on a Randall-Sundrum's braneworld:  the effect of a negative dark radiation term on a Bianchi I brane}

\author{Dagoberto Escobar}
\affiliation{Departamento de F\'isica Universidad de Camaguey, Cuba.}
\email{dagoberto.escobar@reduc.edu.cu}

\author{Carlos R. Fadragas}
\affiliation{Departamento de F\'isica, Universidad Central de Las Villas, CP 54830 Santa Clara, Cuba.}
\email{fadragas@uclv.edu.cu}

\author{Genly Leon}
\affiliation{Instituto de F\'{\i}sica, Pontificia Universidad  Cat\'olica de Valpara\'{\i}so, Casilla 4950, Valpara\'{\i}so, Chile.}
\email{genly.leon@ucv.cl}

\author{Yoelsy Leyva}
\affiliation{Instituto de F\'{\i}sica, Pontificia Universidad  Cat\'olica de Valpara\'{\i}so, Casilla 4950, Valpara\'{\i}so, Chile.}
\affiliation{Divisi\'on de Ciencias e Ingenier\'ia de la  Universidad de Guanajuato,  A.P. 150, 37150,  Le\'on, Guanajuato, M\'exico.}
\email{yoelsy.leyva@ucv.cl}

\begin{abstract}
In this work we present a phase space analysis of
a quintessence field and a perfect fluid trapped in a Randall-
Sundrum's Braneworld of type 2. We consider a homogeneous
but anisotropic Bianchi I brane geometry. Moreover,
we consider the effect of the projection of the fivedimensional
Weyl tensor onto the three-brane in the form
of a negative Dark Radiation term. For the treatment of the
potential we use the ``Method of $f$-devisers'' that allows investigating
arbitrary potentials in a phase space. We present
general conditions on the potential in order to obtain the stability
of standard 4D and non-standard 5D de Sitter solutions,
and we provide the stability conditions for both scalar
field-matter scaling solutions, scalar field-dark radiation solutions
and scalar field-dominated solutions. We find that the
shear-dominated solutions are unstable (particularly, contracting
shear-dominated solutions are of saddle type). As
a main difference with our previous work, the traditionally
ever-expanding models could potentially re-collapse due to
the negativity of the dark radiation. Additionally, our system
admits a large class of static solutions that are of saddle
type. These kinds of solutions are important at intermediate
stages in the evolution of the universe, since they allow the
transition from contracting to expanding models and viceversa.
New features of our scenario are the existence of a
bounce and a turnaround, which lead to cyclic behavior, that
are not allowed in Bianchi I branes with positive dark radiation
term. Finally, as specific examples we consider the
potentials $V\propto\sinh^{-\alpha}(\beta\phi)$ and $V\propto\left[\cosh\left( \xi\phi \right)-1\right]$ which have simple $f$-devisers.
\end{abstract} 
\keywords{Dark Energy; Dark Radiation; Randall-Sundrum; Bianchi I.}
\pacs{04.20.-q, 04.20.Cv, 04.20.Jb, 04.50.Kd, 11.25.-w, 11.25.Wx, 95.36.+x, 98.80.-k, 98.80.Bp,
98.80.Cq, 98.80.Jk}%

\date{\today}

\maketitle

\section{Introduction}

The idea that our Universe is a brane embedded in a 5--dimensional space,  on which the particles of the standard model are confined, while gravity is allowed to propagate not only on the brane but also in the bulk of the higher-dimensional manifold, was proposed by~\citet{Randall1999a,Randall1999}. These braneworld models were baptized as Randall-Sundrum type $1$ (RS1) and type $2$ (RS2).
The original motivation of \citet{Randall1999a} to propose the RS1 scenario was to look for an explanation to the hierarchy problem, whereas the motivation of the RS2 scenario was to propose an alternative mechanism to the Kaluza-Klein compactifications \citep{Randall1999}. 

It is well-known that the cosmological field equations on the RS2 brane  are essentially
different from the standard 4--dimensional cosmology \citep{Binetruy2000,Binetruy2000a,Bowcock2000,Apostolopoulos2005}. In fact, in the case of a flat
Friedmann-Robertson-Walker metric (FRW) the Friedmann equation is modified by:
\be H^{2}= \frac{1}{3}\kappa^{2}\rho\left(1+
\frac{\rho}{2\lambda}\right)\label{2},
\ee
where $\lambda>0$ is the brane tension. These models also have an 
appreciable cosmological
impact in the inflationary scenario \citep{Hawkins2001,Huey2001,Huey2002}. However, is not the study of brane inflation the purpose of the present paper, but the investigation of the bouncing behavior, the study of the asymptotic structure of the model and, also, the investigation of the role of static solutions in the cosmic evolution. 
 
The dynamics of RS2 branes has been widely investigated using the dynamical systems approach \citep{Szydlowski2002,Hoogen2003,Hoogen2003a,Toporensky2003,Savchenko2003,Solomons2006,Haghani2012}. In \citep{Campos2001} was studied the asymptotic behavior of FRW,  Bianchi I
and V brane in the presence of perfect fluid. The same authors analyzed the contribution of the non-vanishing (positive and negative) dark radiation
term, $\cal U$, in the dynamics of FRW and Bianchi I branes in \citep{Campos2001a}. \citet{Goheer2002, Goheer2003} extended the previous study by the inclusion of a scalar field with an exponential potential in FRW and Bianchi I branes. In the first case the authors discussed the changes of the phase space compared with the general relativistic case \citep{Goheer2002} while for the Bianchi I brane, the effects of considering a non vanishing projection of the Weyl tensor on the brane were studied taking into account four different scenarios: a) ${\cal U}\geq 0$, $^3R\leq 0$; b) ${\cal U}\geq 0$, $^3R \geq 0$; c) ${\cal U}\leq 0$, $^3R\leq 0$; d) ${\cal U}\leq 0$, $^3R\geq 0$ \citep{Goheer2003}.  These studies were extended to the case of a scalar field with an arbitrary potential trapped in FRW
branes by \citet{Leyva2009,Escobar2012}, while the impact of a positive dark radiation term (${\cal U}\geq 0$) on the dynamics of a Bianchi I brane, supported by a scalar field with an arbitrary potential, was investigated by \citet{Escobar2012a}.

The term $\cal U$ represents the scalar component of the
electric  (Coulomb)  part    of
the 5-dimensional Weyl tensor  of the bulk \citep{Shiromizu2000,Sasaki2000}. 
$\cal U$ scales just like radiation with a constant $\mu$, i.e., ${\cal U}=\mu a^{-4}$, for that reason it is called the dark radiation. 
The coefficient $\mu$ is a constant of integration obtained by
integrating the 5-dimensional Einstein equations 
\citep{Binetruy2000,Binetruy2000a,Langlois2001,Barrow2002,Kraus1999,Hebecker2001,Ida2000,Vollick2001}.
In FRW brane model, $\mu$ is related with the mass of the black hole in the bulk therefore $\cal U$ is positive defined \citep{Mukohyama2000,Maartens2010,Clifton2012}. For anisotropic models (for instance, Bianchi I models)   both positive and negative $\mu$ are possible \citep{Campos2001a,Goheer2003,Santos2001}. Its magnitude and sign  can depend on the choice of initial conditions when
solving the 5-dimensional Einstein equation. Hence, even the sign of $\mu$ remains an open question \citep{Sasaki2000,Mukohyama2000}.

Dark radiation should strongly affect both big-bang nucleosynthesis (BBN)
and the cosmic microwave background (CMB). Such observations can be used to constrain both the magnitude and sign of the dark radiation as well as other radiation sources (see the works by \citet{Malaney1993,Olive2000}, and by \citet{Ichiki2002,Apostolopoulos2006,Dutta2009,Diamanti2012,Gonzalez-Garcia2013}). The dark radiation restrictions for FRW brane were found in \citep{Ichiki2002, Bratt2002}. In \citep{Ichiki2002}, the constraints on BBN alone allow for $-1.23 \leq \rho_{DR}/\rho_\gamma \leq 0.11$ in  a dark radiation component, where $\rho_\gamma $ is the total energy density in background photons just before the BBN epoch at $T = 1MeV,$ $\rho_{DR}={6{\cal U}}/{\lambda},$ and  $\lambda$ is the tension of the brane.  In order to compute the theoretical prediction of CMB anisotropies exactly, one must eventually solve the perturbations including the contribution from the bulk.  
However, in \citep{Ichiki2002} the CMB power spectrum was used to constrain the dominant expansion-rate effect of the dark-radiation term in the generalized Friedmann equation. If the constraint from effects of dark radiation on the expansion rate are included, the allowed concordance range of dark radiation content reduces to
$-0.41 \leq \rho_{DR}/\rho_\gamma \leq 0.105$ at the 95\% confidence level. Similar result was derived by \citet{Bratt2002}, for high values of the 5-dimensional Planck mass, as $-1<\rho_{DR}/\rho_\nu<0.5$, where $\rho_\nu$ is the energy density contributed by a single, two-component massless neutrino. 

The presence of dark radiation is associated with new degrees of freedom in the relativistic components of our Universe. Indeed, any process able to produce extra (dark) radiation produces the same effects on the background expansion of additional neutrinos, therefore, a larger value for effective number of relativistic degrees of freedom ($N_{eff}$) is obtained  \citep{Archidiacono2011}. These bounds has been improved in the latest analysis by the Wilkinson Microwave Anisotropy
Probe (WMAP) \citep{Bennett2012}, the South Pole Telescope (SPT) \citep{SPT}, the Atacama Cosmology Telescope (ACT) \citep{ACT} and Planck \citep{planck1}. The main results on the effective number of neutrino species are
the following:
\begin{itemize}
\item $N_{eff}=2.79 \pm 0.56$ joint result of WMAP 7 $+$ ACT \citep{ACT},
\item $N_{eff}=3.50 \pm 0.42$ joint result of WMAP 7 $+$ ACT $+$ BAO $+$ HST,
\item $N_{eff}=2.85^{+0.95}_{-0.91}$ joint result of WMAP 9 $+$ ACT \citep{DiValentino2013a},
\citep{ACT},
\item $N_{eff}=3.23^{+0.77}_{-0.76}$ joint result of WMAP 9 $+$ ACT $+$ BAO $+$ HST \citep{DiValentino2013a},
\item $N_{eff}=3.71 \pm 0.35$ joint result of WMAP 7 $+$  South Pole
Telescope (SPT)$+$ BAO $+$HST \citep{SPT},
\item $N_{eff}=3.84 \pm 0.40$ joint result of WMAP 9$+$  eCMB $+$ BAO $+$ HST
\citep{WMAP9},
\end{itemize}
and more recently
\begin{itemize}
\item $N_{eff}=3.83 \pm 0.54$ at $95\%$ confidence level from joint result of
Planck satellite $+$ HST \citep{planck1} and
\item $N_{eff}=3.62^{+0.50}_{-0.48}$ at $95\%$ confidence level from joint
result of Planck satellite $+$ HST $+$ ACT $+$ SPT $+$WMAP 9 \citep{planck1}.
\end{itemize}
These latter bounds indicate the presence of an extra dark radiation
component at the $\approx 2\sigma$  confidence level \citep{DiValentino2013} and the
existence of a tension between the latest ACT and SPT results. So the presence of a dark radiation source, like sterile
neutrinos from the ($3+2$) or ($3+3$) models \citep{DiValentino2013}, or a
(negative) source of dark radiation with an geometrical origin like in our case are not ruled out by
observations. 

In the other hand, observation of the CMB tell us that our Universe is isotropic a great accuracy, to within a part in $10^5$ \citep{Hinshaw2007,Komatsu2011}. The natural framework to approach this highly isotropic Universe we observe today, is to assume that the Universe started in a highly initial anisotropic state and then, a dynamical mechanism get rid of almost all its anisotropy. Several candidates have been proposed to explaining this behavior, among them, inflation mechanisms \citep{Guth1981, Linde1982} is the most popular. In these line, the simplest generalization of a FRW cosmologies are Bianchi cosmologies, the latter provide anisotropic but homogeneous cosmologies \citep{Misner1974}, where the central point of discussion is if the Universe can isotropize without fine-tuning the parameters of the model. The isotropization of Bianchi I braneworld cosmologies has been investigated, from several points of view, in the literature \citep{Maartens2001,Niz2008, Goheer2003,Toporensky2005}. In \citep{Maartens2001}, is shown that large anisotropy does not negatively affect inflation in a Bianchi I braneworld; also it is shown that the initial expansion of the universe is quasi-isotropic if the scalar field possesses a large kinetic term. While considering negative values of dark radiation (${\cal U}<0$) in Bianchi I models lead to very interesting solutions for which the universe can both collapse or isotropize \citep{Campos2001a,Campos2001,Goheer2002,Goheer2003,Toporensky2005}. More recently, the Planck results \citep{planck1, plack2} are rekindled a renewed interest in these Bianchi cosmologies since some anisotropic anomalies seem to appear. At the same time, is required that this isotropization is accompanied with a phase of accelerated expansion in order to be a good candidate to explain the strong results that indicates the current speeding up of the observable universe \citep{Riess1998, Perlmutter1999}. This latter observational fact is approached from two directions: modifying the gravitational sector \citep{Clifton2012} or introducing an hypothetical form of energy baptized as \textit{Dark Energy} \footnote{Dark energy models includes: cosmological constant model, quintessence scalar field, k-essence, tachyon, Chaplygin gas, etc. (see \citep{Copeland2006} for a reviews and references therein).} \citep{Copeland2006}. From this viewpoint, a model in which a dark energy component lives in a Bianchi braneworld combines both approaches. All the above reasons motivate the investigation of a homogeneous but anisotropic brane model with ${\cal U}<0.$ 
 
In this paper, we study the dynamics of a scalar field with an arbitrary potential trapped  in a RS-2 brane-world model. We consider a homogeneous but aniso--tropic Bianchi I (BI) brane filled also with a perfect fluid. Furthermore, we consider the effect of the projection of the five-dimensional Weyl tensor onto the three-brane in the form of a negative dark radiation term. The results presented here complement our previous investigation  \citep{Escobar2012a} where we considered the effect of a positive dark radiation term on the brane.  

For the treatment of the potential we use a  modification of the method  introduced for the investigation of scalar fields in isotropic (FRW) scenarios 
\citep{Fang2009,Matos2009,Leyva2009,Urena-Lopez2012,Copeland2009,Dutta2009}, and that has been generalized to several cosmological contexts in 
\citep{Escobar2012,Escobar2012a,Farajollahi2011,Xiao2011}. The modified method, that we call   ``Method of $f$-devisers'', allows us to perform a phase-space analysis of a cosmological model, without the need for specifying the potential. This is a significant advantage, since one can first perform the analysis for arbitrary potentials and then just substitute the desired forms, instead of repeating the whole procedure for
every distinct potential. This investigation represents a further step in a series of works devoted to the use of the general procedure of $f$-devisers for investigating both FRW and Bianchi I branes, initiated in our previous works \citep{Escobar2012,Escobar2012a}.

A main difference with our results in \citep{Escobar2012a}, is that the
traditionally ever-expanding models could potentially recollapse due to the
negativity of the dark radiation (${\cal U}<0$). New features of our scenario are the possibility of
a bounce and a turnaround, which leads to cyclic behavior. This behavior is
not allowed  in Bianchi I branes with positive dark radiation term
\citep{Escobar2012a}. Observe that in the usual Randall-Sundrum scenario
the tension of the brane is positive. In this case, $H=0$ if and
only if the total matter density satisfy $\rho=0.$ For this reason the bounce is not possible for the
usual braneworld scenario (flat geometry, positive brane tension).
However, it is possible to have a bounce provided the existence of a negative
correction (in our context, the negative ``dark radiation'' component) to the
Friedman equation \eqref{2} compensating the new positive contribution
$\frac{\kappa^2\rho^2}{6\lambda}$ (due to presence of the brane on the higher
dimensional bulk space), still having a total non-negative r.h.s. 

 Also, our system admits a large class of static solutions  that are of saddle type. This kind of
solutions are important at intermediate stages in the evolution of the
universe since they allow the transition from expanding to contracting
models, and viceversa. The interest in static solutions in the cosmological setting goes  back to
the discovering of Einstein static (ES) model. This model was proposed by
\citet{Einstein1917} as an attempt to incorporate Mach's principle into the general relativity (GR)
and also to overcome the boundary conditions of the theory. Several issues of this model have been intensively investigated. \citet{Eddington1930,Harrison1967,Gibbons1987}, considered inhomogeneous and anisotropic perturbations to ES model; the stability issue for  ghost massless
scalar field cosmologies was studied by \citet{Barrow2009}; the case for $f(R)$ theories was studied by \citet{Goswami2008,Goheer2009}, where ES solutions can provide the link between
decelerating/accelerating phases in these theories in the same way as in GR \citep{Barrow2003}. The stability analysis of the ES universe, and other kinds of static
solutions in the context of brane cosmology with dust matter and anisotropic geometry was presented by \citet{Campos2001a,Campos2001}.
Here we present several classes of static solutions for a Bianchi I brane
containing a scalar field with an arbitrary potential. This class of
solutions are more general than the presented by \citet{Campos2001a,Campos2001,Goheer2003}. 

The main objectives of our investigation are: 
\begin{itemize} 
\item[a)] To analyze the bouncing and cyclic behavior of some cosmological
solutions in our set up. 
\item[b)] To obtain qualitative information about the past and future
asymptotic structure of our model, without the need to repeat the whole
procedure each time the scalar field potential is chosen.  Especially, we
want to obtain general conditions for late-time isotropization. The method of
$f$-devisers is the key for this point.
\item[c)] To investigate the viability of static solutions as the link
between contracting and expanding solutions in our set up.
\item[d)] To generalized previous results obtained by us and by other authors
that were discussed in the literature. 
\end{itemize}

The paper is organized as follows. In section \ref{newmethod} we present
our general approach for the investigation of arbitrary potentials. In
section \ref{BianchiI} are presented the cosmological equations of our
model. In section \ref{dyn} we proceed to the dynamical system analysis of
the cosmological model under consideration. For this purpose we use the
method
of $f$- devisers discussed in section \ref{newmethod}, and we introduce
local charts adapted to each of the more interesting singular points. In
this way, we analyze the stability of the shear-dominated solutions, the
stability of the de Sitter solutions and the stability of the solutions
with 5D corrections. In section \ref{discussion} we summarize our
analytical results and proceed to the physical discussion specially to the
comparison with previous results. Particular emphasis is made on static
solutions. In section \ref{Examples} we illustrate our analytical results for two examples: 
 the cosh-like potential $V(\phi)=V_{0}\left[\cosh\left( \xi \phi
\right)-1\right]$ \citep{Ratra1988,Wetterich1988,Matos2000,Sahni2000,Sahni2000a,Lidsey2002,Pavluchenko2003,Matos2009,Copeland2009,Leyva2009} and the inverted sinh-like potential  $V(\phi)=V_{0}\sinh^{-\alpha}(\beta\phi)$ studied by  \citet{Ratra1988,Wetterich1988,Urena-Lopez2000,Sahni2000a,Pavluchenko2003,Copeland2009,Leyva2009}  which have simple
$f$-devisers. Is worthy to mention that this procedure is general and
applies to other potentials different from the cosh-like, the inverted
sinh-like and the exponential one. Finally, in section \ref{conclusions}
are drawn our general conclusions.

\section{``Method of $f$-devisers''}\label{newmethod}

\begin{table*}[ht]
\caption{Explicit form of the $f(s)$-function  for some
quintessence potentials.}
\label{fsform} 
\begin{center}
\begin{minipage}{\textwidth} 
\begin{tabular}{@{\hspace{4pt}}c@{\hspace{
14pt}}c}
\hline
\hline\\[-0.3cm]
Potential & $f(s)$  \\[0.1cm] 
\hline\\[-0.2cm]
$V(\phi)=V_{0}\left[\cosh\left( \xi \phi \right)-1\right]$
\footnote{\citet{Ratra1988,Wetterich1988,Matos2000,Sahni2000,Sahni2000a,Lidsey2002,Pavluchenko2003,Matos2009,Copeland2009,Leyva2009}.} & $-\frac{1}{2}(s^2-\xi^2)$ \\[0.2cm]
$V(\phi)=V_{0}\sinh^{-\alpha}(\beta\phi)$ 
\footnote{\citet{Ratra1988,Wetterich1988,Sahni2000,Urena-Lopez2000,Pavluchenko2003,Copeland2009,Leyva2009}.} &
$\frac{s^2}{\alpha}-\alpha\beta^2$ \\[0.2cm]
$V(\phi)=V_{0}e^{-\lambda\phi}+\Lambda$ 
\footnote{\citet{Pavluchenko2003,Cardenas2003}.} &
  $-s(s-\lambda)$\\[0.2cm]
$V(\phi)=V_{0}\left[e^{\alpha\phi}+e^{\beta\phi}\right]$ 
\footnote{\citet{Barreiro2000,Gonzalez2006,Gonzalez2007}.}
 &
$-(s+\alpha)(s+\beta)$ \\[0.4cm]
\hline \hline
\end{tabular}
\end{minipage}
\end{center}
\end{table*}

In this section we present a  general method, called  ``Method of $f$-devisers'', that allows us to perform a
phase-space analysis of a cosmological model, without the need for
specifying the potential. This procedure consist of a
modification of a method used in isotropic (FRW) scalar field cosmologies
for the treatment of the potential of the scalar field
\citep{Fang2009,Matos2009,Leyva2009,Urena-Lopez2012,Copeland2009}, that has been generalized to several cosmological
contexts by \citet{Farajollahi2011,Xiao2011,Escobar2012,Escobar2012a}. An earlier attempt to introduce the
``Method of $f$-devisers'' as in the present form was by
\citet{Matos2009} in the context of FRW cosmologies. However, there the
authors restricted their attention to a scalar field dark-matter model with
a cosh-like potential, and there was revealed that the late-time attractor
is always the de Sitter solution. In section \ref{Examples} we will take
this potential as an example of the application of the present method in the context of brane cosmology.

The procedure for deriving the ``Method of $f$-devisers'' is as follows.
Let us define the two dynamical variables
\begin{eqnarray}
\label{sdef}
s&=&-\frac{V'(\phi )}{V(\phi
)}\,,\\
f&=&\frac{V''(\phi
   )}{V(\phi )}-\frac{V'(\phi )^2}{V(\phi )^2},
\label{fdef}
\end{eqnarray} while keeping the potential still arbitrary. 

In the formulas  \eqref{sdef} and \eqref{fdef}, the primes denote
differentiation with respect to $\phi$.

Now, for the usual ansatzes of the
cosmological literature, it results that $f$ can be expressed as an
explicit one-valued function of $s$, that is $f=f(s)$ (as can be seen in
table \ref{fsform}), and therefore we obtain a closed dynamical system
for $s$  and a set of normalized-variables. So, instead to consider a fixed potential $V,$ from the beginning, we examine the
asymptotic properties of our cosmological model, by considering suitable
(general) conditions on an arbitrary input function 
\[f:\mathbb{R}\rightarrow \mathbb{R}, \;\; s\rightarrow f(s).\] Now, once we have considered a function $f$ as an input, we can solve the equation 
\be\label{ODE} \frac{V''(\phi
   )}{V(\phi )}-\frac{V'(\phi )^2}{V(\phi )^2}=f\left(-\frac{V'(\phi )}{V(\phi
)}\right),\ee
that gives the solution $V(\phi)$ up to two arbitrary constants, 
provided $f$ is smooth enough such that the Initial Value Problem associated to  \eqref{ODE} is well-posed.
In other words, it is possible to reconstruct the corresponding potential from
a given $f(s)$. 

The ODE \eqref{ODE} is equivalent to the system 
\begin{eqnarray}
&& \frac{ds}{d\phi}=-f(s),
\label{ds-dphi}\\
&& \frac{dV}{d\phi}=-s V.
\label{dV-dphi}
\end{eqnarray}
From this system are deduced the quadratures
\begin{eqnarray}
\phi(s)&=&\phi_0-\int_{s_0}^s \frac{1}{f(K)}\, dK,\label{quadphi}\\
V(s)&=&e^{\int_{s_0}^s \frac{K}{f(K)} \, dK} \bar{V}_0\label{quadV},
\end{eqnarray}
 where the integration constants satisfies $V(s_0)=\bar{V}_0$,
$\phi(s_0)=\phi_0$. The relations \eqref{quadphi} and
\eqref{quadV} are always valid and they provide the potential in an
implicit form. The integrability conditions for \eqref{quadphi} and \eqref{quadV} impose additional constraints to $f$. For the usual cosmological cases of table
\ref{fsform} the potential can be written explicitly, that is $V=V(\phi)$,
after elimination of $s$  between \eqref{quadphi} and \eqref{quadV}.

The direct derivation of the function $f(s)$ from a given $V(\phi)$, as
well as  the reconstruction procedure for obtaining $V(\phi)$ from a given
$f$-function is what we call ``Method of $f$-devisers''. In the section
\ref{dyn} we use this method for investigating  the phase portrait of
quintessence fields trapped in a Randall-Sundrum Braneworld in the case of
an anisotropic Bianchi I brane with a negative dark radiation term.

The method, as introduced here,  has the significant advantage that one can first perform the
analysis for arbitrary potentials, by considering general mathematical
conditions about $f(s),$ and then just substituting the desired forms,
instead
of repeating the whole procedure for every given potential. More importantly is that the method does not
depend on the cosmological scenario, since it is constructed on the scalar
field and its self-interacting potential, and can be generalized to several
scalar fields (for example to quintom models \citep{Leon2012}). The
disadvantages are that such a generalization is only possible if in the
model do not appear functions that contains mixed terms of several
scalar fields.  For example, cosmological models containing an
interaction term given by $V(\phi,\varphi)$ discussed by
\citet{Lazkoz2006} and by \citet{Lazkoz2007}, where $\phi, \varphi$ are
the scalar fields,  cannot be investigated using the method of
$f$-devisers. On the other hand, in the single field case the method cannot
be applied if the model contains more than one ``arbitrary'' function
depending on the scalar field, as for example a cosmological model
containing a potential $V(\phi)$ and coupling function $\chi(\phi)$, studied in the context of non-minimally coupled scalar field cosmologies
by \citet{Leon2009,Leon2010,leon2011a}. In such a case the more convenient
approach is to consider the scalar field itself as a dynamical variable.

\section{The model}\label{BianchiI}

In this paper we follow  \citep{Campos2001a,Goheer2003} where is given a
brief outline of the considerations that lead to the effective Einstein
equations for Bianchi I models. The metric in the brane is given by $ds^2
=-a_{0}^2 (t)dt^2 + \sum a_{i}(t)(dx^i)^2.$

The set up is as follows. Using the Gauss-Codacci equations, relating the
four and
five-dimensional spacetimes, we obtain the modified Einstein equations on the
brane \citep{Shiromizu2000,Sasaki2000}:
\be G_{ab}=-\Lambda_4g_{ab}+\kappa^2T_{ab}+\kappa_{(5)}^4S_{ab}-{\cal
E}_{ab}\label{2.3},
\ee
where $g_{ab}$ is the four-dimensional metric on the brane and
$G_{ab}$ is the Einstein tensor, $\kappa$ is the four-dimensional
gravitational
constant, and $\Lambda_4$ is the cosmological constant induced in the brane.
$S_{ab}$ are quadratic corrections in the matter variables. Finally, the
tensor ${\cal E}_{ab}$ is a  correction to the field equations on the brane 
coming from the extra dimension. This tensor is responsible for the so-called Dark
Radiation, this fact will be more evident in the following developments.

More precisely, ${\cal E}_{ab}$
are the components of the electric part of the 5D Weyl tensor of the bulk
projected on the brane (see \citep{Shiromizu2000} and the review by
\citet{Maartens2010} for more details). Following \citep{Goheer2003,Campos2001a,Maartens2000}, from the energy-momentum tensor conservation
equations ($\nabla_aT^a_b=0$) and equations \eqref{2.3} we get a constraint
on $S_{ab}$ and ${\cal E}_{ab}$:
\begin{equation}
\nabla^a\left({\cal E}_{ab}-\kappa_{(5)}^4S_{ab}\right)=0\label{new4}.
\end{equation}
In general we can decompose ${\cal E}_{ab}$ with respect to a chosen
4-velocity field $u^a$\citep{Maartens2000}
as:
\begin{equation}
 {\cal E}_{ab}=-\left(\frac{\kappa_{(5)}}{\kappa}\right)^4\left[{\cal
U}\left(u_a u_b+\frac{1}{3}h_{ab}\right)+{\cal P}_{ab}+2u_{(a}{\cal Q}_{b)}
\right],
\end{equation}
where 
\begin{equation}
 {\cal P}_{(ab)}={\cal P}_{ab},\;\;\;{\cal P}^a_a=0,\;\;\;{\cal
P}_{ab}u^b=0,\;\;\;{\cal Q}_a u^a=0,
\end{equation}
and
here the scalar component ${\cal U}$ is referred as the ``dark radiation''
energy density due to it has the same form as the energy-momentum tensor of a
radiation perfect fluid. ${\cal Q}_a$ is an spatial vector that corresponds
to an effective nonlocal energy flux on the brane and ${\cal P}_{ab}$ is an
spatial, symmetry and trace-free tensor which is an effective non local
anisotropic stress. Another important point is that the $9$ independent
component in the trace-free ${\cal E}_{ab}$ are reduced to $5$ degrees of
freedom by equation \eqref{new4} \citep{Maartens2010}, i.e., the
constraint equation \eqref{new4} provides evolution equations for ${\cal U}$
and ${\cal Q}_a$, but not for ${\cal P}_{ab}$. 

Taking into account the effective Einstein's equations \eqref{2.3}, the
consequence of having a Bianchi I model  on the brane is \citep{Maartens2001}
\begin{equation}
 {\cal Q}_a=0 \label{condi12}.
\end{equation}
Since we have no information about the dynamics of the tensor ${\cal P}_{ab}$
we assume \citep{Campos2001a,Goheer2003}:
\begin{equation}
 {\cal P}_{ab}=0,\label{condi13}
\end{equation}
this condition, together with \eqref{condi12} and \eqref{new4}, implies:
\begin{equation}
 D_a{\cal U}=0\Leftrightarrow{\cal U}={\cal U}(t).
\end{equation}

Using the above conditions over ${\cal Q}_a$ and ${\cal P}_{ab}$
\eqref{condi12}-\eqref{condi13}, setting the effective cosmological constant in
the brane to zero, i.e., $\Lambda_4=0$ \footnote{The induced cosmological
constant in the brane can be set to $\Lambda_4=0$ by fine tuning the negative
cosmological constant of the $AdS_5$ with the positive brane tension
$\lambda>0$  \citep{Brax2003,Maartens2010}.}, the effective Einstein equations
\eqref{2.3} for Bianchi I models (which have zero 3-curvature, i.e.,
$R^{(3)}=0$)
become:
\be H^{2}= \frac{1}{3}\rho_{T}\left(1+ \frac{\rho_{T}}{2\lambda}\right)+
\frac{1}{3}\sigma^2  + \frac{2{\cal U}}{\lambda},\label{2.4}
\ee
\be \dot{H}=- \frac{1}{2}\left(1+
\frac{\rho_T}{\lambda}\right)\left(\dot{\phi}^2 +\gamma\rho_m
\right)-\frac{4{\cal U}}{\lambda}-\sigma^2,\label{2.5}
\ee
\be \dot{\sigma}=- 3 H\sigma,\label{2.5b}
\ee
\be \dot{\rho}_m+3H(\rho_m+p_m)=0,\label{2.6}
\ee
\be \ddot\phi+3H\dot\phi+\partial_\phi V=0,\label{klein}
\ee
where $H$ is the Hubble factor, $\rho_m$ is the matter energy density, $\phi$ is an scalar field with self-interacting positive potential $V(\phi)$. It is convenient to relate the pressure of the background matter $p_m$
and its energy density $\rho_m$ by
\be p_m=(\gamma-1)\rho_m.
\ee
The parameter $\gamma$ is a constant parameter which is just the barotropic
index of the background matter density $\rho_m.$ $\sigma$ is a measure of the brane anisotropies and $\cal{U}$ denotes the dark radiation term. Finally, $\rho_T$ in the equation \eqref{2.4} denotes the total energy density on the brane (with tension $\lambda>0$), and is given by
${\rho_T}=\frac{1}{2}\dot\phi^2+V(\phi)+\rho_m,$ and it is positive definite. That is, we considered an scalar field density and a
background matter density as our Universe (brane) content, and we have
neglected the radiation in the total matter (that we recall is confined to
the brane).  We have used units in which $\kappa^2=8\pi G=1.$

The dark radiation term in \eqref{2.4}-\eqref{2.5} evolves as ${\cal
U}(t)=\frac{\mu}{a(t)^4},$ where $\mu$ is a constant
parameter \citep{Maartens2000}. From
the brane base formalism, $\mu$ is just an
integration constant that can take any sign (which actually could be
negative, zero or positive, as we will discuss next). However it was shown,
from the bulk based formalism, that for all possible homogeneous and
isotropic solutions on the brane, the bulk spacetime should be
Schwarzschild-AdS. In this case, it is possible to identify $\mu$ with the
mass of a black hole in the bulk ($m=\frac{2 \mu}{\lambda}$). From this fact,
$\mu$ should be positive \citep{Mukohyama2000,Maartens2010,Clifton2012}.
However, for anisotropic models such identification of $\mu$ with a mass-term
is not possible, i.e., $\mu$, and therefore ${\cal U}$, can take any sign
\citep{Campos2001a,Santos2001,Goheer2003}. As shown in Eq. \eqref{2.4},
dark radiation is strictly a correction to the Friedmann equation that scale as
radiation, but its origin is geometric and it has no interaction (e.g., Compton
scattering) with other matter fluids. The total radiation (including photons and neutrinos) that is confined
to the brane (which represents the universe) should be positive. However, it is possible to obtain relevant cosmological results if we allows for
the existence of, even a very tiny, negative dark radiation component.  Interesting, for  ${\cal
U}<0$ the expanding phase could experience the
cosmological turnaround triggered by 
the negativity of this dark radiation, and similarly the contracting
phase can leads to a cosmological bounce and then to expansion.

\section{The bounce and the turnaround}

Before proceeding to the detailed investigation of the model
\eqref{2.4}-\eqref{klein} using dynamical systems tools, let us discuss the
above system using a heuristic reasoning.
Integrating equation \eqref{2.5b} and \eqref{2.6} we obtain that 
$\sigma=\sigma_0 a^{-3}$ and $\rho_m=\rho_{m 0} a^{-3\gamma},$ while as
discussed in the introduction ${\cal U}=\mu a^{-4}.$
Let us examine the case of $\frac{\dot \phi^2}{2}\gg V(\phi),$ thus,
$\rho_\phi\approx\rho_{\phi 0} a^{-6},$ which implies $\rho_T=\rho_{\phi 0}
a^{-6}+\rho_{m 0} a^{-3\gamma}.$ 
Then equation \eqref{2.4} reduces to 
\begin{align}\label{2.4massless}
& 3 H^2=\frac{\rho_{m 0}^2 a^{-6 \gamma }}{2 \lambda }+\frac{\rho_{m 0}
   \rho_{\phi 0} a^{-3 \gamma -6}}{\lambda }+\frac{\rho_{ \phi
0}^2}{2\lambda }  a^{-12} +\nonumber \\ & \ \ \ \ \ \ \ \ \ \ \ \  +\left(\rho_{\phi
0}+\sigma_0^2\right)a^{-6}+\rho_{m 0} a^{-3 \gamma
   }-\frac{6|\mu| }{\lambda }a^{-4}. 
\end{align}

There are several cases depending of the values of $\gamma$. Suppose that
the universe lies currently in the usual expanding phase, i.e.
$H(a=a_0)>0,$ where $a_0$ denotes the scale of the universe at the present
time. We take as a reference universe's state the present expanding one. Thus
a future zero of $H$ will correspond to a turnaround, while a past one
corresponds to a bounce. 

\begin{itemize}
\item For $\gamma> 4/3$ the dark radiation term in equation
\eqref{2.4massless} falls less rapidly than the matter
terms, making the future turnaround inevitable for any initial negative
$\cal U$.
\item For $\gamma=4/3,$ the last two terms in the r.h.s. of
\eqref{2.4massless} are equally important in a low-energy regime,  and the
condition for the future turnaround is 
$$\rho_{m 0}<6 |\mu|/\lambda.$$ 
\item For $1<\gamma<\frac{4}{3},$ the leading terms in the r.h.s. of
\eqref{2.4massless} that are equally important are the last two terms
(since the terms scaling as $a^{-6}$ goes very fast to zero). As a result,
we obtain that the condition $H = 0$ leads to the equation: $$ \rho_{m 0} a^{4-3 \gamma }-\frac{6|\mu| }{\lambda }=0.$$
This equation has the positive root given by 
$$ a_+=\left[\frac{6|\mu| }{\lambda \rho_{m
0}}\right]^{\frac{1}{4-3\gamma}}.$$ 
If $a_+>a_0$ (the scale factor today) it
corresponds to the future of the Universe under consideration and indicates
the point of turnaround. If $a_+<a_0$, will describe the past and
corresponds to a bounce. 

\item For $\gamma=1,$ the leading terms in the r.h.s. of
\eqref{2.4massless} that are equally important are the first and the last
three terms.  As a result, we obtain that the condition $H = 0$ leads to
the cubic equation for the scale factor:\newline
$$ a^3-\frac{6  |\mu|}{\lambda \rho_{m 0}}a^2+\left[\frac{\rho_{m 0} }{2
\lambda }+\frac{\rho_{\phi 0}}{\rho_{m 0} }+\frac{\sigma_{0}^2}{\rho_{m 0}
}\right]=0.$$
Applying Descartes's rule, this equation has always one
negative root and two real positive or either complex conjugated roots. The
condition for the existence of two positive roots is:
$$ \rho_{m 0}^4+2\lambda \rho_{m 0}^3(\rho_{\phi 0}+\sigma_0^2)<\frac{64 
|\mu|^3}{\lambda^2}.$$
This expression reduces to the analogous expression
in  \citep{Toporensky2005} for the case of a perfect fluid under the
assumption $\rho_{\phi 0}=\sigma_0=0.$ A root which is bigger than $a_0$
(the scale factor today) corresponds to the future of the Universe under
consideration and indicates the point of turnaround. A root which is less
than $a_0$, will describe the past and corresponds to a bounce. 

\item For $\frac{2}{3}<\gamma<1,$ the leading terms in the r.h.s. of
\eqref{2.4massless} that are  equally important are the first and the last
two terms. As a result, we obtain that the condition $H = 0$ leads to the
equation:
$$\displaystyle -\frac{6 \mu  a^{6 \gamma -4}}{\lambda }+\rho_{m 0} a^{3 \gamma
}+\frac{\rho_{m 0}^2}{2 \lambda }=0.$$
This equation must be solved
numerically looking for real positive roots.  A root which is bigger than
$a_0$ (the scale factor today) corresponds to the future of the Universe
under consideration and indicates the point of turnaround. A root which is
less than $a_0$, will describe the past and corresponds to a bounce. 

\item For $\gamma=\frac{2}{3},$ the leading terms in the r.h.s. of
\eqref{2.4massless} that are equally important are the first and the last
two terms. As a result, we obtain that the condition $H = 0$ leads to the
quadratic equation for the scale factor:
$$a^{2
   }+\left(\frac{\rho_{m 0} }{2 \lambda }-\frac{6|\mu| }{\lambda \rho_{m 0}
}\right)=0.$$
The condition for the existence of a positive root,
$$a_{+}=\sqrt{\left(\frac{\rho_{m
0} }{2 \lambda }-\frac{6|\mu| }{\lambda \rho_{m 0} }\right)},$$
is $\rho_{m
0}^2<12 |\mu|.$
 If $a_+>a_0$ (the scale factor today) corresponds to the
future of the Universe under consideration and indicates the point of
turnaround. If  $a_+ < a_0$, it describes the
past and corresponds to a bounce. 

\item For $\gamma < 2/3$, even the term $\propto a^{-6 \gamma },$  falls
less rapidly than ${\cal U}$, and the turnaround in the future becomes
impossible.
\end{itemize}

Let us examine the case of $\frac{\dot \phi^2}{2}\ll V(\phi),$ thus,
$\rho_\phi\approx V(\phi(a)).$ Then, $\rho_T=V(\phi(a))+\rho_{m 0}
a^{-3\gamma}.$ 
Then equation \eqref{2.4} reduces to 
\begin{align}\label{2.4const}
&3 H^2=V \left[1+\frac{ V}{2\lambda }\right]+\frac{\rho_{m 0} ^2 a^{-6
\gamma }}{2 \lambda }+\frac{\sigma_0^2}{a^6}+\nonumber \\ & \ \ \ \ \ \ \ \ \ \ \ \ \ \ \ \ \ \ \ \ \ \ \ \ \ \ \ \  +\rho_{m 0}\left[1+\frac{
V}{\lambda }\right] a^{-3 \gamma }-\frac{6 |\mu| }{a^4 \lambda }.
\end{align}
Let consider the simpler case of a potential that tends asymptotically to a
positive  constant $V=V_0>0.$ 
\begin{itemize}
\item For $\gamma>\frac{4}{3},$ the leading terms in  the r.h.s. of 
\eqref{2.4const} are the first and the last one. The condition $H = 0$
leads to the quartic equation for the scale factor:
$$V_0 \left[1+\frac{
V_0}{2\lambda }\right] a^{4}  -\frac{6 |\mu| }{\lambda }=0.$$
This equation
has one negative solution, two purely imaginary solutions, and a positive
one given by:
$$a_+=\frac{\sqrt{2} \sqrt[4]{3} \sqrt[4]{\mu }}{\sqrt[4]{V_0
(2
\lambda +V_0)}}.$$ If  $a_+>a_0$ (the scale factor today), it corresponds to
the future of the Universe under consideration and indicates the point of
turnaround. If $a_+<a_0$, it describes the past and corresponds to a
bounce. 

\item For $\gamma=\frac{4}{3},$ the leading terms in  the r.h.s. of 
\eqref{2.4const} are the first and the last two terms. The condition $H =
0$ leads to the quartic equation for the scale factor:
$$ V_0 \left[1+\frac{
V_0}{2\lambda }\right] a^{4} +\rho_{m 0}\left[1+\frac{ V_0}{\lambda
}\right] -\frac{6 |\mu| }{\lambda }=0.$$
The condition for the future
collapse is:
$$ \rho_{m 0}\left[1+\frac{ V_0}{\lambda }\right] <\frac{6 |\mu|
}{\lambda }.$$
 
\item  For $1<\gamma<\frac{4}{3},$ the leading terms in  the r.h.s. of 
\eqref{2.4const} are the first and the last two terms. The condition $H =
0$ leads to the equation for the scale factor:
$$V_0 \left[1+\frac{
V_0}{2\lambda }\right]a^4+\rho_{m 0}\left[1+\frac{ V_0}{\lambda }\right]
a^{4-3
\gamma }-\frac{6 |\mu| }{\lambda }=0.$$
This equation must be solved
numerically, looking for real positive roots.  A root which is bigger than
$a_0$ (the scale factor today) corresponds to the future of the Universe
under consideration and indicates the point of turnaround. A root which is
less than $a_0$, will describe the past and corresponds to a bounce. 

\item For $\gamma=1,$	the leading terms in  the r.h.s. of 
\eqref{2.4const} are the first and the last two terms. The condition $H =
0$ leads to the quartic equation for the scale factor:
$$ V_0 \left[1+\frac{
V_0}{2\lambda }\right] a^{4} +\rho_{m 0}\left[1+\frac{ V_0}{\lambda
}\right] a -\frac{6 |\mu| }{\lambda }=0.$$
 Applying Descartes's rule this
equation has one positive root, one negative root and two complex
conjugated roots. If the positive root is
bigger than $a_0$ (the scale factor today), it corresponds to the future of
the Universe under consideration
and indicates the point of turnaround. If  it is less than $a_0$, it
describes the past and corresponds to a bounce. 

\item For $\frac{2}{3}< \gamma<1,$ all the terms in  the r.h.s. of 
\eqref{2.4const} but the third, are relevant.  The condition $H = 0$ leads
to the equation for the scale factor:
\begin{align*}
& -\frac{6 \mu  a^{6 \gamma
-4}}{\lambda }+\frac{\rho_{m 0} a^{3 \gamma } (\lambda +V_0)}{\lambda
}+\\ &\ \ \ \ \ \ \ \ \ \ \ \ \ \ \ \ \ \ \ \ \ \ \ \ \ \ \ \ +\frac{V_0 a^{6 \gamma } (2 \lambda +V_0)}{2 \lambda } +\frac{\rho_{m
0}^2}{2 \lambda }=0.
\end{align*}
This equation must be solved numerically looking for
real positive roots.  A root which is bigger than $a_0$ (the scale factor
today) corresponds to the future of the Universe under consideration and
indicates the point of turnaround. A root which is less than $a_0$, will
describe the past and corresponds to a bounce. 

\item For $\gamma=\frac{2}{3},$ all the terms in  the r.h.s. of 
\eqref{2.4const} but the third, are relevant. The condition $H = 0$ leads
to the quartic equation for the scale factor:
$$V_0 \left[1+\frac{
V_0}{2\lambda }\right] a^{4} +\rho_{m 0}\left[1+\frac{ V_0}{\lambda
}\right] a^2 +\frac{\rho_{m 0} ^2}{2 \lambda } -\frac{6 |\mu| }{\lambda
}=0.$$
For $\mu >\frac{\rho_{m 0}^2}{12}$ there is one positive solution
given by: 
{\small $$a_+=\sqrt{\frac{\sqrt{\lambda ^2 \rho_{m 0}^2+12 \mu  V_0 (2
\lambda +V_0)}-\rho_{m 0} (\lambda +V_0)}{V_0 (2 \lambda +V_0)}},$$}
one negative solution given by:
{\small $$a_-=-\sqrt{\frac{\sqrt{\lambda ^2 \rho_{m 0}^2+12 \mu 
V_0 (2 \lambda +V_0)}-\rho_{m 0} (\lambda +V_0)}{V_0 (2 \lambda +V_0)}},$$}
and two complex conjugated ones. Otherwise the four are complex conjugated.
Thus, the condition for the bounce or the collapse is  $\mu >\frac{\rho_{m
0}^2}{12}.$ If the positive root is
bigger than $a_0$ (the scale factor today), it corresponds to the future of
the Universe under consideration
and indicates the point of turnaround. If  it is less than $a_0$, it
describes the past and corresponds to a bounce.
\item For $\gamma < 2/3$, even the term $\propto a^{-6 \gamma },$  falls
less rapidly than ${\cal U}$, and the turnaround in the future becomes
impossible. 
\end{itemize}

Summarizing, using a heuristic reasoning we have obtained several
conditions for the existence of  bounces and turnaround of cosmological
solutions for a massless scalar field and a scalar field with an 
asymptotically constant potential. For the analysis of more general potentials we
submit the reader to the reference \citep{Toporensky2003} where the
authors
discuss conditions for bounces and they provide details of chaotic behavior
in the isotropic brane model containing an scalar field for a negative dark
radiation ${\cal U}<0$.

\section{Asymptotic behavior}\label{dyn}

In this section we recast the equations
\eqref{2.4}-\eqref{klein} as an autonomous system 
\citep{Copeland1998,coley2003dynamical,Tavakol1997,Leon2010,leon2011a}. To avoid ambiguities with the non-compactness at
infinity we can define the compact variables allowing to describe both expanding and
collapsing models
\citep{Campos2001a,Dunsby2004,Goheer2004,Leach2006,Solomons2006,Goheer2007,Goheer2008,Leon2011}:
\begin{eqnarray} 
Q=\frac{H}{D},\,
x=\frac{\dot{\phi}}{\sqrt{6}D}, y= \frac{V}{3H^2}, \Omega_\lambda =
\frac{\rho_{T}^2}{6\lambda D^2},\nonumber\\
\Omega_m =\frac{\rho_m}{3D^2}, \Sigma =\frac{\sigma}{\sqrt{3}D},  \Omega_U
= -\frac{2{\cal U}}{\lambda D^2} \label{var22},
\end{eqnarray} 
where $D=\sqrt{H^2-\frac{2 {\cal U}}{\lambda}}.$

For the scalar potential treatment, we proceed following the method
introduced in section \ref{newmethod}.

Using the Friedmann equation \eqref{2.5} we obtain the following relation
between the variables  \eqref{var22}
\be x^2 +y +\Omega_m +\Omega_{\lambda}+\Sigma^2=1.\label{e2.60}
\ee
The restriction \eqref{e2.60} allows to forget about one of the dynamical
variables, e.g., $y,$  to obtaining a reduced dynamical system. From the
condition $0\leq y \leq1$ we have the following inequality
\be 0\leq x^2  +\Omega_m +\Omega_{\lambda}+\Sigma^2 \leq1.\label{e2.61}
\ee
The definition of $D$ leads to the additional restriction
\be Q^2+ \Omega_{U}=1\label{e2.61b}.
\ee
Thus, restriction \eqref{e2.61b} allows to eliminate another degree of
freedom, namely the variable $\Omega_{U}.$ 

Using the variables \eqref{var22}, the field equations
\eqref{2.4}-\eqref{klein} and the new time variable $d\tau=D dt,$  we
obtain the following autonomous system of ordinary differential equations
(ODE):
\begin{align}
& Q'= \frac{1}{2} \left(Q^2-1\right) \left(3 \gamma  \Omega_m+6
   \Sigma ^2+6 x^2-4+3 \Xi\right),\label{eqQ}
\\
 &x '= \frac{3}{2} Q x \left(\gamma \Omega_m+2
   \Sigma ^2+2 x^2-2+\Xi\right)+\nonumber \\ & +\sqrt{6} s \left(1-x^2-\Omega_m-\Omega_\lambda-
\Sigma^2\right),\label{eqx}
\\
&\Omega_m '=3 Q \Omega_m \left(\gamma  (\Omega_m-1)+2 \Sigma ^2+2
x^2+\Xi\right),\label{eqm}
\\
&\Omega_\lambda '=3 Q \Omega_\lambda  \left(\gamma  \Omega_m+2 \Sigma ^2+2
x^2\right)+3 \Xi  Q (\Omega_\lambda -1), \label{eqlambda}
\\
&\Sigma '=\frac{3}{2} Q \Sigma  \left(\gamma  \Omega_m+2 \Sigma ^2+2
x^2-2+\Xi\right),\label{eqsigma}
\\
& s'=-\sqrt{6} x f(s)\label{eqs}.
\end{align}
Where the comma denotes derivatives with respect to $\tau$, and 
\be \Xi \equiv \frac{\rho_{T}}{\lambda}  \left(\gamma 
   \Omega_m+2 x^2\right),\label{Xi}\ee 
	From  \eqref{e2.60} follows the useful relationship 
\be
\frac{\rho_{T}}{\lambda}=\frac{2\Omega_\lambda}{x^2+y+\Omega_m}=\frac{
2\Omega_{\lambda}}{1-\Omega_{\lambda}-\Sigma^2}.\label{e2.63}
\ee

It is easy to see from \eqref{e2.63} and the definition \eqref{Xi} that 
\be \Xi = \frac{2\Omega_{\lambda} \left(\gamma 
   \Omega_m+2 x^2\right)}{1-\Omega_{\lambda}-\Sigma^2}.\ee

From \eqref{e2.63} follows that the region:
\begin{equation*}
1-\Omega_{\lambda}-\Sigma^2\equiv \Omega_{m}+x^2 +y=0
\end{equation*}
corresponds to
cosmological solutions where $\rho_T \gg \lambda$ (corresponding to the
formal limit $\lambda\rightarrow 0$). Therefore, they are associate to high
energy regions, i.e., to cosmological solutions in a neighborhood of the
initial singularity. We submit the reader to the references \citep{Foster1998,Leon2009} for a classical treatment of cosmological solutions near the
initial singularity in FRW cosmologies. Due to its classic nature, our
model is not appropriate to describing the dynamics near the initial
singularity, where quantum effects appear. However, from the mathematical
viewpoint, this region
($\Omega_{\lambda}+\Sigma^2=1$) is reached asymptotically. In fact, as some
numerical integrations corroborate, there exists an open set of orbits in
the interior of the phase space that tends to the boundary $\Omega_{\lambda}+\Sigma^2=1$
as $\tau\rightarrow-\infty$. Therefore, for mathematical motivations it is
common to attach the boundary $\Omega_{\lambda}+\Sigma^2=1$ to the phase
space.  On the other hand the points with
($\Omega_\lambda=0$) are associated to the standard 4D behavior ($\rho_T
\ll \lambda$ or $\lambda\rightarrow\infty$) and corresponds to the low
energy regime.

From definition \eqref{var22} and from the restriction \eqref{e2.60}, and
taking into account the previous statements, it is enough to investigate to
the flow of \eqref{eqx}-\eqref{eqs} defined in the phase space
\begin{align} & \Psi=\{(Q,x,\Omega_{m},\Omega_{\lambda},\Sigma): 0\leq x^2 
+\Omega_m+\Omega_{\lambda}+\Sigma^2 \leq1, \nonumber\\ 
& -1\leq Q\leq1,  -1\leq x\leq1,  0\leq\Omega_{m}\leq1,
0\leq\Omega_{\lambda}\leq1, \nonumber\\ 
& -1\leq\Sigma\leq1\}\times\left\{s\in\mathbb{R}
\right\}\label{reg}.
\end{align}

The system \eqref{eqx}-\eqref{eqs} admits eighteen classes of\\
(curves of) fixed points corresponding to static solutions ($H=Q=0$). They are displayed in table \ref{tab2a} where we have defined:
\begin{align}
	&f_{1}(\gamma,\Omega_\lambda, \Sigma)=\frac{2 \left(3 \Sigma ^2-2\right)\left(\Omega_\lambda +\Sigma ^2-1\right)}{3 \gamma  \left(\Omega_\lambda-\Sigma ^2+1\right)}\label{f1},\\
  &f_2(\Omega_\lambda, \Sigma)=\frac{4-6 \Sigma ^2}{-3\Sigma ^2+3 \Omega_\lambda +3},\\
  &f_3(\gamma,\Sigma)=2 \left(1-\Sigma^2\right)-\frac{\frac{4}{3}-2 \Sigma ^2}{\gamma },\\
  &f_4(\gamma,\Sigma)=\frac{\frac{4}{3}-2 \Sigma ^2}{\gamma }+\Sigma ^2-1,\\
  &f_5(\Sigma)=\frac{3 \Sigma ^2-2}{3 \Sigma ^2-3},
	\end{align}
	\begin{align}
	&f_6(\gamma,x,\Sigma)=\frac{-6 \Sigma ^2+6 x^2-\sqrt{4 \left(3 (\gamma -1)\Sigma ^2-3 \gamma +2\right)^2+9 (\gamma -2)^2 x^4+12 (\gamma -2) x^2\left(3 (\gamma +1) \Sigma ^2-3 \gamma -2\right)}+4}{6 \gamma },\\
  &f_7(\gamma,x)=\frac{6 x^2+\sqrt{4 (2-3 \gamma )^2+9 (\gamma -2)^2 x^4+12 ((4-3 \gamma ) \gamma +4) x^2}+4}{6 \gamma }\label{f7}.
	\end{align}

The eigenvalues of the linear
perturbation matrix evaluated at each of these critical points are
displayed in the table \ref{autovalores1a}. 
In the appendix \ref{static} are discussed the stability conditions and physical
interpretation for these static solutions. These kind of solutions are important
as intermediate
stages in the evolution of the universe allowing the transition from
expanding to contracting models. 

\begin{table*}
\caption{ Critical points of the system 
\eqref{eqQ}-\eqref{eqs} representing static solutions and  their existence
conditions. We have defined $s^*$ for an $s$-value such that $f(s^*)=0$. The definitions for the $f_{i}$'s are given in \eqref{f1}-\eqref{f7}.}
\label{tab2a}
\begin{center}
{\scriptsize
\begin{tabular}{@{\hspace{4pt}}c@{\hspace{10pt}}c@{\hspace{10pt}}c@{\hspace
{10pt}}c@{\hspace{10pt}}c@{\hspace{10pt}}c@{\hspace{10pt}}c@{\hspace{10pt}}
c}
\hline
\hline\\[0.1cm]
Label &$Q$&$ {x}$& $ {\Omega}_m$& $ {\Omega}_\lambda$& $ {\Sigma}$& $s $&
Existence \\[0.1cm]
\hline\\[0.1cm]
$E_{1}^\pm$&$0$& $0$&$0$ &$\Omega_{\lambda c}\in
\left[0,\frac{1}{3}\right]$ & $\pm\sqrt{\frac{2}{3}}$& $0$& $0<\gamma\leq
2.$ \\[0.1cm]
\hline\\[0.1cm]
$E_2$&$0$& $0$&$0$ &$\cos^2 u$ & $\sin u$& $0$& $\Omega_{\lambda c}\notin
\left\{\pm 1, \pm\sqrt{\frac{2}{3}}\right\},$  \\[0.1cm]
				& & & & & & & $0<\gamma\leq 2, 0\leq u \leq
2\pi .$ \\[0.1cm]
\hline\\[0.1cm]
$E_3$&$0$& $0$&$f_1(\gamma,\Omega_{\lambda_c}, \Sigma_c)$ &$\Omega_{\lambda
c}\in \left[ 0,1-\Sigma_c^2\right)$ &
$\Sigma_c\in\left(-\sqrt{\frac{2}{3}},\sqrt{\frac{2}{3}}\right)$& $0$&
$f_2(\Omega_{\lambda_c}, \Sigma_c)\leq\gamma\leq 2.$ \\[0.1cm]
\hline\\[0.1cm]
$E_4$&$0$& $0$&$f_3(\gamma,\Sigma_c)$ &$f_4(\gamma,\Sigma_c)$ & $\Sigma_c$&
$s^*$& $-1\leq \Sigma_c \leq -\sqrt{\frac{2}{3}},$ 
 \\[0.2cm]
				& & & & & & & $0<\gamma \leq 2$ or
\\[0.2cm]
& & & & & & & $-\sqrt{\frac{2}{3}}<\Sigma_c <\sqrt{\frac{2}{3}},$  \\[0.1cm]
				& & & & & & &
$f_5(\Sigma_c)\leq \gamma \leq 2$ or \\[0.1cm]
& & & & & & & $\sqrt{\frac{2}{3}}\leq \Sigma_c \leq 1, 0<\gamma \leq 2.$
\\[0.1cm]
\hline\\[0.1cm]
$E_5$&$0$& $x_c$&$1-\Sigma_c ^2-{x_c^2}-\Omega_{\lambda c}$
&$-\frac{x_c^2}{2}+f_6(\gamma,x_c,\Sigma_c)$ & $\Sigma_c$& $s^*$&
$x_c^2+\Sigma_c ^2<\frac{2}{3},$  \\[0.2cm]
				& & & & & & & $0<\gamma \leq 2-\frac{2}{3 \left(1-\Sigma_c
^2-x_c^2\right)}.$ \\[0.1cm]
\hline\\[0.1cm]
$E_6^\pm$&$0$& $\pm\sqrt{\frac{2}{3}-\Sigma_c^2}$&$0$ &$0$ & $\Sigma_c$&
$0$& $-\sqrt{\frac{2}{3}}<\Sigma_c<\sqrt{\frac{2}{3}},$  \\[0.1cm]
				& & & & & & & $ 0 \leq \gamma \leq
2.$ \\[0.2cm]\hline\\[0.1cm]
$E_7^\pm$&$0$& $\pm\frac{\Sigma_c}{\sqrt{3}}$&$0$ &$1-2\Sigma_c^2$ &
$\Sigma_c$& $0$& $\Sigma_c^2<\frac{1}{2}, \Sigma_c\neq 0,$  \\[0.1cm]
				& & & & & & & $ 0 \leq \gamma
\leq 2.$ \\[0.1cm]
\hline\\[0.1cm]
$E_8$&$0$& $x_c$&$1-x_c^2-\Omega_{\lambda c}$
&$-\frac{x_c^2}{2}+f_7(\gamma,x_c)$ & $0$& $s^*$& $x_c^2<{\frac{2}{3}},$  \\[0.1cm]
				& & & & & & & $
0<\gamma \leq \frac{4-6 x_c^2}{3-3 x_c^2}.$ \\[0.1cm]
\hline\\[0.1cm]
$E_{9}$&$0$& $0$&$\frac{4 (1-\Omega_{\lambda c} )}{3 \gamma 
(\Omega_{\lambda c} +1)}$ &$\Omega_{\lambda c}$ & $0$& $0$& $0<\gamma \leq
\frac{2}{3}, \Omega_{\lambda c} =1$ or\\[0.1cm]
        & & & & & & & $\frac{2}{3}<\gamma \leq \frac{4}{3},$  \\[0.1cm]
				& & & & & & & $ \frac{4}{3
\gamma }-1\leq \Omega_{\lambda c} \leq 1,$  \\[0.1cm]
				& & & & & & & or $\frac{4}{3}<\gamma \leq 2,$  \\[0.1cm]
				& & & & & & & $
0\leq \Omega_{\lambda c} \leq 1.$ \\[0.1cm]
\hline\\[0.1cm]
$E_{10}^\pm$&$0$&
$\pm\sqrt{\frac{2-3\gamma}{3(2-\gamma)}}\Sigma_c$&$\frac{4\Sigma_c^2}{
3(2-\gamma)}$ &$1-2\Sigma_c^2$ & $\Sigma_c$& $s^*$& $\gamma \neq
\frac{2}{3}, \Sigma_c =0$  \\[0.1cm]
        & & & & & & & or$0\leq\gamma<\frac{2}{3}, $  \\[0.1cm]
				& & & & & & & $ 0<\Sigma_c^2\leq
\frac{1}{2}.$ \\[0.1cm]
\hline\\[0.1cm]
$E_{11}$&$0$& $0$&$\frac{2\Sigma_c^2}{3\gamma}$ &$1-2\Sigma_c^2$ &
$\Sigma_c$& $0$& $0<\gamma< \frac{2}{3}, $  \\[0.1cm]
				& & & & & & & $ \Sigma_c =0$ or  \\[0.1cm]
 & & & & & & & $\frac{2}{3}\leq \gamma \leq 2, $  \\[0.1cm]
				& & & & & & & $\Sigma_c^2\leq \frac{1}{2}.$
\\[0.1cm]
\hline\\[0.1cm]
$E_{12}$&$0$& $0$&$\frac{2(2-3\Sigma_c^2)}{3\gamma}$ &$0$ & $\Sigma_c$&
$0$& $0<\gamma< \frac{4}{3},$  \\[0.2cm]
				& & & & & & & $ 1-\frac{2}{3(2-\gamma)}\leq \Sigma_c^2\leq
\frac{2}{3}$  \\[0.1cm]
				& & & & & & & or $\frac{4}{3}\leq \gamma \leq 2,$  \\[0.1cm]
				& & & & & & & $ \Sigma_c^2\leq
\frac{2}{3}.$ \\[0.1cm]
\hline\\[0.1cm]
$E_{13}^\pm$&$0$&
$\pm\sqrt{\frac{4-3\gamma}{3(2-\gamma)}}$&$\frac{2}{3(2-\gamma)}$ &$0$ &
$0$& $s^*$& $0<\gamma\leq \frac{4}{3}.$ 
\\[0.2cm]\hline \hline
\end{tabular}}\end{center}
\end{table*}

Additionally, the  system \eqref{eqx}-\eqref{eqs} admits twenty four classes of (curves
of) fixed points associated to expanding (contracting solutions). Its
coordinates in the phase space are given in table \ref{tab2}. Note that for
 $U=0$, the models studied in \citep{Escobar2012} are recovered.
In the Appendix \ref{nonstatic} are discussed the stability conditions for the
expanding (contracting) solutions. 

As we commented previously, our model is not applicable near the initial
singularity (that is at the singular surface $1- {\Omega}_\lambda- \Sigma
^2=0$).
However, in principle we can apply a similar approach as in 
\citep{Escobar2012a} to investigate the dynamics on the singular surface.
We submit the reader to the Appendix \ref{section3.2} for further details. 

Summarizing the system \eqref{eqx}-\eqref{eqs} admits forty two classes of
fixed points. For that reason this scenario has a rich cosmological
behavior, including the transition from contracting to expanding solutions
and viceversa. Also is possible  the existence of bouncing solutions and a
turnaround, and even cyclic behavior. 

The possible late-time (stable) attractors are:
\begin{itemize}
\item $Q_4^+(s^*)$ for 
$0<\gamma
<\frac{4}{3},s^*<-\sqrt{3\gamma },f'\left(s^*\right)<0$ 
or $0<\gamma
<\frac{4}{3},s^*>\sqrt{3\gamma },f'\left(s^*\right)>0;$
\item $Q_5^+(s^*)$ for $0<\gamma \leq \frac{4}{3},-\sqrt{3
\gamma }<s^*<0,f'\left(s^*\right)<0$ or $\frac{4}{3}<\gamma \leq
2,-2<s^*<0,f'\left(s^*\right)<0$ or $0<\gamma \leq
\frac{4}{3},0<s^*<\sqrt{3 \gamma },f'\left(s^*\right)>0$ or
$\frac{4}{3}<\gamma \leq 2,0<s^*<2,f'\left(s^*\right)>0;$ 
\item $Q_9^+$ and the line $Q_{10}^+$  for $f(0)\geq 0;$ \footnote{However, if we include the variable $y=\frac{V}{3 H^2}$ in the analysis,
the solutions associated to the line of fixed points $Q_{10}^+$ are
unstable to perturbations along the $y$-direction.}
\item $Q_{11}^-$ for $\gamma>1.$
\item $Q_{12}^+(s^*)$ for $\frac{4}{3}<\gamma \leq 2, s^*>2,
f'\left(s^*\right)>0;$
\end{itemize}
whereas, the possible past (unstable) attractors are:  
\begin{itemize}
\item $Q_4^-(s^*)$, a source for $0<\gamma
<\frac{4}{3},s^*<-\sqrt{3\gamma },f'\left(s^*\right)<0$ or $0<\gamma
<\frac{4}{3},s^*>\sqrt{3\gamma },f'\left(s^*\right)>0;$
\item $Q_5^-(s^*)$, a source for $0<\gamma \leq
\frac{4}{3},-\sqrt{3 \gamma }<s^*<0,f'\left(s^*\right)<0$ or
$\frac{4}{3}<\gamma \leq 2,-2<s^*<0,f'\left(s^*\right)<0$ or $0<\gamma \leq
\frac{4}{3},0<s^*<\sqrt{3 \gamma },f'\left(s^*\right)>0$ or
$\frac{4}{3}<\gamma \leq 2,0<s^*<2,f'\left(s^*\right)>0;$ 
\item $Q_9^-$ and the line $Q_{10}^-$ for $f(0)\geq 0.$
\item $Q_{11}^+$ for $\gamma>1.$
\item $Q_{12}^-(s^*)$, a source for $\frac{4}{3}<\gamma \leq 2, s^*>2,
f'\left(s^*\right)>0.$
\end{itemize}

\begin{table*}
\begin{center}
\caption{Critical points of the system 
\eqref{eqQ}-\eqref{eqs} associated to expanding (contracting) solutions and
 their existence conditions. We use the notation $s^*$ for an $s$-value
such that $f(s^*)=0$ and $s_c$ for an arbitrary $s$-value.}
\label{tab2}
{\begin{tabular}{@{\hspace{4pt}}c@{\hspace{10pt}}c@{\hspace{10pt}}c@{
\hspace{10pt}}c@{\hspace{10pt}}c@{\hspace{10pt}}c@{\hspace{10pt}}c@{\hspace
{10pt}}c}
\hline
\hline\\[-0.3cm]
Label &$Q$&$ {x}$& $ {\Omega}_m$& $ {\Omega}_\lambda$& $ {\Sigma}$& $s $&
Existence \\[0.1cm]
\hline\\[-0.2cm]
$Q_{1}^\pm$&$\pm1$& $0$ &$1$ &$0$& $0$& $s_c\in\mathbb{R}$& always\\[0.2cm]
$Q_{2}^\pm(s^*)$&$\pm 1$& $1$&$0$ &$0$ & $0$& $s^*$&always\\[0.2cm]
$Q_{3}^\pm(s^*)$&$\pm 1$& $-1$&$0$ &$0$ & $0$& $s^*$&always\\[0.2cm]
$Q_{4}^\pm(s^*)$&$\pm 1$& $\pm\sqrt{\frac{3}{2}}\frac{\gamma}{s^*}$
&$1-\frac{3\gamma}{s^{*2}}$ &$0$& $0$& $s^*$&$s^{*2}\geq 3 \gamma$\\[0.2cm]
$Q_{5}^\pm(s^*)$&$\pm1$& $\pm\frac{s^*}{\sqrt{6}}$&$0$ &$0$ & $0$& $s^*$&$
s^{*2}\leq 6$\\[0.2cm]
$Q_{6}^\pm$&$\pm 1$& $0$&$0$ & $0$ & $-1$& $s_c\in\mathbb{R}$&
always\\[0.2cm]
$Q_{7}^\pm$&$\pm 1$& $0$&$0$ & $0$ & $ 1$& $s_c\in\mathbb{R}$& always
\\[0.2cm]
$Q_{8}^\pm(s^*)$&$\pm 1$& $\cos u$&$0$ & $0$ & $\sin u$& $s^*$&
always\\[0.2cm]
$Q_{9}^\pm$&$\pm1$& $0$&$0$&$0$ & $0$& $0$& always\\[0.2cm]
$Q_{10}^\pm$&$\pm1$& $0$& $0$ &$\Omega_{\lambda c}\in(0,1)$ & $0$& $0$&
always\\[0.2cm]
$Q_{11}^\pm$&$\pm 1$& $0$ & $0$ &$1$& $0$& $s_c\in\mathbb{R}$&
always\\[0.2cm]
$Q_{12}^\pm(s^*)$&$\pm\frac{s^*}{2}$& $\pm\sqrt{\frac{2}{3}}$&$0$ &$0$ &
$0$& $s^*$&$ s^{*2}\leq 4$
\\[0.4cm]\hline \hline
\end{tabular}}\end{center}
\end{table*}

\begin{table*}\begin{center}
\caption{Values of the observable parameters $\Omega_{\phi},$
$\omega_{\phi}$, $\omega_{\text{eff}}$, and $q$ evaluated at the critical
points of the system
\eqref{eqQ}-\eqref{eqs} associated to expanding (contracting) solutions. We use the notations $s^*$ for an $s$-value such
that $f(s^*)=0.$}
\label{tab1b} 
\begin{tabular}{@{\hspace{4pt}}c@{\hspace{10pt}}c@{\hspace{10pt}}c@{\hspace
{10pt}}c@{\hspace{10pt}}c}
\hline
\hline\\[-0.3cm]
$P_i$&  $\Omega_\phi$& $\omega_\phi$& $\omega_{\text{eff}}$ & $q$\\[0.1cm]
\hline\\[-0.2cm]
$Q_{1}^\pm$ & 0 & {arbitrary} & $\gamma$& $\frac{3\gamma-2}{2}$\\[0.2cm]
$Q_{2}^\pm(s^*), Q_3^\pm(s^*)$& $1$& $1$& $1$ & $2$\\[0.2cm]
$Q_{4}^\pm(s^*)$& $\frac{3 \gamma }{{s^*}^2}$&$\gamma -1$
&$\gamma -\frac{3 \gamma }{{s^*}^2}$&$\frac{3 \gamma }{2}-1$\\[0.2cm]
$Q_{5}^\pm(s^*)$&$1$ &$\frac{1}{3} \left({s^*}^2-3\right)$&$\frac{1}{3}
\left({s^*}^2-3\right)$&
$\frac{1}{2} \left({s^*}^2-2\right)$\\[0.2cm]
$Q_{6}^\pm$&  $0$&arbitrary& arbitrary&$2$\\[0.2cm]
$Q_{7}^\pm$&  $0$& arbitrary& arbitrary&$2$\\[0.2cm]
$Q_{8}^\pm(s^*)$& $\cos ^2(u)$&$1$&$1$&$2$\\[0.2cm]
$Q_{9}^\pm$& $1$& $-1$&  $-1$ &  $-1$\\[0.2cm]
$Q_{10}^\pm$&$1-\Omega_\lambda$ &$-1$&$-1$&$-1$\\[0.2cm]
$Q_{11}^\pm$&  $0$&arbitrary& arbitrary&$-1$\\[0.2cm]
$Q_{12}^\pm(s^*)$& $\frac{4}{{s^*}^2}$ &$\frac{1}{3}$ &$\frac{1}{3}$&
$1$\\[0.4cm]\hline \hline
\end{tabular}\end{center}
\end{table*}

\section{Physical discussion}\label{discussion}

In order to characterize the cosmological solutions associated to the
singular points it will be helpful to define some observational
parameters in terms of the state variables, namely: the dimensionless
dark energy energy density parameter
$\Omega_\phi$ (which is just the scalar field one):
\be \Omega_\phi\equiv\frac{\rho_{\phi}}{3H^2}=\frac{1- {\Omega}_\lambda- {\Omega}_m-\Sigma ^2}{Q^2}, 
\ee
the dark energy equation of state (EoS) parameter:
\be
\omega_{\phi}\equiv \frac{p_{\phi}}{\rho_{\phi}}=-1+\frac{2 {x}^2}
{1- {\Omega}_\lambda- {\Omega}_m- \Sigma ^2},
\ee
the \textit{total} equation state parameter
\be \omega_{\text{eff}}\equiv \frac{p_T}{\rho_T}=\frac{2 x^2+(\gamma+1) 
\Omega_m+\Omega_\lambda +\Sigma ^2-1}
{1-\Omega_\lambda-\Sigma ^2},
\ee 
and the deceleration parameter
\begin{align}
 & q\equiv -\frac{a \ddot a }{\dot a^2}=-1+\frac{3}{Q^2}\left(\frac{1+ {\Omega}_\lambda+
\Sigma ^2}{1- {\Omega}_\lambda+
\Sigma ^2}\right)\left( {x}^2+\frac{\gamma
 {\Omega}_m}{2}\right)+\nonumber \\ & +2\left(1-\frac{1}{Q^2}\right)+\frac{3\Sigma
^2}{Q^2}.
\end{align}
Observe that these parameters, but $\Omega_\phi,$ blow up as the singular
surfaces ${\Omega}_\lambda+ {\Omega}_m+ \Sigma ^2=1$ and ${\Omega}_\lambda+
 \Sigma ^2=1$ are approached.  Thus, we need to take
the appropriate limits for evaluating at the singular points $Q_{6}^\pm,
Q_{7}^\pm$ and $Q_{11}^\pm.$ 
In table \eqref{tab1b} we present the values of the observable parameters
$\Omega_{\phi},$ $\omega_{\phi}$, $\omega_{\text{eff}}$, and $q$, evaluated
at the critical points of the system
\eqref{eqQ}-\eqref{eqs} that are associated to expanding (contracting)
solutions. 

Observe that, since the variable $Q$ defined in \eqref{var22} is the Hubble
scalar divided by a positive constant, $Q>0$ corresponds to an expanding
universe, while $Q<0$ to a contracting one. Furthermore, as usual, for an
expanding universe, $q<0$ corresponds to accelerating expansion, and $q>0$ to
decelerating expansion. While for a contracting universe, $q<0$
corresponds to decelerating contraction, and $q>0$ to accelerating
contraction.  Lastly, critical points with $\Sigma=0$ correspond to
isotropic universes.

Now, let us comment on the physical interpretation of the critical points
$Q_i$ that are associated to expanding (contracting) solutions. 

The line of singular points $Q_1^\pm$ represents a matter-dominated solution
 ($\Omega_m=1$). As expected, they are transient stages in the evolution of
the universe. 

The singular points $Q_{2}^\pm(s^*)$ and $Q_3^\pm(s^*)$ are solutions
dominated by the kinetic energy of the scalar field. For these solutions
the scalar field mimics stiff matter and $\rho_\phi\propto a^{-6}$ where
$a$ denotes the scale factor and they are transient states in
the evolution of the universe.

The singular point $Q_4^+(s^*)$  represents  matter-scalar field scaling
solutions that are relevant attractors for
$0<\gamma<\frac{4}{3},s^*<-\sqrt{3\gamma },f'\left(s^*\right)<0$ or
$0<\gamma<\frac{4}{3},s^*>\sqrt{3\gamma },f'\left(s^*\right)>0.$ They
satisfy  $\Omega_\phi\sim \Omega_m$ thus they are important for solving or
alleviating the coincidence problem. The corresponding solutions are
accelerating only for $\gamma<\frac{2}{3}.$ Thus, for standard matter
($\gamma>1$), that is satisfying the usual energy conditions, they cannot
represent accurately the present universe, since they are decelerating
solutions. However, the most interesting feature is that the singular point
$Q_4^-(s^*)$, which corresponds to a contracting universe (since
$\text{sgn}(Q)=\text{sgn}(H)=-1$), is a local source under the same
conditions for which $Q_4^+(s^*)$,  which corresponds to an expanding
universe,
is stable. Therefore, in the scenario at hand the transition from
contracting to expanding universes is possible, since for the same values
of the parameters we have a contracting local source and an expanding local
attractor. Such a behavior has very important physical implications.

The singular point $Q_5^+(s^*)$ is the analogous  to $P_5$ in
\citep{Escobar2012a}. It represents a scalar-field dominated solution
($\Omega_\phi=1$) that is a relevant attractor for $0<\gamma \leq
\frac{4}{3},-\sqrt{3 \gamma }<s^*<0,f'\left(s^*\right)<0$ or
$\frac{4}{3}<\gamma \leq 2,-2<s^*<0,f'\left(s^*\right)<0$ or $0<\gamma \leq
\frac{4}{3},0<s^*<\sqrt{3 \gamma },f'\left(s^*\right)>0$ or
$\frac{4}{3}<\gamma \leq 2,0<s^*<2,f'\left(s^*\right)>0.$ These solutions
can represent accurately the late-time accelerating universe if additionally $s^{*2}<2.$ As before, the
singular point $Q_5^-(s^*)$ is a local source under the same conditions for
which $Q_5^+(s^*)$ is stable. This means that we have a large probability
to
have a transition from contracting to expanding solutions. This is one of
the advantages of the present model.   

The singular points $Q_{9}^\pm$ represent an standard 4D cosmological
solution whereas  $Q_{10}^\pm$ represent solutions with 5D-corrections.  In
both cases, from the relationship between $y$ and $\Omega_{\lambda}$,
follows that these solutions are dominated by the potential energy of the
scalar field $\rho_T= V(\phi);$ that is, they are de Sitter-like  solutions where the dark energy behave like a
cosmological constant
($\omega_\phi=- 1$). In this case the Friedmann equation can be expressed
as 
\be 
3H^2 =V\left(1+
\frac{V}{2\lambda}\right).\label{P10.1}
\ee
In the early universe, where $\lambda\ll V,$ the expansion rate for the RS model differs from the GR predictions
\be\frac{H_{RS}}{H_{GR}}= \sqrt{\frac{V}{2\lambda}}.\label{P10.2} \ee
 Contracting de Sitter solutions are associated to the non-hyperbolic fixed
point $Q_9^-,$ and it corresponds to the early-time universe since it
behaves as a local source in the phase-space. 

The singular points $Q_{11}^\epsilon$ represent a 1D set of singular
points, corresponding to isotropic solutions with 5D corrections
($\Omega_\lambda=1$), parametrized by the values of $Q=\epsilon=\pm 1$ and
$s_c.$ They can represent the early (late)-time universe  for $\epsilon=+1$
($\epsilon=-1$) and $\gamma>1.$ 

The  singular points $Q_{12}^\pm(s^*)$ correspond to a scalar field-dark
radiation scaling solution ($\Omega_U\sim \Omega_\phi$), with an effective
equation of state parameter of the total matter, $w_{eff}=\frac{1}{3}$,
corresponding to  radiation. However, although $P_{12}^+$ correspond to a late-time attractors for 
$\frac{4}{3}<\gamma \leq 2,\; s^*>2,\; f'\left(s^*\right)>0,$ it cannot be a good solution for describing the late time accelerating universe. On the other hand, $Q_{12}^-$ is an early-time attractor for $\frac{4}{3}<\gamma \leq 2,\; s^*>2,\; f'\left(s^*\right)>0,$ and in this case it corresponds to a primordial scalar field- dark radiation dominated universe.   

For all the critical points listed above  the isotropization has been
achieved. The existence of such late-time isotropic solutions (e.g.,
$Q_5^+(s^*)$ and $Q_{9}^+,$), that can attract an
initially anisotropic universe, are of significant cosmological interest
and have been obtained and discussed by \citet{Leach2006,Goheer2007}.  The fact that an isotropic solution is
accompanied by acceleration, makes it a very good candidate for the
description of the observable universe. The next important step in the study of the viability of the model is to check, to what extent, the observational constrains of the model parameters allow not only the existence and stability of the above late time attractors, but also a deviation from the concordance model ($\Lambda$CDM). As we mention before, this latter study is absent in the literature for Bianchi I braneworld and goes beyond the objectives of the present manuscript. However, the methods developed by \citep{Sahni2005,Astashenok2012b, Astashenok2013, Astashenok2013a} for FRW branes could be followed to estimate these deviations. In \citep{Sahni2005}, among several cosmological observables, the value of redshift of reionization ($z_{reion}$) is used to distinguish a $\Lambda$CDM cosmology from a FRW brane model. While in \citep{Astashenok2013}, the allowed value for the brane tension, $\lambda$, was constrained for different FRW branes models with a join analysis of SNIa \citep{Amanullah2010}, BAO \citep{Blake2011a}, H(z) \citep{Stern2010} and matter density perturbation \citep{Christopherson2010} finding that, when the brane tension decreases the agreement with cosmological observations becomes worse\footnote{The brane tension have been also constrained in \citep{Holanda2013}, using a similar observational analysis with the extra ingredient of the CMB observations by using the CMB/BAO ratio. While in \citep{Garcia-Aspeitia2013}, the constraints are achieved using several astrophysical methods.}.

On the other hand, there are some anisotropic solutions. For example,
the solutions $Q_{6}^\pm,$ and  $Q_{7}^\pm$ represent shear-dominated
solutions which are always decelerating. $Q_{6}^+$ cannot represent
accurately the late-time universe, but it has a large probability to
represent the early-time universe.  $Q_{6}^-$ is of saddle type. For
$Q_7^\pm$ we have
similar results as for $Q_6^\pm.$  

The circles of critical points  $Q_8 ^\pm$  correspond to scalar
field-anisotropic scaling solutions ($\Omega_{\phi}\sim \Sigma$). Both
solutions represents transient states in the evolution of the universe.
When $x\rightarrow 0$ the anisotropic term in the Friedmann equation
\eqref{2.5} dominates the cosmological dynamics ($\Sigma=\pm 1$).

Summarizing, the possible late-time (stable) attractors are:
\begin{itemize}
\item $Q_4^+(s^*)$ for $0<\gamma
<\frac{4}{3},s^*<-\sqrt{3\gamma },f'\left(s^*\right)<0$ or $0<\gamma
<\frac{4}{3},s^*>\sqrt{3\gamma },f'\left(s^*\right)>0;$
\item $Q_5^+(s^*)$ for $0<\gamma \leq \frac{4}{3},-\sqrt{3
\gamma }<s^*<0,f'\left(s^*\right)<0$ or $\frac{4}{3}<\gamma \leq
2,-2<s^*<0,f'\left(s^*\right)<0$ or $0<\gamma \leq
\frac{4}{3},0<s^*<\sqrt{3 \gamma },f'\left(s^*\right)>0$ or
$\frac{4}{3}<\gamma \leq 2,0<s^*<2,f'\left(s^*\right)>0;$ 
\item $Q_9^+$ and the line $Q_{10}^+$  for $f(0)\geq 0.$
\item $Q_{11}^-$ for $\gamma>1.$
\item $Q_{12}^+(s^*)$ for $\frac{4}{3}<\gamma \leq 2, s^*>2,
f'\left(s^*\right)>0;$
\end{itemize}
whereas, the possible past (unstable) attractors are:  
\begin{itemize}
\item $Q_4^-(s^*)$ for $0<\gamma
<\frac{4}{3},s^*<-\sqrt{3\gamma },f'\left(s^*\right)<0$ or $0<\gamma
<\frac{4}{3},s^*>\sqrt{3\gamma },f'\left(s^*\right)>0;$
\item $Q_5^-(s^*)$ for $0<\gamma \leq
\frac{4}{3},-\sqrt{3 \gamma }<s^*<0,f'\left(s^*\right)<0$ or
$\frac{4}{3}<\gamma \leq 2,-2<s^*<0,f'\left(s^*\right)<0$ or $0<\gamma \leq
\frac{4}{3},0<s^*<\sqrt{3 \gamma },f'\left(s^*\right)>0$ or
$\frac{4}{3}<\gamma \leq 2,0<s^*<2,f'\left(s^*\right)>0;$ 
\item $Q_9^-$ and the line $Q_{10}^-$ for $f(0)\geq 0.$
\item $Q_{11}^+$ for $\gamma>1.$
\item $Q_{12}^-(s^*)$ for $\frac{4}{3}<\gamma \leq 2, s^*>2,
f'\left(s^*\right)>0.$
\end{itemize}

Additionally, our system admits a large class of static  solutions that are
typically of saddle type. 
In fact, the more interesting result for the scenario at hand is that still a large
probability for the transition from contracting to expanding universes,
since for the same values of the parameters we have a contracting local
sources and an expanding local attractors.   The possible re-collapse
in Bianchi I brane worlds has been 
investigated for example by  \citet{Toporensky2005}, but in this case for
a nonstandard effective equation of state of a brane matter $p =
(\gamma-1)\rho$ with $2 <\gamma<4$. \citet{Santos2001} used a similar argument to Wald's no-hair theorem \citep{Wald} for global
anisotropy in the brane world scenarios. There were derived a set of
sufficient conditions which must be satisfied by the brane matter and bulk
metric so that a homogeneous and anisotropic brane asymptotically evolves
to a de Sitter spacetime in the presence of a positive cosmological
constant on the brane. Following that reference, the isotropy is reached
for any initial condition corresponding to ${\cal U} \geq 0$. We have
obtained the same result in \citep{Escobar2012a} for a quintessence field
with positive potential trapped on the brane by without using Wald's
arguments. When ${\cal U}<0$ the brane may not isotropize and can instead
collapse.  The existence of re-collapsing solutions have been also
discussed by \citet{Campos2001a,Goheer2002} for the case of  an scalar
field with exponential potential on the brane. In the present paper we go a
step forward by considering potentials beyond the exponential one. 

An important issue of the late time evolution of every cosmological model is the possible occurrence of future singularities \citep{Nojiri2005, Nojiri2005a, Bamba2012g}. In the case of FRW branes this study was carried on by \citep{Astashenok2012, Astashenok2013} using the equation-of-state formalism. In \citep{Astashenok2012}, the full set of possibilities was obtained for positive and negatives values of the brane tension. This result is quite general and could be extrapolated to Bianchi I braneworld since the contribution of dark radiation and brane anisotropy ($\sigma$) seem not lead to new future singularities solutions. In our model, as we mention before, the late time behavior will be constrained by the attractor nature of several critical points. Among them, $Q_{5}^+$ and $Q_9^+$ correspond to accelerated solutions characterized by $-1<w_{eff}<-1/3$ and $w_{eff}=-1$ respectively. In the first case the asymptotic evolution lies in the quintessence region so in principle singularities of type II \citep{Barrow2004, Nojiri2004,Nojiri2004a}, III and IV may appear (see \citep{Nojiri2005a, Bamba2012g} for the classification). In the latter case, the critical point coincides with an asymptotic de Sitter expansion, $\rho_{T}=\rho_{\phi}=V(\phi)$, if the self interaction potential tends asymptotically to a constant value: $V(\phi)\rightarrow V_0$, which also leads to a non-singular cosmology. But in general, a complete study of future singularities in our scenario, for all possible late time dynamics, is beyond the present study and will be left for a future paper.  

Finally, a very interesting feature of our scenario is that it
allows for bouncing and turnaround, which leads to cyclic behavior
\citep{Allen2004,Biswas2006,Novello2008,Cai2009,Cai2011}.
In fact, using a heuristic reasoning we have obtained several
conditions for the existence of  bounces and turnaround of cosmological
solutions for a massless scalar field and a scalar field with a potential
asymptotically constant. In  \citep{Toporensky2003} the
authors discussed the conditions for bounces and chaotic
behavior in FRW branes with an scalar field and a negative dark
radiation ${\cal U}<0$. Similarly, in \citep{Campos2001a} the results
correspond to a perfect fluid with equation of state $p=(\gamma-1)\rho$
whereas in \citep{Saridakis2009} a scalar field on the brane is
considered.  Bouncing solutions are found to exist both in FRW
$R^n$-gravity
\citep{Carloni2006}, as well as in the Bianchi I and Bianchi III
$R^n$-gravity \citep{Goheer2007} (see also \citep{Barragan2010}) and in
Kantowski-Sachs $R^n$-gravity \citep{Leon2011}. More interesting, these
features have been alternatively obtained in the brane cosmology context
\citep{Khoury2001,Shtanov2003,Mukherji2002,Brandenberger2007,Saridakis2009}.

The complete bouncing and cyclic analysis in our framework (that goes
beyond the heuristic results presented by us in section \ref{BianchiI}), as
well as the corresponding perturbation investigation, will not be
considered in the present work and it is left for a future project.
Strictly speaking, to complete the bounce and  cyclic analysis we have to
additionally examine the second Friedman equation, that is the equation for
$\dot{H}$, since one could have the very improbable case of $H$ transiting from
positive values to become exactly zero and then positive again, without
obtaining negative values at all. In this case we acquire
$H=0$ and $\dot{H}=0$ simultaneously, that is a universe that stops and
starts expanding again without a turnaround.

\section{Examples}\label{Examples}

In this section we want to discuss the cosh-like
potential and  the inverted sinh-like potential in order to illustrate our analytical results.

The cosh-like potential \be\label{potential1}
V(\phi)=V_{0}\left[\cosh\left( \xi \phi \right)-1\right],
\ee has been widely investigated by
\citep{Ratra1988,Wetterich1988,Matos2000,Sahni2000a,Sahni2000,Lidsey2002,Pavluchenko2003,Copeland2009}. It was used to explain the core density problem for disc galaxy halos in the
$\Lambda$CDM model by \citet{Matos2000}, and independently by \citet{Sahni2000a} (see also
\citet{Wetterich1988}, \citet{Ratra1988} and \citet{Copeland2009}). The asymptotic
properties of a cosmological model with a scalar field with cosh-like
potential have been investigated in the general relativistic framework by
\citet{Matos2009} and  for RS2-FRW branes by \citet{Leyva2009}.

The $f$-deviser corresponding to the potential \eqref{potential1} is given by
\be\label{function1}
f(s)=-\frac{1}{2}(s^2-\xi^2),
\ee
which has the zeroes
\be
s^*=\pm\xi\quad f'(s^*)=\mp\xi.
\ee 
The only possible expanding late time attractor of the model with potential
\eqref{potential1} is the Sitter attractor $Q_9^+$.

\begin{figure}
\begin{center}
\subfigure[]{\includegraphics[scale=0.6]{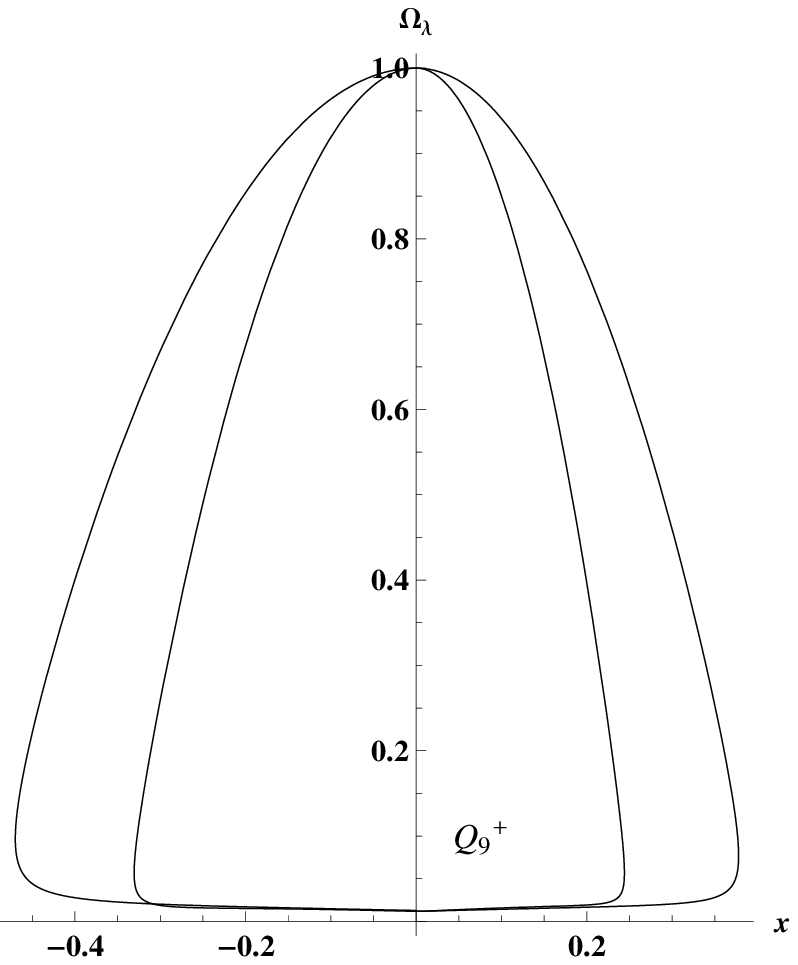}} 
\subfigure[]{\includegraphics[scale=0.6]{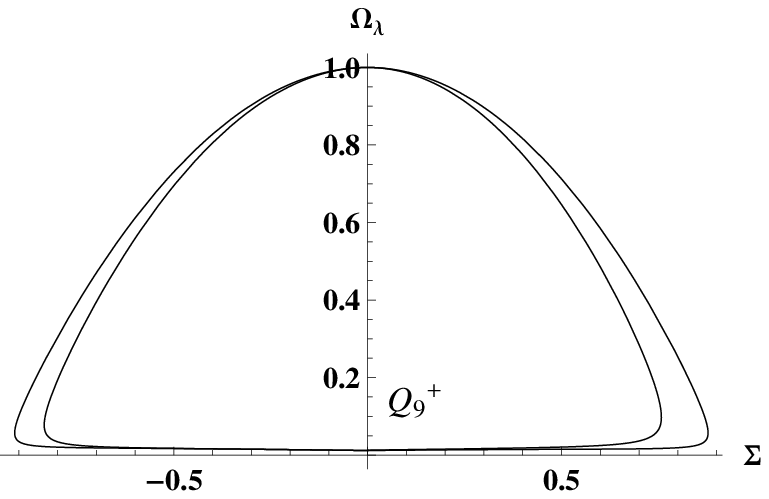}}
\caption{\label{fig1} Projection of some orbits of the system
\eqref{eqQ}-\eqref{eqs} for the potential $V(\phi)=V_{0}\left[\cosh\left(
\xi \phi \right)-1\right]$ with $\xi=1/2$ in the planes (a)  $x-\Omega_\lambda$ 
and (b)  $\Sigma-\Omega_\lambda$ for $\gamma=1$. This numerical
elaboration shows that $Q_9 ^+$ is the local attractor.}
\end{center}
\end{figure}

\begin{figure}
\begin{center}
\includegraphics[scale=0.7]{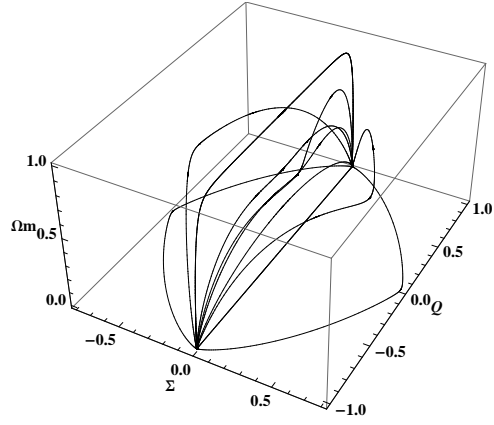}
\caption{\label{fig2}  Projection of some orbits in the space $\Sigma, Q, \Omega_m.$ We choose
the initial states near static solutions.   This numerical elaboration
shows the transition from the contracting de Sitter solution $Q_9 ^-$ (past
attractor) to the expanding de Sitter solution $Q_9^+$ (future attractor).}
\end{center}
\end{figure}

In   figure \ref{fig1} are showed some numerical integrations for the
system \eqref{eqQ}-\eqref{eqs} for the  function \eqref{function1} with
$\xi=1/2$. This numerical elaboration shows that the expanding de Sitter
solution $Q_9^+$ is the future attractor whereas the contracting de Sitter
solution $Q_9^-$ is the past attractor. It is illustrated also the
transition from contracting to expanding solutions. In   figure
\ref{fig2} we choose initial conditions near the static  solutions
presented in table \ref{tab2a}. This kind of solutions allow for the
transition form contracting deceleration to expanding accelerated
solutions.

Now, the inverted sinh-like potential \be\label{potential}
V(\phi)=V_{0}\sinh^{-\alpha}(\beta\phi) ,
\ee has been widely investigated in the literature (e.g.
\citet{Urena-Lopez2000,Sahni2000a,Pavluchenko2003,Copeland2009}).
The asymptotic
properties of a cosmological model with a scalar field with such a
potential have been investigated in the context of FRW brane in
\citep{Leyva2009}. 

The $f$-deviser corresponding to the potential \eqref{potential} is given by
\be\label{function}
f(s)=\frac{s^2}{\alpha}-\alpha\beta^2 .
\ee
The zeroes of this $f(s)$ function are
\be
s^*=\pm\alpha\beta\quad f'(s^*)=\pm 2\beta.
\ee

The sufficient conditions for the existence of late-time attractors are
fulfilled easily
\begin{itemize}
\item The scalar field-matter scaling solution $Q_4 ^+(\alpha\beta)$ is a
late-time attractor provided  $0<\gamma <\frac{4}{3},
\beta<0,\alpha>-\frac{\sqrt{3\gamma}}{\beta},$ or $0<\gamma
<\frac{4}{3},\alpha>\frac{\sqrt{3\gamma }}{\beta},\beta>0.$
\item The scalar field-matter scaling solution $Q_4 ^+(-\alpha\beta)$ is a
late-time attractor provided  $0<\gamma
<\frac{4}{3},\alpha<\frac{\sqrt{3\gamma}}{\beta},\beta>0$ or $0<\gamma
<\frac{4}{3},\alpha>-\frac{\sqrt{3\gamma }}{\beta},\beta<0.$ 
\item The solution dominated by scalar field $Q_5 ^+(\alpha\beta)$ is a
late-time attractor provided $0<\gamma \leq \frac{4}{3},\beta<0,
0<\alpha<-\frac{\sqrt{3 \gamma}}{\beta}$ or $\frac{4}{3}<\gamma \leq 2,
\beta<0,0<\alpha<-\frac{2}{\beta}$ or $0<\gamma  \leq
\frac{4}{3},0<\alpha<\frac{\sqrt{3\gamma}}{\beta},\beta>0$ or
$\frac{4}{3}<\gamma \leq 2,0<\alpha<\frac{2}{\beta},\beta>0;$
\item The solution dominated by scalar field $Q_5 ^+(-\alpha\beta)$ is a
late-time attractor provided $0<\gamma  \leq \frac{4}{3},\beta>0,
0<\alpha<\frac{\sqrt{3 \gamma}}{\beta}$ or $\frac{4}{3}<\gamma \leq 2,
\beta>0,0<\alpha<\frac{2}{\beta}$ or $0<\gamma  \leq
\frac{4}{3},0<\alpha<-\frac{\sqrt{3\gamma}}{\beta},\beta<0$ or
$\frac{4}{3}<\gamma \leq 2,0<\alpha<-\frac{2}{\beta},\beta<0;$ 
\item The de Sitter solution $Q_9^+$ is the late time attractor provided
$f(0)=-\alpha\beta^2>0.$ 
\item $Q_{11}^-$ is stable for $\gamma>1.$
\item The scalar field- dark radiation scaling solution \\ $Q_{12}^+(\alpha
\beta)$ is a late-time attractor for $\frac{4}{3}<\gamma\leq 2,
\alpha\beta>2, \beta>0.$
\item The scalar field- dark radiation scaling solution \\ $Q_{12}^+(-\alpha
\beta)$ is a late-time attractor for $\frac{4}{3}<\gamma\leq 2,
\alpha\beta<-2, \beta<0.$
\end{itemize}

The past attractors of the system are as follows. 
\begin{itemize}
\item The scalar field-matter scaling solution $Q_4 ^-(\alpha\beta)$ is the
past attractor provided  $0<\gamma <\frac{4}{3},
\beta<0,\alpha>-\frac{\sqrt{3\gamma}}{\beta},$ or $0<\gamma
<\frac{4}{3},\alpha>\frac{\sqrt{3\gamma }}{\beta},\beta>0.$
\item The scalar field-matter scaling solution $Q_4 ^-(-\alpha\beta)$ is
the past attractor provided  $0<\gamma
<\frac{4}{3},\alpha<\frac{\sqrt{3\gamma}}{\beta},\beta>0$ or $0<\gamma
<\frac{4}{3},\alpha>-\frac{\sqrt{3\gamma }}{\beta},\beta<0.$ 
\item The solution dominated by scalar field $Q_5 ^-(\alpha\beta)$ is the
past attractor provided $0<\gamma \leq \frac{4}{3},\beta<0,
0<\alpha<-\frac{\sqrt{3 \gamma}}{\beta}$ or $\frac{4}{3}<\gamma \leq 2,
\beta<0,0<\alpha<-\frac{2}{\beta}$ or $0<\gamma  \leq
\frac{4}{3},0<\alpha<\frac{\sqrt{3\gamma}}{\beta},\beta>0$ or
$\frac{4}{3}<\gamma \leq 2,0<\alpha<\frac{2}{\beta},\beta>0;$
\item The solution dominated by scalar field $Q_5 ^-(-\alpha\beta)$ is the
past attractor provided $0<\gamma  \leq \frac{4}{3},\beta>0,
0<\alpha<\frac{\sqrt{3 \gamma}}{\beta}$ or $\frac{4}{3}<\gamma \leq 2,
\beta>0,0<\alpha<\frac{2}{\beta}$ or $0<\gamma  \leq
\frac{4}{3},0<\alpha<-\frac{\sqrt{3\gamma}}{\beta},\beta<0$ or
$\frac{4}{3}<\gamma \leq 2,0<\alpha<-\frac{2}{\beta},\beta<0;$ 
\item The de Sitter solution $Q_9^-$ is the past attractor provided
$f(0)=-\alpha\beta^2>0.$ 
\item $Q_{11}^+$ is unstable for $\gamma>1.$
\item The scalar field- dark radiation scaling solution \\ $Q_{12}^-(\alpha
\beta)$ is a past attractor for $\frac{4}{3}<\gamma\leq 2, \alpha\beta>2,
\beta>0.$
\item The scalar field- dark radiation scaling solution \\ $Q_{12}^-(-\alpha
\beta)$ is the past attractor for $\frac{4}{3}<\gamma\leq 2,
\alpha\beta<-2, \beta<0.$
\end{itemize}

\begin{figure}
\begin{center}
\subfigure[]{\includegraphics[scale=0.6]{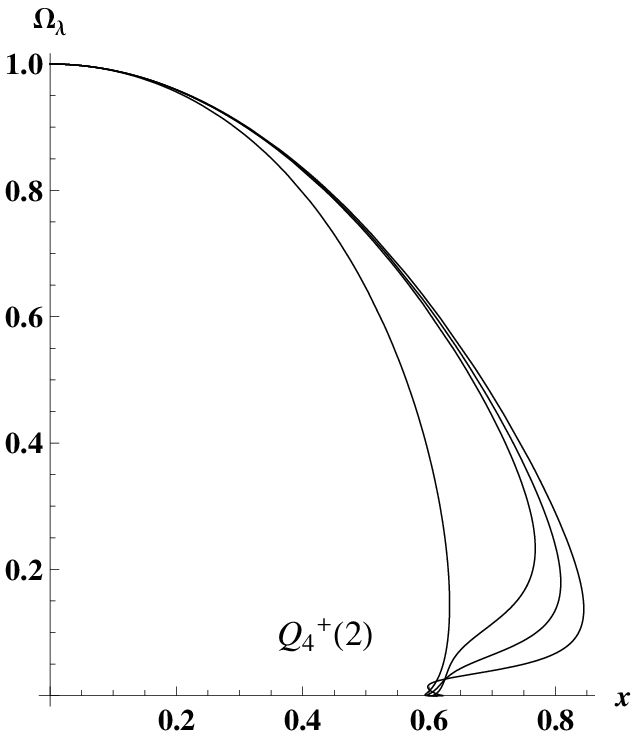}}
\subfigure[]{\includegraphics[scale=0.8]{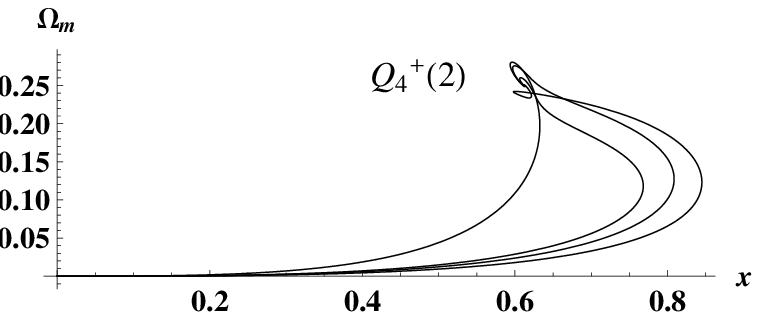}}
\caption{\label{Fig3} Projection of some orbits in the phase space of
\eqref{eqQ}-\eqref{eqs} for the potential
$V(\phi)=V_{0}\sinh^{-\alpha}(\beta\phi)$ with $\alpha=1/2, \beta=4$ and
$\gamma=1$  in the planes (a) $x-\Omega_\lambda$ and (b) $x-\Omega_m$.  We set
$Q=+1$ for the numerical simulation. This numerical elaboration shows that
$Q_4 ^+(2)$ is the future attractor.}
\end{center}
\end{figure}

\begin{figure}
\begin{center}
\vspace{-1cm}
\includegraphics[scale=0.8]{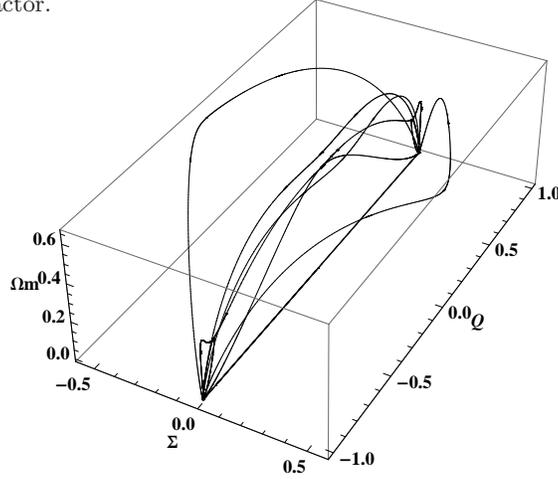}
\caption{\label{Fig4} Projection of some orbits in the phase space of
\eqref{eqQ}-\eqref{eqs} for the potential
$V(\phi)=V_{0}\sinh^{-\alpha}(\beta\phi)$ with $\alpha=1/2, \beta=4$ and
$\gamma=1$  in the subspace $(\Sigma, Q,\Omega_m)$. We choose the
initial states near static solutions. This numerical elaboration illustrate
the transition for the contracting scalar-field matter scaling solution
$Q_4 ^-(2)$ (past attractor) to the expanding one  $Q_4 ^+(2)$ (future
attractor).}
\end{center}
\end{figure}

\begin{figure}
\begin{center}
\subfigure[]{\includegraphics[scale=0.6]{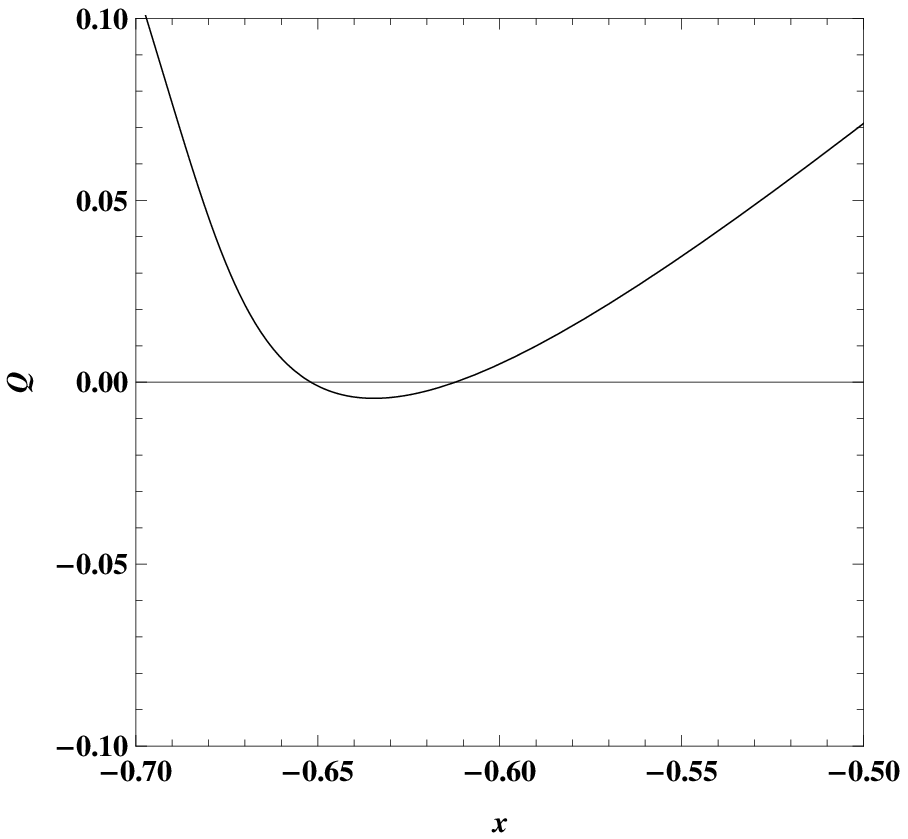}}
\subfigure[]{\includegraphics[scale=0.6]{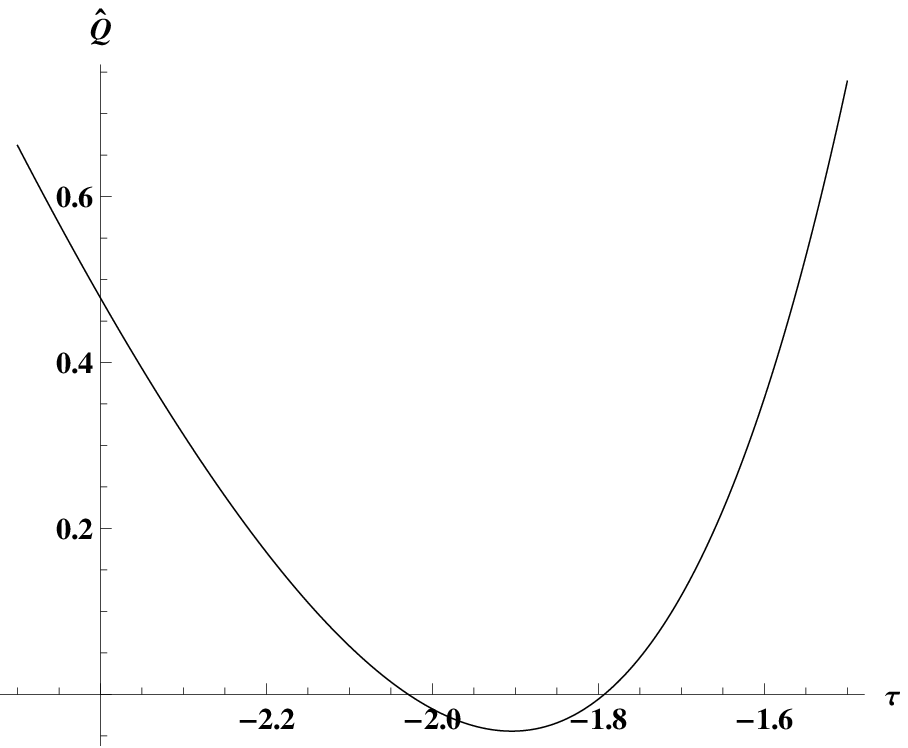}}
\caption{\label{Fig5} Transition from expansion ($Q>0$) to contraction ($Q<0$) and then back
to expansion ($Q>0$), that is a cosmological turnaround and
a bounce for the potential
$V(\phi)=V_{0}\sinh^{-\alpha}(\beta\phi)$ with $\alpha=1/2, \beta=4$. The left panel corresponds to the
projection of an orbit in the plane $x$-$Q$, while the right panel shows
$\hat{Q}\equiv 10 Q$ vs $\tau$, in order to illustrate that $Q$ change its
sign twice during the evolution. We choose the initial
condition $Q(0)=0.9, x(0)=0.6, \Omega_m(0)=0.25,\Omega_\lambda (0)=0,
\Sigma (0)=0,s(0)=2.0.$}
\end{center}
\end{figure}

In the figures \ref{Fig3}, \ref{Fig4} and \ref{Fig5} are presented some orbits in the phase space of
\eqref{eqQ}-\eqref{eqs} for the potential 
$V(\phi)=V_{0}\sinh^{-\alpha}(\beta\phi)$ with $\alpha=1/2, \beta=4$ and
$\gamma=1$. In   figure \ref{Fig3}  we set $Q=+1$ for the numerical
simulation. This numerical elaboration reveals that $Q_4 ^+(2)$ is the
future attractor. In   figure  \ref{Fig4} we present a projection of
some orbits in the subspace $(\Sigma, Q,\Omega_m)$. We choose the initial
states near the static solutions. This numerical elaboration illustrates the
transition for the contracting scalar-field matter scaling solution $Q_4
^-(2)$ (past attractor) to the expanding one  $Q_4 ^+(2)$ (future
attractor). As can be seen in the figures, the static solutions play an
important role in the transition from contracting to expanding solutions
and viceversa. In   figure \ref{Fig5} we show the transition from
expansion to contraction and then back to expansion, that is a
cosmological turnaround and a cosmological bounce, for the solutions of
\eqref{eqQ}-\eqref{eqs}, for the potential
$V(\phi)=V_{0}\sinh^{-\alpha}(\beta\phi)$ with $\alpha=1/2, \beta=4$ and
$\gamma=1$. The left panel corresponds to the projection of an orbit in the
plane $x$-$Q$, while the right panel shows $\hat{Q}\equiv 10 Q$   vs $\tau$
in order to illustrate that $Q$ change its sign twice during the evolution.
We choose the initial condition $Q(0)=0.9, x(0)=0.6,
\Omega_m(0)=0.25,\Omega_\lambda (0)=0, \Sigma (0)=0,s(0)=2.0.$

\begin{figure}
\begin{center}
\subfigure[]{\includegraphics[scale=0.5]{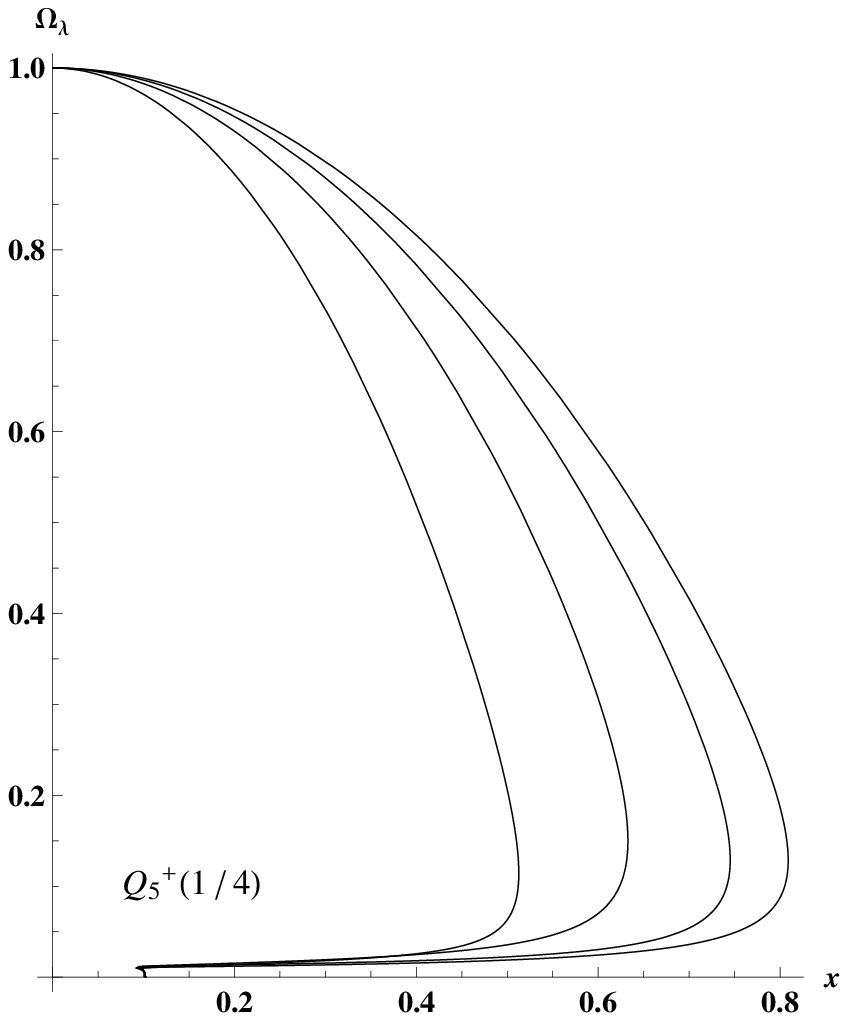}} 
\subfigure[]{\includegraphics[scale=0.4]{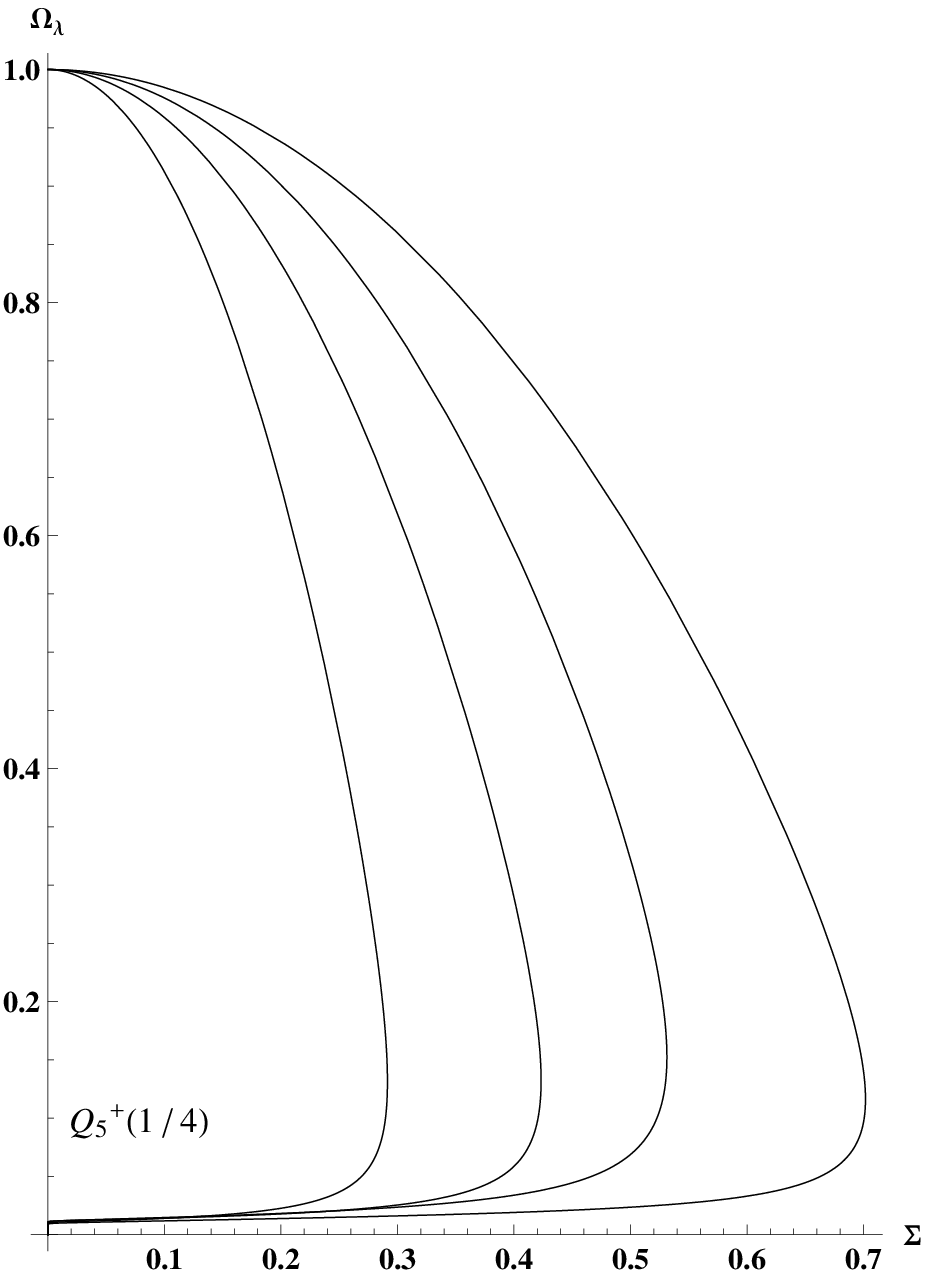}}
\caption{\label{Fig6}  Some orbits of  the phase space of
\eqref{eqQ}-\eqref{eqs} for the potential
$V(\phi)=V_{0}\sinh^{-\alpha}(\beta\phi)$ with $\alpha=\beta=1/2$ and
$\gamma=1$.  Projection in the planes (a) $x-\Omega_\lambda$ and
(b) $\Sigma-\Omega_\lambda$ (We set $Q=+1$ for the numerical
simulation). }
\end{center}
\end{figure}

\begin{figure}[t]
\begin{center}
\includegraphics[scale=0.8]{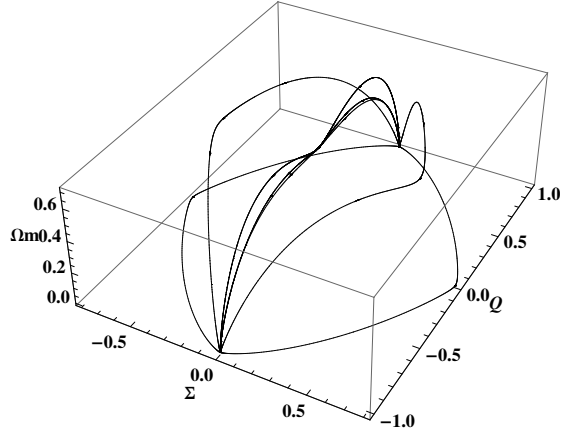}
\caption{\label{Fig7}  Some orbits in the projection $(\Sigma, Q, \Omega_m)$ of  the phase space of
\eqref{eqQ}-\eqref{eqs} for the potential
$V(\phi)=V_{0}\sinh^{-\alpha}(\beta\phi)$ with $\alpha=\beta=1/2$ and
$\gamma=1$.  We choose initial states near the static solutions. This numerical elaboration shows
the transition from the contracting scalar field dominated solution   $Q_5
^-(1/4)$ to the expanding one  $Q_5 ^+(1/4)$. }
\end{center}
\end{figure}

In   figures \ref{Fig6} and \ref{Fig7} are presented some orbits in the phase space of
\eqref{eqQ}-\eqref{eqs} for the potential 
$V(\phi)=V_{0}\sinh^{-\alpha}(\beta\phi)$ with $\alpha=\beta=1/2$ and
$\gamma=1$. In figure \ref{Fig6} we set $Q=+1$ for the numerical
simulation and we present the projection in the planes $x-\Omega_\lambda$
and $\Sigma-\Omega_\lambda.$ In this case the future attractor is
the isotropic solution dominated by scalar  field $Q_5 ^+(1/4)$, also this
solution corresponds to an accelerated expansion rate since
$s^*=1/4<\sqrt{2}$.
In   figure \ref{Fig7} are presented some orbits in the projection
$(\Sigma, Q, \Omega_m)$. This numerical elaboration shows the transition
from the contracting scalar field dominated solution   $Q_5 ^-(1/4)$ to the
expanding one  $Q_5 ^+(1/4)$. We choose the initial states near the static
solutions. 

\begin{figure}
\begin{center}
\subfigure{\includegraphics[scale=0.7]{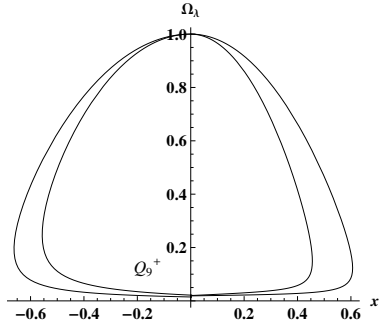}}
\begin{center} (a) \end{center}
\subfigure{\includegraphics[scale=0.6]{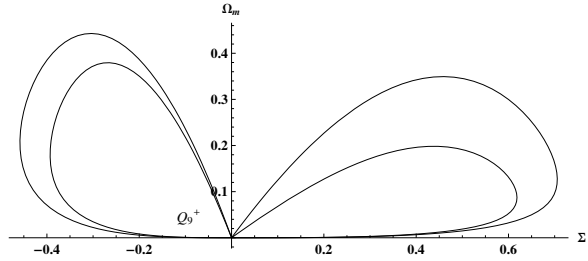}}
\begin{center} (b) \end{center}
\caption{\label{Fig8}  Some orbits of  the phase space of
\eqref{eqQ}-\eqref{eqs} for the potential
$V(\phi)=V_{0}\sinh^{-\alpha}(\beta\phi)$ with $\alpha=-1/2, \beta=-1/4$
and $\gamma=1$. Projection in the planes (a) $x-\Omega_\lambda$ and (b) $\Sigma-\Omega_m.$  We
set $Q=+1$ for the numerical simulation.}
\end{center}
\end{figure}

\begin{figure}
\begin{center}
\vspace{-1cm}
\includegraphics[scale=0.8]{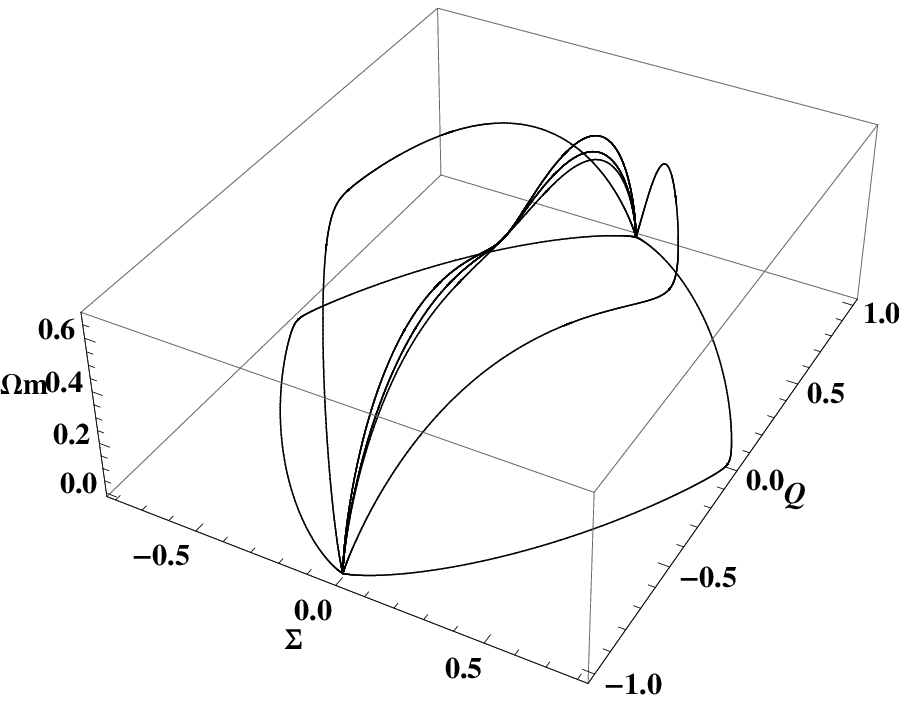}
\caption{\label{Fig9}  Some orbits in the subspace $(\Sigma, Q,\Omega_m)$, $Q\neq +1$ of  the phase space of
\eqref{eqQ}-\eqref{eqs} for the potential
$V(\phi)=V_{0}\sinh^{-\alpha}(\beta\phi)$ with $\alpha=-1/2, \beta=-1/4$
and $\gamma=1$. We choose the initial states near static solutions. This numerical elaboration
illustrate the transition for the contracting de Sitter solution $Q_9^-$
(past attractor) to the expanding de Sitter solution  $Q_9^+$ (future
attractor).}
\end{center}
\end{figure}

In   figures \ref{Fig8} and \ref{Fig9} are showed some numerical integrations for the
system
\eqref{eqQ}-\eqref{eqs} for the  function \eqref{function} with
$\alpha=-1/2,\beta=-1/4$. In this case the local attractor is the solution
dominated by the potential energy of the scalar field $Q_9 ^+$ (de Sitter
attractor). The past attractor is $Q_9^-.$ This numerical elaboration
illustrate the transition from contracting to expanding solutions and
viceversa that is not allowed for Bianchi I branes with positive Dark
Radiation term (${\cal U}>0$).

\section{Conclusions}\label{conclusions}

In this paper we have presented a full phase space analysis of a model consisting of a
quintessence field with an arbitrary potential and a perfect fluid trapped in
a Randall-Sundrum's Braneworld of type 2. We have considered a homogeneous but anisotropic
Bianchi I brane geometry.   Moreover, we have also included the effect of the
projection of the five-dimensional Weyl tensor onto the three-brane in the form of a
negative dark radiation term. 

We have discussed a general method for the treatment of the potential,
called  ``Method of $f$-devisers''. Combining this method with some tools from the Theory of Dynamical Systems,
we have obtained general conditions which have to be satisfied by the
potential (encoded in the mathematical properties of the key
$f(s)$-functions) in order to obtain the stability conditions of standard
4D and non-standard 5D de Sitter solutions.

We have presented general  conditions under the potential
for the stability of standard 4D de Sitter solutions. We proved that the de
Sitter solutions with 5D corrections are stable against the perturbations
introduced here, but are unstable against small perturbations of $V/(3H^2).$
This fact does not conflict with our previous results for BI
branes with positive dark radiation term. Also, we presented the stability
conditions for both scalar field-matter scaling solutions, for scalar
field-dark radiation scaling solutions and scalar field-dominated
solutions. The shear-dominated solutions are always unstable (contracting
shear-dominated solutions are of saddle type). Also, we have shown that for all possible late time stable expanding solutions (attractors) the isotropization has been archived independently of the initial conditions, in all cases, with observational parameters in concordance with observations. 

The main ansatz of our research is the assumption over the sign of Dark Radiation. If was shown that for ${\cal U}<0$,
the ever-expanding models could potentially re-collapse. Additionally, our system admits a large
class of static solutions  that are of saddle type allowing the transition from contracting to
expanding models and viceversa. Thus, a new feature of this scenario is
the existence of a bounce and a turnaround, which leads to cyclic behavior. This features are not
allowed in Bianchi I branes with positive dark radiation term. This is an important cosmological result in our scenario, that is a consequence of the negativeness of the dark radiation. In this direction, it is worth exploring in detail the current magnitudes and sign of the Dark Radiation, for a Bianchi I braneworld, allowed by cosmological observations since it is an study that is absent, to our knowledge, in the literature. For this latter purpose, not only the well-establish tests of SNe, BAO, H(z), CMB and BBN would provide significant constraints, but also the recent results from multi-wavelength observation could improved the observational study. More precisely, the recent detection of the cosmic $\gamma$-ray horizon from multi-wavelength observation of blazars \citep{Dominguez2013}, using the data from the Fermi satellite and the Imaging Atmospheric Cherenkov
Telescopes, allowed a new measurement of the Hubble constant \citep{Dominguez2013a} and opening a new door to, not only, control systematic uncertainties associated with determinations of the Hubble constant \citep{Suyu2012, Dominguez2013a},  but also to impose new constraints over dark energy and neutrino physics \citep{Weinberg2013, Suyu2012, Freedman2010}. Is expected that in the near future, with more data from the upcoming generation of multi-wavelength observation/ instrumentation facilities planned worldwide such as the Cherenkov Telescope Array, this kind of estimation will become more competitive at the level of other techniques such as CMB, BAO etc.

Finally, in order to be more transparent, we have illustrated our main
results for the specific potentials $V\propto\sinh^{-\alpha}(\beta\phi)$
and $V\propto\left[\cosh\left( \xi \phi \right)-1\right]$ which have simple
$f$-devisers.

\begin{acknowledgements}
This work was partially supported by PROMEP, DAIP, and by CONACyT,
M\'exico, under grant 167335 (YL);  by  MECESUP FSM0806, from Ministerio de
Educaci\'on, Chile (GL) and by PUCV through Proyecto DI Postdoctorado 2013 (GL, YL). GL and YL are grateful to the Instituto de F\'{\i}sica,
Pontificia Universidad Cat\'olica de Valpara\'{\i}so, Chile, for their kind
hospitality and their joint support for a research visit. YL is also grateful to
the Departamento de F\'isica and the CA de Gravitaci\'on
y F\'isica Matem\'atica for their kind hospitality and their joint support
for a postdoctoral fellowship. The authors would like to thank  
E. N. Saridakis for reading the original manuscript and for giving helpful
comments concerning bouncing solutions. I. Quiros is acknowledged for helpful suggestions. The authors whish to thank to two anonymous referees for their useful comments and criticisms.  
\end{acknowledgements}

\begin{appendix}

\section{Stability of static solutions}\label{static}

As we commented before,
the system \eqref{eqx}-\eqref{eqs} admits eighteen classes of (curves of)
fixed points corresponding to static solutions, i.e.,  having $Q=0,$ that
is with $H=0.$ Their coordinates in the phase space and their existence
conditions are displayed in table \eqref{tab2a}. In the table \ref{autovalores1a} are presented the corresponding eigenvalues, where $\lambda_1(\epsilon,  \Sigma_c),\lambda_2(\epsilon,  \Sigma_c),
\lambda_3(\epsilon,  \Sigma_c),$ $\lambda_4(\epsilon,  \Sigma_c)$ are
the roots of the equation 
\begin{align*}
& \lambda ^4+\lambda ^3 \sqrt{4-6 \Sigma_c^2} \epsilon 
f'(0)+\lambda ^2\left[f(0)-4\right]  -4 \lambda  \sqrt{4-6 \Sigma_c^2}
\epsilon  f'(0)-6 f(0) \Sigma_c^2=0,
\end{align*}
	and 
	$\mu_1(\epsilon,  \Sigma_c),\mu_2(\epsilon,  \Sigma_c),
\mu_3(\epsilon,  \Sigma_c),$  $\mu_4(\epsilon,  \Sigma_c)$ are the roots
of the equation 
\begin{align*}
& \mu ^4+\sqrt{2} \mu ^3 \Sigma_c \epsilon  f'(0)+\mu ^2 \left[2
(f(0)+4) {\Sigma   _c}^2-8\right]+8 \sqrt{2} \mu  \Sigma_c
\left(\Sigma_c^2-1\right) \epsilon  f'(0)+4 f(0) \Sigma_c^2
\left(\Sigma_c^2-2\right)=0.
\end{align*}

\begin{table*}[ht]
\caption{Eigenvalues for the critical points in table \eqref{tab2a}. Are used the notations $ g_1(\gamma,
\Omega_\lambda,\Sigma)=\frac{\sqrt{-3 \gamma  \left(3 \Sigma ^2-2\right)
\left(-\Sigma ^2+\Omega_\lambda +1\right)^2+18 \Sigma ^6-44 \Sigma ^4+2
\Sigma ^2 (\Omega_\lambda  (9 \Omega_\lambda +2)+17)-8
   \left(\Omega_\lambda ^2+1\right)}}{-\Sigma ^2+\Omega_\lambda +1},$ 
 $g_2(\gamma, \Omega_\lambda,\Sigma)=\frac{\sqrt{f(0)} \sqrt{\Sigma
^2+\Omega_\lambda -1} \sqrt{3 \gamma  \left(-\Sigma ^2+\Omega_\lambda
+1\right)+6 \Sigma ^2-4}}{\sqrt{\gamma } \sqrt{-\Sigma ^2+\Omega_\lambda
   +1}},$ 
	$g_3(\gamma, x, \Omega_\lambda,\Sigma)=-\frac{9}{2} \left(\gamma ^2
(\Omega_\lambda -1) \Omega_\lambda +(\gamma -2)^2 \Sigma ^4-(\gamma -2)
\Sigma ^2 (2 \gamma  \Omega_\lambda +\gamma -2)\right)-\frac{9 (\gamma
-2)^2
   x^4 \left(\Sigma ^4-2 \Sigma ^2 (\Omega_\lambda +1)+\Omega_\lambda
^2+1\right)}{2 \left(\Sigma ^2+\Omega_\lambda -1\right)^2}-\frac{9 (\gamma
-2) x^2 \left(-\Omega_\lambda 
   \left(4 (\gamma -1) \Sigma ^2+\gamma +2\right)+(\gamma -2) \left(2
\Sigma ^4-3 \Sigma ^2+1\right)+2 \gamma  \Omega_\lambda ^2\right)}{2
\left(\Sigma ^2+\Omega_\lambda
   -1\right)},$ \newline $g_4(\gamma, x, \Omega_\lambda)=\frac{3 \sqrt{-\gamma ^2
(\Omega_\lambda -1)^3 \Omega_\lambda -(\gamma -2)^2 x^4
\left(\Omega_\lambda ^2+1\right)-(\gamma -2) x^2 (\Omega_\lambda -1)
\left(2 \gamma  \Omega_\lambda ^2-(\gamma +2) \Omega_\lambda +\gamma -2\right)}}{\sqrt{2} (\Omega_\lambda -1)}.$}
\label{autovalores1a}
\begin{center}
{\footnotesize
\begin{tabular}{@{\hspace{4pt}}c@{\hspace{14pt}}c@{\hspace{14pt}}c}
\hline
\hline\\[-0.3cm]
Labels & Eigenvalues\\[0.1cm]
\hline\\[-0.2cm]
$E_{1}^\pm$& $\left\{-2,2,0,0,-\sqrt{f(0)} \sqrt{3 \Omega_{\lambda c}
-1},\sqrt{f(0)} \sqrt{3 \Omega_{\lambda c} -1}\right\}$\\[0.2cm]
$E_{2}$& $\left\{0,0,0,0,-\frac{3}{2} \sqrt{1-\cos (4 u)},\frac{3}{2}
\sqrt{1-\cos (4 u)}\right\}$\\[0.2cm]
$E_{3}$& $\left\{0,0,-g_1(\gamma, \Omega_{\lambda c},\Sigma_c), g_1(\gamma,
\Omega_{\lambda c},\Sigma_c),-g_2(\gamma, \Omega_{\lambda c},\Sigma_c),
g_2(\gamma, \Omega_{\lambda c},\Sigma_c)\right\}$\\[0.2cm]
$E_4$ & $\left\{0,0,0,0,-\sqrt{-9 \gamma ^2+9 (\gamma -2) (\gamma -1)
\Sigma_c^2+18 \gamma -8},\sqrt{-9 \gamma ^2+9 (\gamma -2) (\gamma -1)
\Sigma_c^2+18 \gamma
   -8}\right\}$\\[0.2cm]
$E_5$ & $\left\{0,0,-\sqrt{g_3(\gamma, x_c, \Omega_{\lambda
c},\Sigma_c)},\sqrt{g_3(\gamma, x_c, \Omega_{\lambda
c},\Sigma_c)},-\sqrt{6} s^* x_c,-\sqrt{6} x_c
f'\left(s^*\right)\right\}$\\[0.2cm]
$E_6^\epsilon$ & $\left\{0,0,	\lambda_1(\epsilon, 
\Sigma_c),\lambda_2(\epsilon,  \Sigma_c), \lambda_3(\epsilon,  \Sigma_c),
\lambda_4(\epsilon,  \Sigma_c)\right\}$\\[0.2cm]
$E_7^\epsilon$ & $\left\{0,0,	\mu_1(\epsilon,  \Sigma_c),\mu_2(\epsilon, 
\Sigma_c), \mu_3(\epsilon,  \Sigma_c), \mu_4(\epsilon, 
\Sigma_c)\right\}$\\[0.2cm]
$E_8$ & $\left\{0,0,-g_4(\gamma, x_c,\Omega_{\lambda c}),g_4(\gamma,
x_c,\Omega_{\lambda c}),-\sqrt{6} s^* x_c,-\sqrt{6} x_c
f'\left(s^*\right)\right\}$\\[0.2cm]
$E_{9}$ & $\left\{0,0,-\frac{\sqrt{6 \gamma 
(\Omega_{\lambda}+1)^2-8 \left(\Omega_{\lambda
}^2+1\right)}}{\Omega_{\lambda }+1},
\frac{\sqrt{6 \gamma 
(\Omega_{\lambda}+1)^2-8 \left(\Omega_{\lambda
}^2+1\right)}}{\Omega_{\lambda }+1},-\frac{\sqrt{f(0)}
\sqrt{\Omega_{\lambda}-1} \sqrt{3
   \gamma  (\Omega_{\lambda}+1)-4}}{\sqrt{\gamma }
\sqrt{\Omega_{\lambda}+1}},\frac{\sqrt{f(0)}
\sqrt{\Omega_{\lambda}-1} \sqrt{3
   \gamma  (\Omega_{\lambda}+1)-4}}{\sqrt{\gamma }
\sqrt{\Omega_{\lambda}+1}}\right\}$\\[0.2cm]
$E_{10}^\epsilon$ & $\left\{0,0,-\epsilon\sqrt{2} \sqrt{\frac{3 \gamma
-2}{\gamma -2}} s^* \Sigma_c ,\sqrt{2} \sqrt{\gamma  \left(9
\Sigma^2-6\right)-4 \Sigma^2+4}, -\sqrt{2} \sqrt{\gamma  \left(9
\Sigma^2-6\right)-4 \Sigma^2+4},-\epsilon\sqrt{2} \sqrt{\frac{3 \gamma -2}{\gamma -2}} \Sigma_c 
   f'\left(s^*\right)\right\}$\\[0.2cm]
$E_{11}$ & $\left\{0,0,-\sqrt{\gamma  \left(6-9 \Sigma_c ^2\right)+10
\Sigma_c ^2-4},\sqrt{\gamma  \left(6-9 \Sigma_c ^2\right)+10 \Sigma_c
^2-4},-\frac{\sqrt{2-3 \gamma } \sqrt{f(0)} \Sigma_c
   }{\sqrt{\gamma }},\frac{\sqrt{2-3 \gamma } \sqrt{f(0)} \Sigma_c
}{\sqrt{\gamma }}\right\}$\\[0.2cm]
$E_{12}$ & $\left\{0,0,-\sqrt{\gamma  \left(6-9 \Sigma ^2\right)+18
\Sigma ^2-8},\sqrt{\gamma  \left(6-9 \Sigma ^2\right)+18
\Sigma ^2-8},-\frac{\sqrt{f(0)} \sqrt{3 (\gamma -2) \Sigma
^2-3
   \gamma +4}}{\sqrt{\gamma }},\frac{\sqrt{f(0)} \sqrt{3 (\gamma -2) \Sigma
^2-3
   \gamma +4}}{\sqrt{\gamma }}\right\}$\\[0.2cm]
$E_{13}^\epsilon$ & $\left\{0,0,\sqrt{4-3 \gamma },-\sqrt{4-3 \gamma
},-\frac{\sqrt{2} \sqrt{4-3 \gamma } s^* \epsilon }{\sqrt{2-\gamma
}},-\sqrt{2} \sqrt{\frac{3 \gamma -4}{\gamma -2}} \epsilon 
   f'\left(s^*\right)\right\}$
\\[0.4cm]\hline \hline
\end{tabular}
}\end{center}
\end{table*}

Now let us comment on the stability of the first order perturbations of
\eqref{eqQ}-\eqref{eqs} near the critical points showed in table
\ref{tab2a}.

The line of fixed  points $E_1$ is non-hyperbolic. However, the points
located at the curve behaves as saddle points since its matrix of
perturbations admits at least two real eigenvalues of different signs. 

The one-parametric line of fixed  points $E_2$ has two real eigenvalues of
different  signs provided $\cos (4 u)\neq 1.$ In this case, although
non-hyperbolic, it behaves as a set of saddle points. For
$u\in\left\{\frac{\pi}{2},\frac{3\pi}{2}\right\}$ all the eigenvalues are
zero. In this case we need to resort to a numerical elaboration. 

$g_1(\gamma, \Omega_\lambda,\Sigma)$ is always  real-valued for the allowed
values of the phase space variables and the allowed range for the free
parameters. Then, although non-hyperbolic, the 2D set of fixed points $E_3$
is of saddle type.  

The eigenvalues associated to the one-parametric curve of fixed points
$E_4$ are always reals for the allowed values of the phase space variables
and the allowed range for the free parameters. Thus, although
non-hyperbolic, it behaves as a saddle point. 
For the allowed values of the phase space variables and the allowed range
for the free parameters the expression  $g_3(\gamma, x_c, \Omega_{\lambda c},\Sigma_c)\geq
0.$ If it is strictly positive, then the fixed points located in the 2D
invariant set $E_5$ behaves as  saddle points. 

The eigenvalues of the perturbation matrix associated to $E_6^\epsilon$ are
  $\left\{0,0,	\lambda_1(\epsilon,  \Sigma_c),\lambda_2(\epsilon, 
\Sigma_c), \lambda_3(\epsilon,  \Sigma_c), \lambda_4(\epsilon, 
\Sigma_c)\right\}$ where $\lambda_1(\epsilon, 
\Sigma_c),\lambda_2(\epsilon,  \Sigma_c), \lambda_3(\epsilon,  \Sigma_c)$
and $\lambda_4(\epsilon,  \Sigma_c)$ are the roots of the polynomial
equation with real coefficients: 
	\begin{align}\label{eqLambda}
	&\lambda ^4+\lambda ^3 \sqrt{4-6 \Sigma_c^2} \epsilon  f'(0)+\lambda
^2\left[f(0)-4\right]  -4 \lambda  \sqrt{4-6 \Sigma_c^2} \epsilon  f'(0)-6
f(0) \Sigma_c^2=0.
		\end{align}

\begin{itemize}

\item For $f'(0)\neq 0, f(0)<0,$ equation \eqref{eqLambda} has only two
changes of signs in the sequence of its coefficients. Hence, using the
Descartes's rule, we conclude that for this range of the parameters, there
is zero or two positive roots of \eqref{eqLambda}. Substituting $\lambda$
by $-\lambda$ in \eqref{eqLambda} and applying the same rule we have zero
or two negative roots of \eqref{eqLambda} for $ f'(0)\neq 0, f(0)<0.$ If
all of them are complex conjugated, we need to resort to numerical
investigation. If none of them are complex conjugated, then the curve of
fixed points $E_6^\epsilon$ consists of saddle points. 

\item For $\epsilon f'(0)>0, f(0)>0,$ equation \eqref{eqLambda} has only
one change of sign in the sequence of its coefficients. Hence, using the
Descartes's rule, we conclude that for this range of the parameters, there
is one positive root of \eqref{eqLambda}. Substituting $\lambda$ by
$-\lambda$ in \eqref{eqLambda} and applying the same rule we have three
negative roots of \eqref{eqLambda} for $\epsilon f'(0)>0,f(0)>0.$ In
summary, for
$\epsilon f'(0)>0,f(0)>0,$ at least two eigenvalues of the linear
perturbation matrix of $E_6^+$ are real of different signs. In this case
the curve of fixed points $E_6^\epsilon$ consists of saddle points. 

\item For $\epsilon f'(0)<0, f(0)>0,$ equation \eqref{eqLambda}  has three
changes of sign in the sequence of its coefficients. Hence, using the
Descartes's rule, we conclude that for this range of the parameters, there
are three positive roots of \eqref{eqLambda}. Substituting $\lambda$ by
$-\lambda$ in \eqref{eqLambda} and applying the same rule we have only one
negative root of \eqref{eqLambda} for $\epsilon f'(0)<0,f(0)>0.$ In this
case the
curve of fixed points $E_6^\epsilon$ consists of saddle points.

\item For $f'(0)=f(0)=0$ the non null eigenvalues are $-2,2;$   for
$f'(0)=0, f(0)\neq 0,$ the non null eigenvalues are:\newline
$\displaystyle \pm\frac{\sqrt{-\sqrt{24 f(0)
\Sigma_c^2+(f(0)-4)^2}-f(0)+4}}{\sqrt{2}},\;\;\;\;\pm \frac{\sqrt{\sqrt{24 f(0)
\Sigma_c^2+(f(0)-4)^2}-f(0)+4}}{\sqrt{2}};$\newline
and for $f'(0)\neq 0, f(0)=0,$
the non null eigenvalues are $-2, 2, -\epsilon \sqrt{4 - 6
\Sigma_c^2}f'(0).$  Thus, in both cases $E_6^\pm$ consists of saddle
points. 

\end{itemize}

The eigenvalues of the perturbation matrix associated to $E_7^\epsilon$ are
  $\left\{0,0,	\mu_1(\epsilon,  \Sigma_c),\mu_2(\epsilon,  \Sigma_c),
\mu_3(\epsilon,  \Sigma_c), \mu_4(\epsilon,  \Sigma_c)\right\}$ where
$\mu_1(\epsilon,  \Sigma_c),\mu_2(\epsilon,  \Sigma_c), \mu_3(\epsilon, 
\Sigma_c)$ and $\mu_4(\epsilon,  \Sigma_c)$ are the roots of the polynomial
equation with real coefficients: 
\begin{align}
\label{eqmu}
&\mu ^4+\sqrt{2} \mu ^3 \Sigma_c \epsilon  f'(0)+\mu ^2 \left[2 (f(0)+4)
{\Sigma   _c}^2-8\right] +8 \sqrt{2} \mu  \Sigma_c \left(\Sigma_c^2-1\right)
\epsilon  f'(0)+4 f(0) \Sigma_c^2 \left(\Sigma_c^2-2\right)=0.
\end{align}

\begin{itemize}
\item For $\epsilon  f'(0)<0, f(0)<0, -\frac{1}{\sqrt{2}}<\Sigma_c <0$ or
$\epsilon f'(0)>0, f(0)<0, 0<\Sigma_c <\frac{1}{\sqrt{2}},$ or $\epsilon
f'(0)<0, f(0)<0, 0<\Sigma_c <\frac{1}{\sqrt{2}},$ or  $\epsilon  f'(0)>0,
f(0)<0, -\frac{1}{\sqrt{2}}<\Sigma_c <0,$ the  equation \eqref{eqmu} has
only two changes of signs in the sequence of its coefficients. Hence, using
the Descartes's rule, we conclude that for this range of the parameters,
there is zero or two positive roots of \eqref{eqmu}. Substituting $\mu$ by
$-\mu$ in \eqref{eqmu} and applying the same rule we have zero or two
negative roots of \eqref{eqmu} for the same values of the free parameters.
If all of them are complex conjugated, we need to resort to numerical
investigation. If none of them are complex conjugated, then the curve of
fixed points $E_7^\epsilon$ consists of saddle points. 

\item For $f(0)>0, \epsilon  f'(0)>0, 0<\Sigma_c <\frac{1}{\sqrt{2}}$ or
$f(0)>0, \epsilon  f'(0)<0, \frac{1}{\sqrt{2}}<\Sigma_c <0,$ equation
\eqref{eqmu} has only one change of sign in the sequence of its
coefficients. Hence, using the Descartes's rule, we conclude that for this
range of the parameters, there is one positive root of \eqref{eqmu}.
Substituting $\mu$ by $-\mu$ in \eqref{eqmu} and applying the same rule we
have three negative roots of \eqref{eqmu} for the same values of the
parameters. In this case the curve of fixed points $E_7^\epsilon$ consists
of saddle points. 

\item For $f(0)>0, 0<\Sigma_c <\frac{1}{\sqrt{2}},  \epsilon f'(0)<0$ or
$f(0)>0, -\frac{1}{\sqrt{2}}<\Sigma_c <0,  \epsilon f'(0)>0$ equation
\eqref{eqmu} has three changes of sign in the sequence of its coefficients.
Hence, using the Descartes's rule, we conclude that for this range of the
parameters, there are three positive roots of \eqref{eqmu}. Substituting
$\mu$ by $-\mu$ in \eqref{eqmu} and applying the same rule we have only one
negative root of \eqref{eqmu} for the same values of the parameters. In
this case the curve of fixed points $E_7^\epsilon$ consists of saddle
points. 

\item For $f'(0)=f(0)=0$ the non null eigenvalues are  $2 \sqrt{2}
\sqrt{1-\Sigma_c^2},-2 \sqrt{2} \sqrt{1-\Sigma_c^2};$   for $f'(0)=0,
f(0)\neq 0,$ the non null eigenvalues are $\pm\sqrt{4-(f(0)+4)
\Sigma_c^2-\sqrt{\Delta}}$, and
$\pm\sqrt{4-(f(0)+4) \Sigma_c^2+\sqrt{\Delta}},$ where $\Delta=(f(0) (f(0)+4)+16) \Sigma_c^4-32 \Sigma_c^2+16$;  and for $f'(0)\neq 0, f(0)=0,$ the non null eigenvalues
are  $-\sqrt{2} \Sigma_c \epsilon  f'(0),\pm 2 \sqrt{2-2 \Sigma_c^2}.$ Thus, in both cases $E_7^\pm$ consists of saddle
points. 

\end{itemize}

Observe that $g_4(\gamma, x, \Omega_\lambda)$ is always real-valued for the
allowed values of the phase space variables and the allowed range for the
free parameters. Then, although non-hyperbolic, $E_8$ behaves as a set of
saddle points.

When $\displaystyle \Omega_{\lambda c}\neq 1,\;\;\;
 \frac{\sqrt{6 \gamma 
(\Omega_{\lambda}+1)^2-8 \left(\Omega_{\lambda
}^2+1\right)}}{\Omega_{\lambda }+1}$ is real
valued, in this case the fixed points in the line $E_9$ behaves as saddle
points. 

For $\Sigma_c\neq 0,$ $\sqrt{2} \sqrt{\gamma  \left(9
\Sigma^2-6\right)-4 \Sigma^2+4}$ is real-valued, in
this case the fixed points in the line $E_{10}$ behaves as saddle points. 

For $\Sigma_c\neq 0,$ the linear perturbation matrix evaluated at $E_{11}$
has at least two real eigenvalues of different signs, thus, the fixed
points in the line $E_{11}$ behaves as saddle points. 

Observe that the line $E_9$ and the line $E_{11}$ contains the special
point with coordinates $F: (Q=0, x=0, \Omega_m=0, \Omega_\lambda=1,
\Sigma=0)$ in the first case when $\Omega_{\lambda c}\rightarrow 1$ whereas
in the second case for $\Sigma_c\rightarrow 0.$ Taking in both cases the
proper limits we have that the eigenvalues of the linearization for $F$ are
$\left\{0,0,-\sqrt{6 \gamma -4},\sqrt{6 \gamma -4},0,0\right\}.$  Hence,
for $\gamma> \frac{2}{3},$ $F$ is of saddle type, whereas for  $\gamma<
\frac{2}{3},$ there are two purely imaginary eigenvalues and the rest of
the eigenvalues are zero. In this case we cannot say anything about its
stability from the linearization and we need to resort to numerical
inspection.

Observe that $\sqrt{\gamma  \left(6-9 \Sigma ^2\right)+18
\Sigma ^2-8}$ is real-valued for the allowed values of
the phase space variables and the allowed range for the free parameters.
Thus, all the points located at the line $E_{12}$ behaves as saddle
points. 

For $0<\gamma<\frac{4}{3},$ $E_{13}^\pm$ behaves as a saddle point.

\section{Stability of expanding (contracting)
solutions}\label{nonstatic}

Now let us comment on the stability of the first order perturbations of
\eqref{eqQ}-\eqref{eqs} near the critical points showed in table
\eqref{tab2}. 

The line of singular points $Q_1^\pm$, although it is non-hyperbolic
(actually, normally hyperbolic), behaves like a saddle point in the
phase space of the RS model, since
they have both nonempty stable and unstable manifolds (see the table
\ref{autovalores1}). The class $Q_1^+$ is the analogous to the line denoted
by $P_1$ in \citep{Escobar2012a}.

The singular points $Q_{2}^\pm(s^*)$ and $Q_3^\pm(s^*)$ are non-hyperbolic
(these points are related to $P_3^\pm$ investigated in
\citep{Escobar2012a}), however they behave as saddle points since they
have both nonempty stable and unstable manifolds (see the table
\ref{autovalores1}).

The singular point $Q_4^+(s^*)$ is the analogous to $P_4$ in
\citep{Escobar2012a}. It is a stable node in the cases 
$0<\gamma \leq \frac{2}{9},s^*<-\sqrt{3\gamma },f'\left(s^*\right)<0$ or
$\frac{2}{9}<\gamma <\frac{4}{3},-\frac{2 \sqrt{6} \gamma }{\sqrt{9 \gamma
-2}}\leq s^*<-\sqrt{3\gamma },
   f'\left(s^*\right)<0,$ or $0<\gamma \leq \frac{2}{9},s^*>\sqrt{3\gamma
}, f'\left(s^*\right)>0,$ or $\frac{2}{9}<\gamma <\frac{4}{3},\sqrt{3\gamma
}<s^*\leq \frac{2 \sqrt{6} \gamma }{\sqrt{9 \gamma -2}},
   f'\left(s^*\right)>0.$ It is a spiral stable point for \\ $\frac{2}{9}<\gamma
<\frac{4}{3},s^*<-\frac{2 \sqrt{6} \gamma }{\sqrt{9 \gamma
-2}},f'\left(s^*\right)<0$ or $\frac{2}{9}<\gamma <\frac{4}{3},s^*>\frac{2
\sqrt{6} \gamma }{\sqrt{9 \gamma -2}},f'\left(s^*\right)>0.$ In summary, it is stable for $0<\gamma
<\frac{4}{3},s^*<-\sqrt{3\gamma },f'\left(s^*\right)<0$ or $0<\gamma
<\frac{4}{3},s^*>\sqrt{3\gamma },f'\left(s^*\right)>0.$ Otherwise, it is a
saddle point.

The singular point $Q_4^-(s^*)$ is a local source under the same conditions
for which $Q_4^+(s^*)$ is stable.

The singular point $Q_5^+(s^*)$ is the analogous to $P_5$ in
\citep{Escobar2012a}. It is not hyperbolic for
$s^*\in\left\{0,\pm\sqrt{6},\pm\sqrt{3\gamma},2\right\}$ or $f'(s^*)=0$. In the hyperbolic case, $Q_5^+(s^*)$ is a stable node for   $0<\gamma
\leq \frac{4}{3},-\sqrt{3 \gamma }<s^*<0,f'\left(s^*\right)<0$ or
$\frac{4}{3}<\gamma \leq 2,-2<s^*<0,f'\left(s^*\right)<0$ or $0<\gamma \leq
\frac{4}{3},0<s^*<\sqrt{3 \gamma },f'\left(s^*\right)>0$ or
$\frac{4}{3}<\gamma \leq 2,0<s^*<2,f'\left(s^*\right)>0;$ otherwise, it is
a saddle point.

The singular point $Q_5^-(s^*)$ is a local source under the same conditions
for which $Q_5^+(s^*)$ is stable.

For evaluating the Jacobian matrix, and obtaining their eigenvalues at the
singular points $Q_{6}^\pm,$ $Q_{7}^\pm$ and $Q_{11}^\pm$ we need to take
the appropriate limits.  The details on the stability analysis for
$Q_{6}^+$ and $Q_{11}^+$ are offered in the Appendix \ref{section3.2}.
The analysis of $Q_7^+$ is essentially the same as for $Q_6^+.$ 

The circles of critical points  $Q_8 ^\pm$ are non-hyperbolic. Both
solutions represents transient states in the evolution of the universe. 

The singular point $Q_{9}^\pm$ and the line of critical points $Q_{10}^\pm$
are non-hyperbolic  ($Q_{9}^+$ is the analogous of $P_{10}$ for 
$\Omega_{\lambda}=0$  while $Q_{10}^+$ is the analogous of $P_{10}$ for
$\Omega_{\lambda}\neq 0$ in \citep{Escobar2012a}).  In the Appendix
\ref{dSsection} we  analyze the stability of $Q_9^+$ by introducing a local
set of coordinates adapted to the singular points.  The non-hyperbolic
fixed point $Q_9^-,$ behaves as a past attractor. 

The singular point $Q_{11}^+$ is the analogous of $P_{11}$ in
\citep{Escobar2012a}, this point represents a 1D set of singular points
such that ($\Omega_\lambda=1$), parametrized by the values of  $s_c.$ It
is normally hyperbolic. 

The singular points $Q_{12}^\pm(s^*)$ are non-hyperbolic for $s^*=0,\pm2$,
$\gamma=4/3$ or $f'(s^*)=0$. $Q_{12}^+(s^*)$ is a stable node for 
$\frac{4}{3}<\gamma \leq 2,2<s^*\leq \frac{8}{\sqrt{15}},
f'\left(s^*\right)>0,$ a stable spiral for  $\frac{4}{3}<\gamma \leq 2,
s^*>\frac{8}{\sqrt{15}}, f'\left(s^*\right)>0.$ In summary, $Q_{12}^+(s^*)$
is  stable for $\frac{4}{3}<\gamma \leq 2, s^*>2, f'\left(s^*\right)>0.$
Otherwise it is a saddle point. $Q_{12}^-(s^*)$ is unstable for the same
conditions for which  $Q_{12}^+(s^*)$ is stable.

The eigenvalues of the linear perturbation matrix evaluated at each of
these critical points are displayed in the table \ref{autovalores1}.

\begin{table*}[tbp]
\begin{center}
\caption{ Eigenvalues for the critical points in table
\eqref{tab2}. We use the notations
$K^\pm(\gamma, s^*)=-\frac{3}{4} \left(2-\gamma\pm \frac{\sqrt{2-\gamma }
\sqrt{{s^*}^2 (2-9 \gamma
   )+24 \gamma ^2}}{s^*}\right),$ 
			$L^\pm(\Omega_\lambda)=-\frac{1}{2}  \left(3\pm
\sqrt{12 f(0) (\Omega_\lambda -1)+9}\right)$ and 
	$M^\pm(s^*)=\frac{1}{4} \left(-s^*\pm\sqrt{64-15
\left(s^*\right)^2}\right) ,$
where $s^*$ denotes an $s$-value
such that $f(s^*)=0.$}
\label{autovalores1}
{\begin{tabular}{@{\hspace{4pt}}c@{\hspace{14pt}}c@{\hspace{14pt}}c}
\hline
\hline\\[-0.3cm]
Label & Eigenvalues\\[0.1cm]
\hline\\[-0.2cm]
$Q_{1}^\pm$& $\left\{(3 \gamma -4)\epsilon,\frac{3}{2} (\gamma -2) \epsilon
,\frac{3}{2} (\gamma -2) \epsilon ,-3 \gamma  \epsilon ,3 \gamma 
\epsilon,0 \right\}$\\[0.2cm]
$Q_{2}^\pm(s^*)$& $\left\{2\epsilon,0,-6 \epsilon ,-3 (\gamma -2) \epsilon
,6 \epsilon -\sqrt{6} s^*,-\sqrt{6} f'\left(s^*\right)\right\}$\\[0.2cm]
$Q_{3}^\pm(s^*)$&$\left\{2\epsilon,0,-6 \epsilon ,-3 (\gamma -2) \epsilon
,6 \epsilon \sqrt{6} s^* ,\sqrt{6} f'\left(s^*\right)\right\}$\\[0.2cm]
$Q_{4}^\pm(s^*)$& $\left\{-3 \gamma  \epsilon ,\frac{3}{2} (\gamma -2)
\epsilon ,(3 \gamma -4)\epsilon, \epsilon  K^+(\gamma, s^*),\epsilon 
K^-(\gamma, s^*),-\frac{3 \gamma  \epsilon 
f'\left(s^*\right)}{s^*}\right\}$\\[0.2cm]
$Q_{5}^\pm(s^*)$&  $\left\{-\epsilon\left( s^*\right)^2,\frac{1}{2}
\left({s^*}^2-6\right) \epsilon ,\frac{1}{2} \left({s^*}^2-6\right)
\epsilon ,
   \left({s^*}^2-3 \gamma \right)\epsilon,({s^*}^2-4)\epsilon,-\epsilon s^*
 f'\left(s^*\right)\right\}$\\[0.2cm]
$Q_{6}^\pm$& $\{6 \epsilon ,6 \epsilon ,2 \epsilon ,0,0,3 (2-\gamma )
\epsilon \}$\\[0.2cm]
$Q_{7}^\pm$& $\{6 \epsilon ,6 \epsilon ,2 \epsilon ,0,0,3 (2-\gamma )
\epsilon \}$\\[0.2cm]
$Q_{8}^\pm(s^*)$& $\left\{0,6 \epsilon -\sqrt{6} s \cos (u),-3 (\gamma -2)
\epsilon ,2 \epsilon ,-6 \epsilon ,-\sqrt{6} \cos (u)
f'(s)\right\}$\\[0.2cm]
$Q_{9}^\pm$& $\left\{-4\epsilon,-3\epsilon,0,0,\frac{1}{2}
\left(-\sqrt{9-12 f(0)} -3\right)\epsilon,\frac{1}{2}
   \left(\sqrt{9-12 f(0)}-3\right)\epsilon\right\}$\\[0.2cm]
$Q_{10}^\pm$& $\left\{-4\epsilon,-3\epsilon,0,-3
\gamma\epsilon\Omega_\lambda ,\epsilon L^+(\Omega_\lambda),\epsilon
L^-(\Omega_\lambda)\right\}$\\[0.2cm]
$Q_{11}^\pm$& $\left\{2(3\gamma -2)\epsilon, 3(\gamma-1)\epsilon, 3\gamma
\epsilon, 3\gamma\epsilon, 3(\gamma-1)\epsilon, 0\right\}$\\[0.2cm]
$Q_{12}^\pm(s^*)$& $\left\{-2 s^* \epsilon ,-\frac{s^* \epsilon
}{2},\frac{1}{2} (4-3 \gamma ) s^* \epsilon ,\epsilon M^+(s^*) , \epsilon
M^-(s^*)  ,-2 \epsilon  f'\left(s^*\right)\right\}$\\[0.4cm]\hline \hline
\end{tabular}
}\end{center}
\end{table*}

\subsection{De Sitter solutions.}\label{dSsection}

To investigate the de Sitter solution $Q_{9}^+,$ we introduce the local
coordinates:
\begin{align}
& \left\{x,\Omega_\lambda,Q-1,\Omega_m,\Sigma,s\right\}=  \epsilon\left\{\hat{x
},\hat{\Omega}_\lambda,\hat{Q},\hat{\Omega}_m,\hat{\Omega}_\sigma,\hat{s}
\right\}+{\cal O}(\epsilon)^2,
\end{align}
where $\epsilon$ is a small constant $\epsilon\ll 1.$ 
Then, we obtain the approximated system 
\begin{align}\label{DS}
&\hat{Q}'=-4 \hat{Q}, \; \hat{x}'=\sqrt{\frac{3}{2}} \hat{s}-3 \hat{x}, \; \hat{\Omega}_\lambda'=0, \; \hat{\Omega}_m'=-3 \gamma 
   \hat{\Omega}_m, \; 
	\hat{\Omega}_\sigma'=-3 
   \hat{\Omega}_\sigma, \;
	\hat{s}'=-\sqrt{6} f(0) \hat{x}.
\end{align}
The system \eqref{DS} admits the exact solution passing by ($\hat{Q}_0,\hat{x}_0,\hat{\Omega}_{\lambda0},\hat{\Omega}_{m0},\hat{\Omega
}_{\sigma0},\hat{s}_0$) at $\tau=0$
given by
\begin{align}
&\hat{Q}(\tau)=\hat{Q}_0 e^{-4 \tau }, \; \hat{x}(\tau)=\frac{1}{2} \hat{x}_0 e^{-\frac{1}{2} \tau 
   (\vartheta +3)} \left(e^{\tau  \vartheta
}+1\right) -\frac{e^{-\frac{1}{2} \tau  (\vartheta +3)} \left(e^{\tau 
\vartheta }-1\right)
   \left(\sqrt{6} \hat{s}_0 (u_c-1)+3 \hat{x}_0\right)}{2 \vartheta }, \nonumber \\
& \hat{\Omega}_\lambda(\tau)=\hat{\Omega}_{\lambda0},\;
\hat{\Omega}_m(\tau)=\hat{\Omega}_{m0} e^{-3\gamma \tau},
\hat{\Omega}_\sigma(\tau)=\hat{\Omega}_{\sigma0}e^{-3\tau},\nonumber\\&
\hat{s}(\tau)= \hat{s}_0 e^{-\frac{3}{2} \tau} \cosh (\beta  \tau )+\frac{3 e^{-\frac{3}{2} \tau} \left(2 \hat{s}_0-\sqrt{6} \hat{x}_0\right) \sinh (\beta  \tau
)}{4 \beta }+ \sqrt{\frac{2}{3}} \beta  e^{-\frac{3}{2} \tau}
   \hat{x}_0 \sinh (\beta  \tau ),
\end{align}
where $\beta=\frac{1}{2} \sqrt{9-12 f(0)}.$

For the choice $\beta^2<\frac{9}{4},$ i.e., $f(0)>0,$ all the perturbations
$\hat{Q},\hat{x} ,\hat{\Omega}_{m },\hat{\Omega}_{\sigma},\hat{s}$ shrink
to zero as $\tau\rightarrow +\infty,$ and $\hat{\Omega}_{\lambda }$
converges to a constant value. 
For $\beta=\pm \frac{3}{2},$ i.e., for $f(0)=0,$ $\hat{x}\rightarrow
\frac{\hat{s}_0}{\sqrt{6}}$ and $\hat{s}\rightarrow \hat{s}_0,$ and the
other perturbations shrink to zero as $\tau\rightarrow +\infty.$ Combining
the above arguments we obtain that for $f(0)\geq 0,$ $Q_{9}^+$ is stable,
but not asymptotically stable.
For $\beta^2>\frac{9}{4},$ i.e., $f(0)<0,$ all the perturbation values, but
$\hat{x}$ and $\hat{s}$, which diverges, shrink to zero as $\tau\rightarrow
+\infty.$ Thus, $Q_{9}^+$ is a saddle.

For analyzing the curve of singular points $Q_{10}^+$ we consider an
arbitrary value $\Omega_\lambda=u_c,$ $0<u_c<1,$ and introduce the local
coordinates 
\begin{align}
&\left\{x,\Omega_\lambda-u_c,Q-1,\Omega_m,\Sigma,s\right\}=  \epsilon\left\{
\hat{x},\hat{\Omega}_\lambda,\hat{Q},\hat{\Omega}_m,\hat{\Omega}_\sigma,
\hat{s}\right\}+{\cal O}(\epsilon)^2,
\end{align}
where $\epsilon$ is a small constant $\epsilon\ll 1.$ 

Then, we obtain the approximated system 
\begin{align}\label{DS2}
&\hat{Q}'=-4 \hat{Q}, \; \hat{x}'=
-3 \hat{x}-\sqrt{\frac{3}{2}} \hat{s} (u_c-1), \; \hat{\Omega}_\lambda'=-{3
\gamma 
   u_c \hat{\Omega}_m}, \; \hat{\Omega}_m'=-3 \gamma 
   \hat{\Omega}_m, \; \hat{\Omega}_\sigma'=-3 
   \hat{\Omega}_\sigma, \; \hat{s}'=-\sqrt{6} f(0) \hat{x}.
\end{align}
 
The system \eqref{DS2} admits the exact solution passing by 
($\hat{Q}_0,\hat{x}_0,\hat{\Omega}_{\lambda0},\hat{\Omega}_{m0},\hat{\Omega
}_{\sigma0},\hat{s}_0$) at $\tau=0$
given by
\begin{align}
&\hat{Q}(\tau)=\hat{Q}_0 e^{-4 \tau }, \; \hat{x}(\tau)=\frac{e^{-\frac{3}{2} \tau} \left(\sqrt{6} \hat{s}_0-3
\hat{x}_0\right) \sinh (\beta  \tau )}{2 \beta }+e^{-\frac{3}{2} \tau} \hat{x}_0
\cosh (\beta 
   \tau ), \; \hat{\Omega}_\lambda(\tau)=u_c \hat{\Omega}_{m0}
\left(e^{-3 \gamma  \tau }-1\right)+\hat{\Omega}_ {\lambda0},\nonumber\\&
\hat{\Omega}_m(\tau)=\hat{\Omega}_{m0} e^{-3\gamma \tau},
\hat{\Omega}_\sigma(\tau)=\hat{\Omega}_{\sigma0}e^{-3\tau},\nonumber\\&
\hat{s}(\tau)= \frac{3 e^{-\frac{1}{2} \tau  (\vartheta +3)} \left(e^{\tau 
   \vartheta }-1\right) \left(2 \hat{s}_0 (u_c-1)+\sqrt{6}
\hat{x}_0\right)}{4 (u_c-1) \vartheta }+\frac{1}{2} \hat{s}_0
   e^{-\frac{1}{2} \tau  (\vartheta +3)} \left(e^{\tau  \vartheta
}+1\right)-\frac{\hat{x}_0 \vartheta  e^{-\frac{1}{2} \tau 
(\vartheta
   +3)} \left(e^{\tau  \vartheta }-1\right)}{2 \sqrt{6} (u_c-1)},
\end{align}
where $\vartheta=\sqrt{12 f(0) (u_c-1)+9}.$
It is easy to see that for $f(0)\geq 0$ the solution is stable, but not
asymptotically stable. For $f(0)<0$ is of saddle type. 

It is worthy to mention that the above results could be proved by noticing
that $Q_{10}^+$ -as  1D set- is normaly hyperbolic since the
eigen-direction
associated with the null eigenvalue, $(0,0,1,0,0,0)^T,$ that is, the
$\Omega_\lambda$-axis, is tangent to the set. Thus, the stability issue can
be resolved by analyzing the signs of the non-null eigenvalues.
However, let us comment that if we include the variable $y=\frac{V}{3
H^2},$ in the analysis, the solution is unstable to perturbations along the
$y$-direction.

\section{Asymptotic analysis on the singular surface
$\Omega_{\lambda}+\Sigma^2=1.$}\label{section3.2}

In this section we analyze the asymptotic structures of the system
\eqref{eqQ}-\eqref{eqs} near the critical points at the singular surface 
$\Omega_{\lambda}+\Sigma^2=1.$

\subsection{Shear-dominated solution.}

Observe that the one of the singular points at the surface
$\Omega_{\lambda}+\Sigma^2=1$ that represents a shear-dominated solution is
$Q_{6}^+.$
To investigate its stability we introduce the local coordinates:
\begin{align}
&\left\{x,\Omega_\lambda,Q-1,\Omega_m,\Sigma+1,s-s_c\right\}=   \epsilon\left\{
\hat{x},\hat{\Omega}_\lambda,\hat{Q},\hat{\Omega}_m,\hat{\Omega}_\sigma,
\hat{s}\right\} +{\cal O}(\epsilon)^2,
\end{align}
where $\epsilon$ is a constant satisfying $\epsilon<<1$ and $ s_c$ is an
arbitrary real value for $s.$ Taylor expanding the resulting system and
truncating the second order terms we obtain the approximated system
\begin{align}
&\hat{Q}'=2\hat{Q},\; 
\hat{x}'=-\sqrt{\frac{3}{2}}s_c \left(-2
\hat{\Omega}_\sigma+{\hat{\Omega}_\lambda}+{\hat{\Omega}_m}\right),\;
\hat{\Omega}_\lambda'=6 \hat{\Omega}_\lambda  \left(\frac{\gamma 
\hat{\Omega}_m}{\hat{\Omega}_\lambda-2 \hat{\Omega}_ \sigma}+1\right),\; \hat{\Omega}_m'=3(2-\gamma)\hat{\Omega}_m,\nonumber \\ & 
\hat{\Omega}_\sigma'=\frac{3 \gamma  \hat{\Omega}_\lambda
\hat{\Omega}_m}{\hat{\Omega}_\lambda-2 \hat{\Omega}_\sigma}+\frac{3}{2}
(4\hat{\Omega}_\sigma
-\gamma\hat{\Omega}_m),\;
\hat{s}'=-\sqrt{6}\hat{x} f(s_c)\label{3.6}.
\end{align}

Let us define the rates:
\be
\{r_m,r_\sigma\}=\frac{1}{\hat{\Omega}_\lambda}\{\hat{\Omega}_m,
\hat{\Omega}_\sigma\}.
\ee

Let us assume $\gamma>0.$ With the above assumptions we can define the new
time variable $N=\gamma \ln a$ which preserves the time arrow.
Then we deduce the differential equations
\be  r_m'= \frac{6 r_m^2}{2 r_\sigma -1}-3 r_m, \;   r_\sigma'=\frac{3  r_m}{2},\ee where the comma denotes
derivatives with respect to $N.$

The system admits the general solution:
\be
 r_m=\frac{1}{-2 c_1 \cosh \left(3 N-c_2\right)-2 c_1}, \; r_\sigma=\frac{2
c_1-1+\tanh \left(\frac{1}{2} \left(3 N-c_2\right)\right)}{4 c_1},
\ee 
or 
\be r_m=\frac{1}{-2 c_1 \cosh \left(3 N-c_2\right)-2 c_1}, \; r_\sigma=\frac{2
c_1-1-\tanh \left(\frac{1}{2} \left(3 N-c_2\right)\right)}{4 c_1},
\ee 
where $c_1\neq 0$ and $c_2$ are arbitrary constants. 
In the general case we have that $r_m\rightarrow 0$ and $r_\sigma$ tends to
a constant as $N\rightarrow -\infty.$

Thus, as $\tau\rightarrow -\infty$ the system
\eqref{3.6} has the asymptotic structure:
\begin{align}
&\hat {Q}'=2\hat{Q},
\hat{x}'=-\sqrt{\frac{3}{2}} s_c (\hat{\Omega}_\lambda -2
\hat{\Omega}_\sigma ), \;
\hat{\Omega}_\lambda'=6\hat{\Omega}_\lambda,
\hat{\Omega}_m'=3(2-\gamma)\hat{\Omega}_m, \;
\hat{\Omega}_\sigma'=6\hat{\Omega}_\sigma,
\hat{s}'=-\sqrt{6}\hat{x} f(s_c).\label{4.6}
\end{align} 
 
The system \eqref{4.6} admits the exact solution passing by
(${Q}_0,\hat{x}_0,\hat{\Omega}_{\lambda0},\hat{\Omega}_{m0},\hat{\Omega}_{
\sigma0},\hat{s}_0$) at $\tau=0$
given by
\begin{align}
&{Q}(\tau)=\hat{Q}_0 e^{2\tau},
\hat{x}(\tau)=\hat{x}_0-\frac{s_c \left(e^{6 \tau }-1\right)
(\hat{\Omega}_{\lambda 0}-2 \hat{\Omega}_{\sigma 0})}{2 \sqrt{6}}, \;
\hat{\Omega}_\lambda(\tau)=\hat{\Omega}_{\lambda0}e^{6\tau},\;
\hat{\Omega}_m(\tau)=\hat{\Omega}_{m0} e^{3(2-\gamma)\tau},
\hat{\Omega}_\sigma(\tau)=\hat{\Omega}_{\sigma0}e^{6\tau},\nonumber\\&
\hat{s}(\tau)= \frac{1}{12} s_c \left(-6 \tau +e^{6 \tau }-1\right) f(s_c)
(\hat{\Omega}_{\lambda 0}-2 \hat{\Omega}_{\sigma 0}) -\sqrt{6} \tau 
\hat{x}_0
   f(s_c)+\hat{s}_0.
\end{align}
Thus, the energy density perturbations goes to zero,  and $s$ diverge  as
$\tau\rightarrow -\infty$. For that reason, $Q_{6}^+$ cannot be the future
attractor of the system, thus $Q_{6}^+$ has a large probability to be a
past attractor. Using the same approach for $Q_{6}^-$ we obtain that as
$\tau\rightarrow +\infty,$ the energy density perturbations goes to zero, 
but $s$ diverge. Thus, although $Q_{6}^-$ have a large probability to be a
late time attractor, actually it is of saddle type. For $Q_7^\pm$ we have
similar results as for $Q_6^\pm.$ Since the procedure is essentially the
same as for $Q_6^\pm,$ we omit the details. 

\subsection{Solution with 5D-corrections.}

Another singular points at the surface $\Omega_{\lambda}+\Sigma^2=1$ are the isotropic solutions with  $\Omega_\lambda=1$ (critical points $Q_{11}^\pm$). 
For the stability analysis of $Q_{11}^+$ , we introduce the local coordinates:
\begin{align}
&\left\{x,\Omega_\lambda-1,Q-1,\Omega_m,\Sigma,s-s_c\right\}=  \epsilon\left\{\hat{x},\hat{\Omega}_\lambda,\hat{Q},\hat{\Omega}_m,\hat{\Omega}_\sigma,\hat{s}\right\}+{\cal O}(\epsilon)^2,
\end{align}
where $\epsilon$ is a constant satisfying $\epsilon<<1$ and $Q_c, s_c$ are arbitrary real values of $Q, s$ respectively.
Then, we obtain the approximated system 
\begin{align}
& \hat{Q}'=-2 \hat{Q} (3 \gamma  r+2), \; \hat{x}'=\frac{1}{2} \left(-6 \hat{x} (\gamma  r+1)-\sqrt{6} s_c (\hat{\Omega}_\lambda+\hat{\Omega}_m)\right),
\;
\hat{\Omega}_\lambda'=-3 \gamma  r
   \hat{\Omega}_\lambda, \nonumber \\ &
\hat{\Omega}_m'=-3 \gamma  (2 r+1) \hat{\Omega}_m,
\;
\hat{\Omega}_\sigma'=-3 \hat{\Omega}_\sigma (\gamma  r+1),
\hat{s}'=-\sqrt{6} \hat{x} f(s_c)\label{P12.6}.
\end{align}
where we have defined the rate $r=\frac{\hat{\Omega}_m}{\hat{\Omega}_\lambda}.$
The evolution equation for $r$ is given by 
\be\label{evolr} r'=-3 \gamma  r (r+1).
\ee
The equation \eqref{evolr} has two singular points $r=0$ with eigenvalue $ \frac{dr'}{dr}|_{r=0}=-3\gamma$ and  $r=-1$ with eigenvalue $ \frac{dr'}{dr}|_{r=-1}=3\gamma.$ Then, assuming $\gamma>0$ we have that  $r\rightarrow -1 $ as $\tau\rightarrow -\infty.$
This argument allow us to prove that as as $\tau\rightarrow -\infty,$ the system \eqref{P12.6} has the asymptotic structure 
\begin{align}
&\hat{Q}'=2 (3 \gamma -2)\hat{Q}, 
\hat{x}'=3 \hat{x} (\gamma-1), \;
\hat{\Omega}_\lambda'=3  \gamma  \hat{\Omega}_\lambda, 
\hat{\Omega}_m'=3 \gamma  \hat{\Omega}_m, \;
\hat{\Omega}_\sigma'=3 \hat{\Omega}_\sigma (\gamma -1),
\hat{s}'=-\sqrt{6}\hat{x} f(s_c)\label{asymptoticP12.6}.
\end{align}

The system \eqref{asymptoticP12.6} admits the exact solution passing by (${Q}_0,\hat{x}_0,\hat{\Omega}_{\lambda0},\hat{\Omega}_{m0},\hat{\Omega}_{\sigma0},\hat{s}_0$) at $\tau=0$
given by
\begin{align}
&\hat{Q}(\tau)=\hat{Q}_0 e^{2(3\gamma-2)\tau},
\hat{x}(\tau)=\hat{x}_0 e^{3 (\gamma-1)\tau}, \hat{\Omega}_\lambda(\tau)=\hat{\Omega}_{\lambda0}e^{3\gamma\tau},\; \hat{\Omega}_m(\tau)=\hat{\Omega}_{m0} e^{3\gamma\tau},
\hat{\Omega}_\sigma(\tau)=\hat{\Omega}_{\sigma0}e^{3(\gamma-1)\tau},\nonumber \\ & \hat{s}(\tau)= \hat{s}_0-\frac{\sqrt{\frac{2}{3}} \hat{x}_0 f(s_c) \left(e^{3 (\gamma -1) 
   \tau }-1\right)}{(\gamma -1)}.
\end{align}
 Hence, for $ \gamma>1,$ the perturbations $\hat{Q},\hat{x}, \hat{\Omega}_\lambda,\hat{\Omega}_m,\hat{\Omega}_\sigma$ shrink to zero as $\tau\rightarrow -\infty.$ In that limit  $s$ tends to a finite value. In this case, the solution under investigation has a 4D unstable manifold. The numerical simulations suggest that the singular point at the surface $\Omega_{\lambda}+\Sigma^2=1$ with  $\Omega_\lambda=1$ is a local source. 
On the other hand, for $\gamma<1,$ the solution behave as a saddle point, since the perturbation values  $\hat{x}$ and $\hat{\Omega}_\sigma$ do not tend to zero as $\tau\rightarrow -\infty.$ Using the same approach for $Q_{11}^-$ we obtain that for $\gamma>1,$ $Q_{11}^-$ is stable as $\tau\rightarrow +\infty,$ whereas for $\gamma<1$ it is of saddle type. 

These results for $Q_{11}^+$ and for $Q_{11}^-$ are in agreement with the analogous results for $m_+$ and for $m_-$ in the reference \citep{Campos2001a}. 

\end{appendix}






\begin{thebibliography}{141}
\expandafter\ifx\csname natexlab\endcsname\relax\def\natexlab#1{#1}\fi
\expandafter\ifx\csname bibnamefont\endcsname\relax
  \def\bibnamefont#1{#1}\fi
\expandafter\ifx\csname bibfnamefont\endcsname\relax
  \def\bibfnamefont#1{#1}\fi
\expandafter\ifx\csname citenamefont\endcsname\relax
  \def\citenamefont#1{#1}\fi
\expandafter\ifx\csname url\endcsname\relax
  \def\url#1{\texttt{#1}}\fi
\expandafter\ifx\csname urlprefix\endcsname\relax\def\urlprefix{URL }\fi
\providecommand{\bibinfo}[2]{#2}
\providecommand{\eprint}[2][]{\url{#2}}

\bibitem[{\citenamefont{Randall and
  Sundrum}(1999{\natexlab{a}})}]{Randall1999a}
\bibinfo{author}{\bibfnamefont{L.}~\bibnamefont{Randall}} \bibnamefont{and}
  \bibinfo{author}{\bibfnamefont{R.}~\bibnamefont{Sundrum}},
  \bibinfo{journal}{Phys.Rev.Lett.} \textbf{\bibinfo{volume}{83}},
  \bibinfo{pages}{3370} (\bibinfo{year}{1999}{\natexlab{a}}),
  \eprint{hep-ph/9905221}.

\bibitem[{\citenamefont{Randall and Sundrum}(1999{\natexlab{b}})}]{Randall1999}
\bibinfo{author}{\bibfnamefont{L.}~\bibnamefont{Randall}} \bibnamefont{and}
  \bibinfo{author}{\bibfnamefont{R.}~\bibnamefont{Sundrum}},
  \bibinfo{journal}{Phys.Rev.Lett.} \textbf{\bibinfo{volume}{83}},
  \bibinfo{pages}{4690} (\bibinfo{year}{1999}{\natexlab{b}}),
  \eprint{hep-th/9906064}.

\bibitem[{\citenamefont{Binetruy
  et~al.}(2000{\natexlab{a}})\citenamefont{Binetruy, Deffayet, and
  Langlois}}]{Binetruy2000}
\bibinfo{author}{\bibfnamefont{P.}~\bibnamefont{Binetruy}},
  \bibinfo{author}{\bibfnamefont{C.}~\bibnamefont{Deffayet}}, \bibnamefont{and}
  \bibinfo{author}{\bibfnamefont{D.}~\bibnamefont{Langlois}},
  \bibinfo{journal}{Nucl.Phys.} \textbf{\bibinfo{volume}{B565}},
  \bibinfo{pages}{269} (\bibinfo{year}{2000}{\natexlab{a}}),
  \eprint{hep-th/9905012}.

\bibitem[{\citenamefont{Binetruy
  et~al.}(2000{\natexlab{b}})\citenamefont{Binetruy, Deffayet, Ellwanger, and
  Langlois}}]{Binetruy2000a}
\bibinfo{author}{\bibfnamefont{P.}~\bibnamefont{Binetruy}},
  \bibinfo{author}{\bibfnamefont{C.}~\bibnamefont{Deffayet}},
  \bibinfo{author}{\bibfnamefont{U.}~\bibnamefont{Ellwanger}},
  \bibnamefont{and} \bibinfo{author}{\bibfnamefont{D.}~\bibnamefont{Langlois}},
  \bibinfo{journal}{Phys.Lett.} \textbf{\bibinfo{volume}{B477}},
  \bibinfo{pages}{285} (\bibinfo{year}{2000}{\natexlab{b}}),
  \eprint{hep-th/9910219}.

\bibitem[{\citenamefont{Bowcock et~al.}(2000)\citenamefont{Bowcock, Charmousis,
  and Gregory}}]{Bowcock2000}
\bibinfo{author}{\bibfnamefont{P.}~\bibnamefont{Bowcock}},
  \bibinfo{author}{\bibfnamefont{C.}~\bibnamefont{Charmousis}},
  \bibnamefont{and} \bibinfo{author}{\bibfnamefont{R.}~\bibnamefont{Gregory}},
  \bibinfo{journal}{Class.Quant.Grav.} \textbf{\bibinfo{volume}{17}},
  \bibinfo{pages}{4745} (\bibinfo{year}{2000}), \eprint{hep-th/0007177}.

\bibitem[{\citenamefont{Apostolopoulos
  et~al.}(2005)\citenamefont{Apostolopoulos, Brouzakis, Saridakis, and
  Tetradis}}]{Apostolopoulos2005}
\bibinfo{author}{\bibfnamefont{P.~S.} \bibnamefont{Apostolopoulos}},
  \bibinfo{author}{\bibfnamefont{N.}~\bibnamefont{Brouzakis}},
  \bibinfo{author}{\bibfnamefont{E.~N.} \bibnamefont{Saridakis}},
  \bibnamefont{and} \bibinfo{author}{\bibfnamefont{N.}~\bibnamefont{Tetradis}},
  \bibinfo{journal}{Phys.Rev.} \textbf{\bibinfo{volume}{D72}},
  \bibinfo{pages}{044013} (\bibinfo{year}{2005}), \eprint{hep-th/0502115}.

\bibitem[{\citenamefont{Hawkins and Lidsey}(2001)}]{Hawkins2001}
\bibinfo{author}{\bibfnamefont{R.~M.} \bibnamefont{Hawkins}} \bibnamefont{and}
  \bibinfo{author}{\bibfnamefont{J.~E.} \bibnamefont{Lidsey}},
  \bibinfo{journal}{Phys.Rev.} \textbf{\bibinfo{volume}{D63}},
  \bibinfo{pages}{041301} (\bibinfo{year}{2001}), \eprint{gr-qc/0011060}.

\bibitem[{\citenamefont{Huey and Lidsey}(2001)}]{Huey2001}
\bibinfo{author}{\bibfnamefont{G.}~\bibnamefont{Huey}} \bibnamefont{and}
  \bibinfo{author}{\bibfnamefont{J.~E.} \bibnamefont{Lidsey}},
  \bibinfo{journal}{Phys.Lett.} \textbf{\bibinfo{volume}{B514}},
  \bibinfo{pages}{217} (\bibinfo{year}{2001}), \eprint{astro-ph/0104006}.

\bibitem[{\citenamefont{Huey and Lidsey}(2002)}]{Huey2002}
\bibinfo{author}{\bibfnamefont{G.}~\bibnamefont{Huey}} \bibnamefont{and}
  \bibinfo{author}{\bibfnamefont{J.~E.} \bibnamefont{Lidsey}},
  \bibinfo{journal}{Phys.Rev.} \textbf{\bibinfo{volume}{D66}},
  \bibinfo{pages}{043514} (\bibinfo{year}{2002}), \eprint{astro-ph/0205236}.

\bibitem[{\citenamefont{Szydlowski et~al.}(2002)\citenamefont{Szydlowski,
  Dabrowski, and Krawiec}}]{Szydlowski2002}
\bibinfo{author}{\bibfnamefont{M.}~\bibnamefont{Szydlowski}},
  \bibinfo{author}{\bibfnamefont{M.~P.} \bibnamefont{Dabrowski}},
  \bibnamefont{and} \bibinfo{author}{\bibfnamefont{A.}~\bibnamefont{Krawiec}},
  \bibinfo{journal}{Phys.Rev.} \textbf{\bibinfo{volume}{D66}},
  \bibinfo{pages}{064003} (\bibinfo{year}{2002}), \eprint{hep-th/0201066}.

\bibitem[{\citenamefont{van~den Hoogen et~al.}(2003)\citenamefont{van~den
  Hoogen, Coley, and He}}]{Hoogen2003}
\bibinfo{author}{\bibfnamefont{R.}~\bibnamefont{van~den Hoogen}},
  \bibinfo{author}{\bibfnamefont{A.}~\bibnamefont{Coley}}, \bibnamefont{and}
  \bibinfo{author}{\bibfnamefont{Y.}~\bibnamefont{He}},
  \bibinfo{journal}{Phys.Rev.} \textbf{\bibinfo{volume}{D68}},
  \bibinfo{pages}{023502} (\bibinfo{year}{2003}), \eprint{gr-qc/0212094}.

\bibitem[{\citenamefont{van~den Hoogen and Ibanez}(2003)}]{Hoogen2003a}
\bibinfo{author}{\bibfnamefont{R.}~\bibnamefont{van~den Hoogen}}
  \bibnamefont{and} \bibinfo{author}{\bibfnamefont{J.}~\bibnamefont{Ibanez}},
  \bibinfo{journal}{Phys.Rev.} \textbf{\bibinfo{volume}{D67}},
  \bibinfo{pages}{083510} (\bibinfo{year}{2003}), \eprint{gr-qc/0212095}.

\bibitem[{\citenamefont{Toporensky et~al.}(2003)\citenamefont{Toporensky,
  Tretyakov, and Ustiansky}}]{Toporensky2003}
\bibinfo{author}{\bibfnamefont{A.}~\bibnamefont{Toporensky}},
  \bibinfo{author}{\bibfnamefont{P.}~\bibnamefont{Tretyakov}},
  \bibnamefont{and}
  \bibinfo{author}{\bibfnamefont{V.}~\bibnamefont{Ustiansky}},
  \bibinfo{journal}{Astron.Lett.} \textbf{\bibinfo{volume}{29}},
  \bibinfo{pages}{1} (\bibinfo{year}{2003}), \eprint{gr-qc/0207091}.

\bibitem[{\citenamefont{Savchenko and Toporensky}(2003)}]{Savchenko2003}
\bibinfo{author}{\bibfnamefont{N.~Y.} \bibnamefont{Savchenko}}
  \bibnamefont{and}
  \bibinfo{author}{\bibfnamefont{A.}~\bibnamefont{Toporensky}},
  \bibinfo{journal}{Class.Quant.Grav.} \textbf{\bibinfo{volume}{20}},
  \bibinfo{pages}{2553} (\bibinfo{year}{2003}), \eprint{gr-qc/0212104}.

\bibitem[{\citenamefont{Solomons et~al.}(2006)\citenamefont{Solomons, Dunsby,
  and Ellis}}]{Solomons2006}
\bibinfo{author}{\bibfnamefont{D.~M.} \bibnamefont{Solomons}},
  \bibinfo{author}{\bibfnamefont{P.}~\bibnamefont{Dunsby}}, \bibnamefont{and}
  \bibinfo{author}{\bibfnamefont{G.}~\bibnamefont{Ellis}},
  \bibinfo{journal}{Class.Quant.Grav.} \textbf{\bibinfo{volume}{23}},
  \bibinfo{pages}{6585} (\bibinfo{year}{2006}), \eprint{gr-qc/0103087}.

\bibitem[{\citenamefont{Haghani et~al.}(2012)\citenamefont{Haghani, Sepangi,
  and Shahidi}}]{Haghani2012}
\bibinfo{author}{\bibfnamefont{Z.}~\bibnamefont{Haghani}},
  \bibinfo{author}{\bibfnamefont{H.~R.} \bibnamefont{Sepangi}},
  \bibnamefont{and} \bibinfo{author}{\bibfnamefont{S.}~\bibnamefont{Shahidi}},
  \bibinfo{journal}{JCAP} \textbf{\bibinfo{volume}{1202}}, \bibinfo{pages}{031}
  (\bibinfo{year}{2012}), \eprint{1201.6448}.

\bibitem[{\citenamefont{Campos and Sopuerta}(2001{\natexlab{a}})}]{Campos2001}
\bibinfo{author}{\bibfnamefont{A.}~\bibnamefont{Campos}} \bibnamefont{and}
  \bibinfo{author}{\bibfnamefont{C.~F.} \bibnamefont{Sopuerta}},
  \bibinfo{journal}{Phys.Rev.} \textbf{\bibinfo{volume}{D63}},
  \bibinfo{pages}{104012} (\bibinfo{year}{2001}{\natexlab{a}}),
  \eprint{hep-th/0101060}.

\bibitem[{\citenamefont{Campos and Sopuerta}(2001{\natexlab{b}})}]{Campos2001a}
\bibinfo{author}{\bibfnamefont{A.}~\bibnamefont{Campos}} \bibnamefont{and}
  \bibinfo{author}{\bibfnamefont{C.~F.} \bibnamefont{Sopuerta}},
  \bibinfo{journal}{Phys.Rev.} \textbf{\bibinfo{volume}{D64}},
  \bibinfo{pages}{104011} (\bibinfo{year}{2001}{\natexlab{b}}),
  \eprint{hep-th/0105100}.

\bibitem[{\citenamefont{Goheer and Dunsby}(2002)}]{Goheer2002}
\bibinfo{author}{\bibfnamefont{N.}~\bibnamefont{Goheer}} \bibnamefont{and}
  \bibinfo{author}{\bibfnamefont{P.}~\bibnamefont{Dunsby}},
  \bibinfo{journal}{Phys.Rev.} \textbf{\bibinfo{volume}{D66}},
  \bibinfo{pages}{043527} (\bibinfo{year}{2002}), \eprint{gr-qc/0204059}.

\bibitem[{\citenamefont{Goheer and Dunsby}(2003)}]{Goheer2003}
\bibinfo{author}{\bibfnamefont{N.}~\bibnamefont{Goheer}} \bibnamefont{and}
  \bibinfo{author}{\bibfnamefont{P.~K.} \bibnamefont{Dunsby}},
  \bibinfo{journal}{Phys.Rev.} \textbf{\bibinfo{volume}{D67}},
  \bibinfo{pages}{103513} (\bibinfo{year}{2003}), \eprint{gr-qc/0211020}.

\bibitem[{\citenamefont{Leyva et~al.}(2009)\citenamefont{Leyva, Gonzalez,
  Gonzalez, Matos, and Quiros}}]{Leyva2009}
\bibinfo{author}{\bibfnamefont{Y.}~\bibnamefont{Leyva}},
  \bibinfo{author}{\bibfnamefont{D.}~\bibnamefont{Gonzalez}},
  \bibinfo{author}{\bibfnamefont{T.}~\bibnamefont{Gonzalez}},
  \bibinfo{author}{\bibfnamefont{T.}~\bibnamefont{Matos}}, \bibnamefont{and}
  \bibinfo{author}{\bibfnamefont{I.}~\bibnamefont{Quiros}},
  \bibinfo{journal}{Phys.Rev.} \textbf{\bibinfo{volume}{D80}},
  \bibinfo{pages}{044026} (\bibinfo{year}{2009}), \eprint{0909.0281}.

\bibitem[{\citenamefont{Escobar
  et~al.}(2012{\natexlab{a}})\citenamefont{Escobar, Fadragas, Leon, and
  Leyva}}]{Escobar2012}
\bibinfo{author}{\bibfnamefont{D.}~\bibnamefont{Escobar}},
  \bibinfo{author}{\bibfnamefont{C.~R.} \bibnamefont{Fadragas}},
  \bibinfo{author}{\bibfnamefont{G.}~\bibnamefont{Leon}}, \bibnamefont{and}
  \bibinfo{author}{\bibfnamefont{Y.}~\bibnamefont{Leyva}},
  \bibinfo{journal}{Class.Quant.Grav.} \textbf{\bibinfo{volume}{29}},
  \bibinfo{pages}{175005} (\bibinfo{year}{2012}{\natexlab{a}}),
  \eprint{1110.1736}.

\bibitem[{\citenamefont{Escobar
  et~al.}(2012{\natexlab{b}})\citenamefont{Escobar, Fadragas, Leon, and
  Leyva}}]{Escobar2012a}
\bibinfo{author}{\bibfnamefont{D.}~\bibnamefont{Escobar}},
  \bibinfo{author}{\bibfnamefont{C.~R.} \bibnamefont{Fadragas}},
  \bibinfo{author}{\bibfnamefont{G.}~\bibnamefont{Leon}}, \bibnamefont{and}
  \bibinfo{author}{\bibfnamefont{Y.}~\bibnamefont{Leyva}},
  \bibinfo{journal}{Class.Quant.Grav.} \textbf{\bibinfo{volume}{29}},
  \bibinfo{pages}{175006} (\bibinfo{year}{2012}{\natexlab{b}}),
  \eprint{1201.5672}.

\bibitem[{\citenamefont{Shiromizu et~al.}(2000)\citenamefont{Shiromizu, Maeda,
  and Sasaki}}]{Shiromizu2000}
\bibinfo{author}{\bibfnamefont{T.}~\bibnamefont{Shiromizu}},
  \bibinfo{author}{\bibfnamefont{K.-i.} \bibnamefont{Maeda}}, \bibnamefont{and}
  \bibinfo{author}{\bibfnamefont{M.}~\bibnamefont{Sasaki}},
  \bibinfo{journal}{Phys.Rev.} \textbf{\bibinfo{volume}{D62}},
  \bibinfo{pages}{024012} (\bibinfo{year}{2000}), \eprint{gr-qc/9910076}.

\bibitem[{\citenamefont{Sasaki et~al.}(2000)\citenamefont{Sasaki, Shiromizu,
  and Maeda}}]{Sasaki2000}
\bibinfo{author}{\bibfnamefont{M.}~\bibnamefont{Sasaki}},
  \bibinfo{author}{\bibfnamefont{T.}~\bibnamefont{Shiromizu}},
  \bibnamefont{and} \bibinfo{author}{\bibfnamefont{K.-i.} \bibnamefont{Maeda}},
  \bibinfo{journal}{Phys.Rev.} \textbf{\bibinfo{volume}{D62}},
  \bibinfo{pages}{024008} (\bibinfo{year}{2000}), \eprint{hep-th/9912233}.

\bibitem[{\citenamefont{Langlois et~al.}(2001)\citenamefont{Langlois, Maartens,
  Sasaki, and Wands}}]{Langlois2001}
\bibinfo{author}{\bibfnamefont{D.}~\bibnamefont{Langlois}},
  \bibinfo{author}{\bibfnamefont{R.}~\bibnamefont{Maartens}},
  \bibinfo{author}{\bibfnamefont{M.}~\bibnamefont{Sasaki}}, \bibnamefont{and}
  \bibinfo{author}{\bibfnamefont{D.}~\bibnamefont{Wands}},
  \bibinfo{journal}{Phys.Rev.} \textbf{\bibinfo{volume}{D63}},
  \bibinfo{pages}{084009} (\bibinfo{year}{2001}), \eprint{hep-th/0012044}.

\bibitem[{\citenamefont{Barrow and Maartens}(2002)}]{Barrow2002}
\bibinfo{author}{\bibfnamefont{J.~D.} \bibnamefont{Barrow}} \bibnamefont{and}
  \bibinfo{author}{\bibfnamefont{R.}~\bibnamefont{Maartens}},
  \bibinfo{journal}{Phys.Lett.} \textbf{\bibinfo{volume}{B532}},
  \bibinfo{pages}{153} (\bibinfo{year}{2002}), \eprint{gr-qc/0108073}.

\bibitem[{\citenamefont{Kraus}(1999)}]{Kraus1999}
\bibinfo{author}{\bibfnamefont{P.}~\bibnamefont{Kraus}},
  \bibinfo{journal}{JHEP} \textbf{\bibinfo{volume}{9912}}, \bibinfo{pages}{011}
  (\bibinfo{year}{1999}), \eprint{hep-th/9910149}.

\bibitem[{\citenamefont{Hebecker and March-Russell}(2001)}]{Hebecker2001}
\bibinfo{author}{\bibfnamefont{A.}~\bibnamefont{Hebecker}} \bibnamefont{and}
  \bibinfo{author}{\bibfnamefont{J.}~\bibnamefont{March-Russell}},
  \bibinfo{journal}{Nucl.Phys.} \textbf{\bibinfo{volume}{B608}},
  \bibinfo{pages}{375} (\bibinfo{year}{2001}), \eprint{hep-ph/0103214}.

\bibitem[{\citenamefont{Ida}(2000)}]{Ida2000}
\bibinfo{author}{\bibfnamefont{D.}~\bibnamefont{Ida}}, \bibinfo{journal}{JHEP}
  \textbf{\bibinfo{volume}{0009}}, \bibinfo{pages}{014} (\bibinfo{year}{2000}),
  \eprint{gr-qc/9912002}.

\bibitem[{\citenamefont{Vollick}(2001)}]{Vollick2001}
\bibinfo{author}{\bibfnamefont{D.~N.} \bibnamefont{Vollick}},
  \bibinfo{journal}{Class.Quant.Grav.} \textbf{\bibinfo{volume}{18}},
  \bibinfo{pages}{1} (\bibinfo{year}{2001}), \eprint{hep-th/9911181}.

\bibitem[{\citenamefont{Mukohyama et~al.}(2000)\citenamefont{Mukohyama,
  Shiromizu, and Maeda}}]{Mukohyama2000}
\bibinfo{author}{\bibfnamefont{S.}~\bibnamefont{Mukohyama}},
  \bibinfo{author}{\bibfnamefont{T.}~\bibnamefont{Shiromizu}},
  \bibnamefont{and} \bibinfo{author}{\bibfnamefont{K.-i.} \bibnamefont{Maeda}},
  \bibinfo{journal}{Phys.Rev.} \textbf{\bibinfo{volume}{D62}},
  \bibinfo{pages}{024028} (\bibinfo{year}{2000}), \eprint{hep-th/9912287}.

\bibitem[{\citenamefont{Maartens and Koyama}(2010)}]{Maartens2010}
\bibinfo{author}{\bibfnamefont{R.}~\bibnamefont{Maartens}} \bibnamefont{and}
  \bibinfo{author}{\bibfnamefont{K.}~\bibnamefont{Koyama}},
  \bibinfo{journal}{Living Rev.Rel.} \textbf{\bibinfo{volume}{13}},
  \bibinfo{pages}{5} (\bibinfo{year}{2010}), \eprint{1004.3962}.

\bibitem[{\citenamefont{Clifton et~al.}(2012)\citenamefont{Clifton, Ferreira,
  Padilla, and Skordis}}]{Clifton2012}
\bibinfo{author}{\bibfnamefont{T.}~\bibnamefont{Clifton}},
  \bibinfo{author}{\bibfnamefont{P.~G.} \bibnamefont{Ferreira}},
  \bibinfo{author}{\bibfnamefont{A.}~\bibnamefont{Padilla}}, \bibnamefont{and}
  \bibinfo{author}{\bibfnamefont{C.}~\bibnamefont{Skordis}},
  \bibinfo{journal}{Phys.Rept.} \textbf{\bibinfo{volume}{513}},
  \bibinfo{pages}{1} (\bibinfo{year}{2012}), \eprint{1106.2476}.

\bibitem[{\citenamefont{Santos et~al.}(2001)\citenamefont{Santos, Vernizzi, and
  Ferreira}}]{Santos2001}
\bibinfo{author}{\bibfnamefont{M.}~\bibnamefont{Santos}},
  \bibinfo{author}{\bibfnamefont{F.}~\bibnamefont{Vernizzi}}, \bibnamefont{and}
  \bibinfo{author}{\bibfnamefont{P.}~\bibnamefont{Ferreira}},
  \bibinfo{journal}{Phys.Rev.} \textbf{\bibinfo{volume}{D64}},
  \bibinfo{pages}{063506} (\bibinfo{year}{2001}), \eprint{hep-ph/0103112}.

\bibitem[{\citenamefont{Malaney and Mathews}(1993)}]{Malaney1993}
\bibinfo{author}{\bibfnamefont{R.}~\bibnamefont{Malaney}} \bibnamefont{and}
  \bibinfo{author}{\bibfnamefont{G.}~\bibnamefont{Mathews}},
  \bibinfo{journal}{Phys.Rept.} \textbf{\bibinfo{volume}{229}},
  \bibinfo{pages}{145} (\bibinfo{year}{1993}).

\bibitem[{\citenamefont{Olive et~al.}(2000)\citenamefont{Olive, Steigman, and
  Walker}}]{Olive2000}
\bibinfo{author}{\bibfnamefont{K.~A.} \bibnamefont{Olive}},
  \bibinfo{author}{\bibfnamefont{G.}~\bibnamefont{Steigman}}, \bibnamefont{and}
  \bibinfo{author}{\bibfnamefont{T.~P.} \bibnamefont{Walker}},
  \bibinfo{journal}{Phys.Rept.} \textbf{\bibinfo{volume}{333}},
  \bibinfo{pages}{389} (\bibinfo{year}{2000}), \eprint{astro-ph/9905320}.

\bibitem[{\citenamefont{Ichiki et~al.}(2002)\citenamefont{Ichiki, Yahiro,
  Kajino, Orito, and Mathews}}]{Ichiki2002}
\bibinfo{author}{\bibfnamefont{K.}~\bibnamefont{Ichiki}},
  \bibinfo{author}{\bibfnamefont{M.}~\bibnamefont{Yahiro}},
  \bibinfo{author}{\bibfnamefont{T.}~\bibnamefont{Kajino}},
  \bibinfo{author}{\bibfnamefont{M.}~\bibnamefont{Orito}}, \bibnamefont{and}
  \bibinfo{author}{\bibfnamefont{G.}~\bibnamefont{Mathews}},
  \bibinfo{journal}{Phys.Rev.} \textbf{\bibinfo{volume}{D66}},
  \bibinfo{pages}{043521} (\bibinfo{year}{2002}), \eprint{astro-ph/0203272}.

\bibitem[{\citenamefont{Apostolopoulos and
  Tetradis}(2006)}]{Apostolopoulos2006}
\bibinfo{author}{\bibfnamefont{P.~S.} \bibnamefont{Apostolopoulos}}
  \bibnamefont{and} \bibinfo{author}{\bibfnamefont{N.}~\bibnamefont{Tetradis}},
  \bibinfo{journal}{Phys.Lett.} \textbf{\bibinfo{volume}{B633}},
  \bibinfo{pages}{409} (\bibinfo{year}{2006}), \eprint{hep-th/0509182}.

\bibitem[{\citenamefont{Dutta et~al.}(2009)\citenamefont{Dutta, Saridakis, and
  Scherrer}}]{Dutta2009}
\bibinfo{author}{\bibfnamefont{S.}~\bibnamefont{Dutta}},
  \bibinfo{author}{\bibfnamefont{E.~N.} \bibnamefont{Saridakis}},
  \bibnamefont{and} \bibinfo{author}{\bibfnamefont{R.~J.}
  \bibnamefont{Scherrer}}, \bibinfo{journal}{Phys.Rev.}
  \textbf{\bibinfo{volume}{D79}}, \bibinfo{pages}{103005}
  (\bibinfo{year}{2009}), \eprint{0903.3412}.

\bibitem[{\citenamefont{Diamanti et~al.}(2012)\citenamefont{Diamanti, Giusarma,
  Mena, Archidiacono, and Melchiorri}}]{Diamanti2012}
\bibinfo{author}{\bibfnamefont{R.}~\bibnamefont{Diamanti}},
  \bibinfo{author}{\bibfnamefont{E.}~\bibnamefont{Giusarma}},
  \bibinfo{author}{\bibfnamefont{O.}~\bibnamefont{Mena}},
  \bibinfo{author}{\bibfnamefont{M.}~\bibnamefont{Archidiacono}},
  \bibnamefont{and}
  \bibinfo{author}{\bibfnamefont{A.}~\bibnamefont{Melchiorri}}
  (\bibinfo{year}{2012}), \eprint{1212.6007}.

\bibitem[{\citenamefont{Gonzalez-Garcia
  et~al.}(2013)\citenamefont{Gonzalez-Garcia, Niro, and
  Salvado}}]{Gonzalez-Garcia2013}
\bibinfo{author}{\bibfnamefont{M.}~\bibnamefont{Gonzalez-Garcia}},
  \bibinfo{author}{\bibfnamefont{V.}~\bibnamefont{Niro}}, \bibnamefont{and}
  \bibinfo{author}{\bibfnamefont{J.}~\bibnamefont{Salvado}},
  \bibinfo{journal}{JHEP} \textbf{\bibinfo{volume}{1304}}, \bibinfo{pages}{052}
  (\bibinfo{year}{2013}), \eprint{1212.1472}.

\bibitem[{\citenamefont{Bratt et~al.}(2002)\citenamefont{Bratt, Gault,
  Scherrer, and Walker}}]{Bratt2002}
\bibinfo{author}{\bibfnamefont{J.~D.} \bibnamefont{Bratt}},
  \bibinfo{author}{\bibfnamefont{A.}~\bibnamefont{Gault}},
  \bibinfo{author}{\bibfnamefont{R.~J.} \bibnamefont{Scherrer}},
  \bibnamefont{and} \bibinfo{author}{\bibfnamefont{T.}~\bibnamefont{Walker}},
  \bibinfo{journal}{Phys.Lett.} \textbf{\bibinfo{volume}{B546}},
  \bibinfo{pages}{19} (\bibinfo{year}{2002}), \eprint{astro-ph/0208133}.

\bibitem[{\citenamefont{Archidiacono et~al.}(2011)\citenamefont{Archidiacono,
  Calabrese, and Melchiorri}}]{Archidiacono2011}
\bibinfo{author}{\bibfnamefont{M.}~\bibnamefont{Archidiacono}},
  \bibinfo{author}{\bibfnamefont{E.}~\bibnamefont{Calabrese}},
  \bibnamefont{and}
  \bibinfo{author}{\bibfnamefont{A.}~\bibnamefont{Melchiorri}},
  \bibinfo{journal}{Phys.Rev.} \textbf{\bibinfo{volume}{D84}},
  \bibinfo{pages}{123008} (\bibinfo{year}{2011}), \eprint{1109.2767}.

\bibitem[{\citenamefont{Bennett et~al.}(2012)\citenamefont{Bennett, Larson,
  Weiland, Jarosik, Hinshaw et~al.}}]{Bennett2012}
\bibinfo{author}{\bibfnamefont{C.}~\bibnamefont{Bennett}},
  \bibinfo{author}{\bibfnamefont{D.}~\bibnamefont{Larson}},
  \bibinfo{author}{\bibfnamefont{J.}~\bibnamefont{Weiland}},
  \bibinfo{author}{\bibfnamefont{N.}~\bibnamefont{Jarosik}},
  \bibinfo{author}{\bibfnamefont{G.}~\bibnamefont{Hinshaw}},
  \bibnamefont{et~al.} (\bibinfo{year}{2012}), \eprint{1212.5225}.

\bibitem[{\citenamefont{Hou et~al.}(2012)\citenamefont{Hou, Reichardt, Story,
  Follin, Keisler et~al.}}]{SPT}
\bibinfo{author}{\bibfnamefont{Z.}~\bibnamefont{Hou}},
  \bibinfo{author}{\bibfnamefont{C.}~\bibnamefont{Reichardt}},
  \bibinfo{author}{\bibfnamefont{K.}~\bibnamefont{Story}},
  \bibinfo{author}{\bibfnamefont{B.}~\bibnamefont{Follin}},
  \bibinfo{author}{\bibfnamefont{R.}~\bibnamefont{Keisler}},
  \bibnamefont{et~al.} (\bibinfo{year}{2012}), \eprint{1212.6267}.

\bibitem[{\citenamefont{Sievers et~al.}(2013)\citenamefont{Sievers, Hlozek,
  Nolta, Acquaviva, Addison et~al.}}]{ACT}
\bibinfo{author}{\bibfnamefont{J.~L.} \bibnamefont{Sievers}},
  \bibinfo{author}{\bibfnamefont{R.~A.} \bibnamefont{Hlozek}},
  \bibinfo{author}{\bibfnamefont{M.~R.} \bibnamefont{Nolta}},
  \bibinfo{author}{\bibfnamefont{V.}~\bibnamefont{Acquaviva}},
  \bibinfo{author}{\bibfnamefont{G.~E.} \bibnamefont{Addison}},
  \bibnamefont{et~al.} (\bibinfo{year}{2013}), \eprint{1301.0824}.

\bibitem[{\citenamefont{Ade et~al.}(2013{\natexlab{a}})}]{planck1}
\bibinfo{author}{\bibfnamefont{P.}~\bibnamefont{Ade}} \bibnamefont{et~al.}
  (\bibinfo{collaboration}{Planck Collaboration})
  (\bibinfo{year}{2013}{\natexlab{a}}), \eprint{1303.5076}.

\bibitem[{\citenamefont{Di~Valentino
  et~al.}(2013{\natexlab{a}})\citenamefont{Di~Valentino, Galli, Lattanzi,
  Melchiorri, Natoli et~al.}}]{DiValentino2013a}
\bibinfo{author}{\bibfnamefont{E.}~\bibnamefont{Di~Valentino}},
  \bibinfo{author}{\bibfnamefont{S.}~\bibnamefont{Galli}},
  \bibinfo{author}{\bibfnamefont{M.}~\bibnamefont{Lattanzi}},
  \bibinfo{author}{\bibfnamefont{A.}~\bibnamefont{Melchiorri}},
  \bibinfo{author}{\bibfnamefont{P.}~\bibnamefont{Natoli}},
  \bibnamefont{et~al.}, \bibinfo{journal}{Phys.Rev.}
  \textbf{\bibinfo{volume}{D88}}, \bibinfo{pages}{023501}
  (\bibinfo{year}{2013}{\natexlab{a}}), \eprint{1301.7343}.

\bibitem[{\citenamefont{Hinshaw et~al.}(2012)}]{WMAP9}
\bibinfo{author}{\bibfnamefont{G.}~\bibnamefont{Hinshaw}} \bibnamefont{et~al.}
  (\bibinfo{collaboration}{WMAP Collaboration}) (\bibinfo{year}{2012}),
  \eprint{1212.5226}.

\bibitem[{\citenamefont{Di~Valentino
  et~al.}(2013{\natexlab{b}})\citenamefont{Di~Valentino, Melchiorri, and
  Mena}}]{DiValentino2013}
\bibinfo{author}{\bibfnamefont{E.}~\bibnamefont{Di~Valentino}},
  \bibinfo{author}{\bibfnamefont{A.}~\bibnamefont{Melchiorri}},
  \bibnamefont{and} \bibinfo{author}{\bibfnamefont{O.}~\bibnamefont{Mena}}
  (\bibinfo{year}{2013}{\natexlab{b}}), \eprint{1304.5981}.

\bibitem[{\citenamefont{Hinshaw et~al.}(2007)}]{Hinshaw2007}
\bibinfo{author}{\bibfnamefont{G.}~\bibnamefont{Hinshaw}} \bibnamefont{et~al.}
  (\bibinfo{collaboration}{WMAP Collaboration}),
  \bibinfo{journal}{Astrophys.J.Suppl.} \textbf{\bibinfo{volume}{170}},
  \bibinfo{pages}{288} (\bibinfo{year}{2007}), \eprint{astro-ph/0603451}.

\bibitem[{\citenamefont{Komatsu et~al.}(2011)}]{Komatsu2011}
\bibinfo{author}{\bibfnamefont{E.}~\bibnamefont{Komatsu}} \bibnamefont{et~al.}
  (\bibinfo{collaboration}{WMAP Collaboration}),
  \bibinfo{journal}{Astrophys.J.Suppl.} \textbf{\bibinfo{volume}{192}},
  \bibinfo{pages}{18} (\bibinfo{year}{2011}), \eprint{1001.4538}.

\bibitem[{\citenamefont{Guth}(1981)}]{Guth1981}
\bibinfo{author}{\bibfnamefont{A.~H.} \bibnamefont{Guth}},
  \bibinfo{journal}{Phys.Rev.} \textbf{\bibinfo{volume}{D23}},
  \bibinfo{pages}{347} (\bibinfo{year}{1981}).

\bibitem[{\citenamefont{Linde}(1982)}]{Linde1982}
\bibinfo{author}{\bibfnamefont{A.~D.} \bibnamefont{Linde}},
  \bibinfo{journal}{Phys.Lett.} \textbf{\bibinfo{volume}{B108}},
  \bibinfo{pages}{389} (\bibinfo{year}{1982}).

\bibitem[{\citenamefont{Misner et~al.}(1974)\citenamefont{Misner, Thorne, and
  Wheeler}}]{Misner1974}
\bibinfo{author}{\bibfnamefont{C.~W.} \bibnamefont{Misner}},
  \bibinfo{author}{\bibfnamefont{K.}~\bibnamefont{Thorne}}, \bibnamefont{and}
  \bibinfo{author}{\bibfnamefont{J.}~\bibnamefont{Wheeler}}
  (\bibinfo{year}{1974}).

\bibitem[{\citenamefont{Maartens et~al.}(2001)\citenamefont{Maartens, Sahni,
  and Saini}}]{Maartens2001}
\bibinfo{author}{\bibfnamefont{R.}~\bibnamefont{Maartens}},
  \bibinfo{author}{\bibfnamefont{V.}~\bibnamefont{Sahni}}, \bibnamefont{and}
  \bibinfo{author}{\bibfnamefont{T.~D.} \bibnamefont{Saini}},
  \bibinfo{journal}{Phys.Rev.} \textbf{\bibinfo{volume}{D63}},
  \bibinfo{pages}{063509} (\bibinfo{year}{2001}), \eprint{gr-qc/0011105}.

\bibitem[{\citenamefont{Niz et~al.}(2008)\citenamefont{Niz, Padilla, and
  Kunduri}}]{Niz2008}
\bibinfo{author}{\bibfnamefont{G.}~\bibnamefont{Niz}},
  \bibinfo{author}{\bibfnamefont{A.}~\bibnamefont{Padilla}}, \bibnamefont{and}
  \bibinfo{author}{\bibfnamefont{H.~K.} \bibnamefont{Kunduri}},
  \bibinfo{journal}{JCAP} \textbf{\bibinfo{volume}{0804}}, \bibinfo{pages}{012}
  (\bibinfo{year}{2008}), \eprint{0801.3462}.

\bibitem[{\citenamefont{Toporensky and Tretyakov}(2005)}]{Toporensky2005}
\bibinfo{author}{\bibfnamefont{A.}~\bibnamefont{Toporensky}} \bibnamefont{and}
  \bibinfo{author}{\bibfnamefont{P.}~\bibnamefont{Tretyakov}},
  \bibinfo{journal}{Grav.Cosmol.} \textbf{\bibinfo{volume}{11}},
  \bibinfo{pages}{226} (\bibinfo{year}{2005}), \eprint{gr-qc/0510025}.

\bibitem[{\citenamefont{Ade et~al.}(2013{\natexlab{b}})}]{plack2}
\bibinfo{author}{\bibfnamefont{P.}~\bibnamefont{Ade}} \bibnamefont{et~al.}
  (\bibinfo{collaboration}{Planck Collaboration})
  (\bibinfo{year}{2013}{\natexlab{b}}), \eprint{1303.5086}.

\bibitem[{\citenamefont{Riess et~al.}(1998)}]{Riess1998}
\bibinfo{author}{\bibfnamefont{A.~G.} \bibnamefont{Riess}} \bibnamefont{et~al.}
  (\bibinfo{collaboration}{Supernova Search Team}),
  \bibinfo{journal}{Astron.J.} \textbf{\bibinfo{volume}{116}},
  \bibinfo{pages}{1009} (\bibinfo{year}{1998}), \eprint{astro-ph/9805201}.

\bibitem[{\citenamefont{Perlmutter et~al.}(1999)}]{Perlmutter1999}
\bibinfo{author}{\bibfnamefont{S.}~\bibnamefont{Perlmutter}}
  \bibnamefont{et~al.} (\bibinfo{collaboration}{Supernova Cosmology Project}),
  \bibinfo{journal}{Astrophys.J.} \textbf{\bibinfo{volume}{517}},
  \bibinfo{pages}{565} (\bibinfo{year}{1999}), \eprint{astro-ph/9812133}.

\bibitem[{\citenamefont{Copeland et~al.}(2006)\citenamefont{Copeland, Sami, and
  Tsujikawa}}]{Copeland2006}
\bibinfo{author}{\bibfnamefont{E.~J.} \bibnamefont{Copeland}},
  \bibinfo{author}{\bibfnamefont{M.}~\bibnamefont{Sami}}, \bibnamefont{and}
  \bibinfo{author}{\bibfnamefont{S.}~\bibnamefont{Tsujikawa}},
  \bibinfo{journal}{Int.J.Mod.Phys.} \textbf{\bibinfo{volume}{D15}},
  \bibinfo{pages}{1753} (\bibinfo{year}{2006}), \eprint{hep-th/0603057}.

\bibitem[{\citenamefont{Fang et~al.}(2009)\citenamefont{Fang, Li, Zhang, and
  Lu}}]{Fang2009}
\bibinfo{author}{\bibfnamefont{W.}~\bibnamefont{Fang}},
  \bibinfo{author}{\bibfnamefont{Y.}~\bibnamefont{Li}},
  \bibinfo{author}{\bibfnamefont{K.}~\bibnamefont{Zhang}}, \bibnamefont{and}
  \bibinfo{author}{\bibfnamefont{H.-Q.} \bibnamefont{Lu}},
  \bibinfo{journal}{Class.Quant.Grav.} \textbf{\bibinfo{volume}{26}},
  \bibinfo{pages}{155005} (\bibinfo{year}{2009}), \eprint{0810.4193}.

\bibitem[{\citenamefont{Matos et~al.}(2009)\citenamefont{Matos, Luevano,
  Quiros, Urena-Lopez, and Vazquez}}]{Matos2009}
\bibinfo{author}{\bibfnamefont{T.}~\bibnamefont{Matos}},
  \bibinfo{author}{\bibfnamefont{J.-R.} \bibnamefont{Luevano}},
  \bibinfo{author}{\bibfnamefont{I.}~\bibnamefont{Quiros}},
  \bibinfo{author}{\bibfnamefont{L.~A.} \bibnamefont{Urena-Lopez}},
  \bibnamefont{and} \bibinfo{author}{\bibfnamefont{J.~A.}
  \bibnamefont{Vazquez}}, \bibinfo{journal}{Phys.Rev.}
  \textbf{\bibinfo{volume}{D80}}, \bibinfo{pages}{123521}
  (\bibinfo{year}{2009}), \eprint{0906.0396}.

\bibitem[{\citenamefont{Urena-Lopez}(2012)}]{Urena-Lopez2012}
\bibinfo{author}{\bibfnamefont{L.~A.} \bibnamefont{Urena-Lopez}},
  \bibinfo{journal}{JCAP} \textbf{\bibinfo{volume}{1203}}, \bibinfo{pages}{035}
  (\bibinfo{year}{2012}), \eprint{1108.4712}.

\bibitem[{\citenamefont{Copeland et~al.}(2009)\citenamefont{Copeland, Mizuno,
  and Shaeri}}]{Copeland2009}
\bibinfo{author}{\bibfnamefont{E.~J.} \bibnamefont{Copeland}},
  \bibinfo{author}{\bibfnamefont{S.}~\bibnamefont{Mizuno}}, \bibnamefont{and}
  \bibinfo{author}{\bibfnamefont{M.}~\bibnamefont{Shaeri}},
  \bibinfo{journal}{Phys.Rev.} \textbf{\bibinfo{volume}{D79}},
  \bibinfo{pages}{103515} (\bibinfo{year}{2009}), \eprint{0904.0877}.

\bibitem[{\citenamefont{Farajollahi et~al.}(2011)\citenamefont{Farajollahi,
  Salehi, Tayebi, and Ravanpak}}]{Farajollahi2011}
\bibinfo{author}{\bibfnamefont{H.}~\bibnamefont{Farajollahi}},
  \bibinfo{author}{\bibfnamefont{A.}~\bibnamefont{Salehi}},
  \bibinfo{author}{\bibfnamefont{F.}~\bibnamefont{Tayebi}}, \bibnamefont{and}
  \bibinfo{author}{\bibfnamefont{A.}~\bibnamefont{Ravanpak}},
  \bibinfo{journal}{JCAP} \textbf{\bibinfo{volume}{1105}}, \bibinfo{pages}{017}
  (\bibinfo{year}{2011}), \eprint{1105.4045}.

\bibitem[{\citenamefont{Xiao and Zhu}(2011)}]{Xiao2011}
\bibinfo{author}{\bibfnamefont{K.}~\bibnamefont{Xiao}} \bibnamefont{and}
  \bibinfo{author}{\bibfnamefont{J.-Y.} \bibnamefont{Zhu}},
  \bibinfo{journal}{Phys.Rev.} \textbf{\bibinfo{volume}{D83}},
  \bibinfo{pages}{083501} (\bibinfo{year}{2011}), \eprint{1102.2695}.

\bibitem[{\citenamefont{Einstein}(1917)}]{Einstein1917}
\bibinfo{author}{\bibfnamefont{A.}~\bibnamefont{Einstein}},
  \bibinfo{journal}{Sitzungsber.Preuss.Akad.Wiss.Berlin (Math.Phys.)}
  \textbf{\bibinfo{volume}{1917}}, \bibinfo{pages}{142} (\bibinfo{year}{1917}).

\bibitem[{\citenamefont{Eddington}(1930)}]{Eddington1930}
\bibinfo{author}{\bibfnamefont{A.}~\bibnamefont{Eddington}},
  \bibinfo{journal}{Mon.Not.Roy.Astron.Soc.} \textbf{\bibinfo{volume}{90}},
  \bibinfo{pages}{668} (\bibinfo{year}{1930}).

\bibitem[{\citenamefont{Harrison}(1967)}]{Harrison1967}
\bibinfo{author}{\bibfnamefont{E.}~\bibnamefont{Harrison}},
  \bibinfo{journal}{Rev.Mod.Phys.} \textbf{\bibinfo{volume}{39}},
  \bibinfo{pages}{862} (\bibinfo{year}{1967}).

\bibitem[{\citenamefont{Gibbons}(1987)}]{Gibbons1987}
\bibinfo{author}{\bibfnamefont{G.}~\bibnamefont{Gibbons}},
  \bibinfo{journal}{Nucl.Phys.} \textbf{\bibinfo{volume}{B292}},
  \bibinfo{pages}{784} (\bibinfo{year}{1987}).

\bibitem[{\citenamefont{Barrow and Tsagas}(2009)}]{Barrow2009}
\bibinfo{author}{\bibfnamefont{J.~D.} \bibnamefont{Barrow}} \bibnamefont{and}
  \bibinfo{author}{\bibfnamefont{C.~G.} \bibnamefont{Tsagas}},
  \bibinfo{journal}{Class.Quant.Grav.} \textbf{\bibinfo{volume}{26}},
  \bibinfo{pages}{195003} (\bibinfo{year}{2009}), \eprint{0904.1340}.

\bibitem[{\citenamefont{Goswami et~al.}(2008)\citenamefont{Goswami, Goheer, and
  Dunsby}}]{Goswami2008}
\bibinfo{author}{\bibfnamefont{R.}~\bibnamefont{Goswami}},
  \bibinfo{author}{\bibfnamefont{N.}~\bibnamefont{Goheer}}, \bibnamefont{and}
  \bibinfo{author}{\bibfnamefont{P.~K.} \bibnamefont{Dunsby}},
  \bibinfo{journal}{Phys.Rev.} \textbf{\bibinfo{volume}{D78}},
  \bibinfo{pages}{044011} (\bibinfo{year}{2008}), \eprint{0804.3528}.

\bibitem[{\citenamefont{Goheer et~al.}(2009)\citenamefont{Goheer, Goswami, and
  Dunsby}}]{Goheer2009}
\bibinfo{author}{\bibfnamefont{N.}~\bibnamefont{Goheer}},
  \bibinfo{author}{\bibfnamefont{R.}~\bibnamefont{Goswami}}, \bibnamefont{and}
  \bibinfo{author}{\bibfnamefont{P.~K.} \bibnamefont{Dunsby}},
  \bibinfo{journal}{Class.Quant.Grav.} \textbf{\bibinfo{volume}{26}},
  \bibinfo{pages}{105003} (\bibinfo{year}{2009}), \eprint{0809.5247}.

\bibitem[{\citenamefont{Barrow et~al.}(2003)\citenamefont{Barrow, Ellis,
  Maartens, and Tsagas}}]{Barrow2003}
\bibinfo{author}{\bibfnamefont{J.~D.} \bibnamefont{Barrow}},
  \bibinfo{author}{\bibfnamefont{G.~F.} \bibnamefont{Ellis}},
  \bibinfo{author}{\bibfnamefont{R.}~\bibnamefont{Maartens}}, \bibnamefont{and}
  \bibinfo{author}{\bibfnamefont{C.~G.} \bibnamefont{Tsagas}},
  \bibinfo{journal}{Class.Quant.Grav.} \textbf{\bibinfo{volume}{20}},
  \bibinfo{pages}{L155} (\bibinfo{year}{2003}), \eprint{gr-qc/0302094}.

\bibitem[{\citenamefont{Ratra and Peebles}(1988)}]{Ratra1988}
\bibinfo{author}{\bibfnamefont{B.}~\bibnamefont{Ratra}} \bibnamefont{and}
  \bibinfo{author}{\bibfnamefont{P.}~\bibnamefont{Peebles}},
  \bibinfo{journal}{Phys.Rev.} \textbf{\bibinfo{volume}{D37}},
  \bibinfo{pages}{3406} (\bibinfo{year}{1988}).

\bibitem[{\citenamefont{Wetterich}(1988)}]{Wetterich1988}
\bibinfo{author}{\bibfnamefont{C.}~\bibnamefont{Wetterich}},
  \bibinfo{journal}{Nucl.Phys.} \textbf{\bibinfo{volume}{B302}},
  \bibinfo{pages}{668} (\bibinfo{year}{1988}).

\bibitem[{\citenamefont{Matos and Urena-Lopez}(2000)}]{Matos2000}
\bibinfo{author}{\bibfnamefont{T.}~\bibnamefont{Matos}} \bibnamefont{and}
  \bibinfo{author}{\bibfnamefont{L.~A.} \bibnamefont{Urena-Lopez}},
  \bibinfo{journal}{Class.Quant.Grav.} \textbf{\bibinfo{volume}{17}},
  \bibinfo{pages}{L75} (\bibinfo{year}{2000}), \eprint{astro-ph/0004332}.

\bibitem[{\citenamefont{Sahni and Wang}(2000)}]{Sahni2000}
\bibinfo{author}{\bibfnamefont{V.}~\bibnamefont{Sahni}} \bibnamefont{and}
  \bibinfo{author}{\bibfnamefont{L.-M.} \bibnamefont{Wang}},
  \bibinfo{journal}{Phys.Rev.} \textbf{\bibinfo{volume}{D62}},
  \bibinfo{pages}{103517} (\bibinfo{year}{2000}), \eprint{astro-ph/9910097}.

\bibitem[{\citenamefont{Sahni and Starobinsky}(2000)}]{Sahni2000a}
\bibinfo{author}{\bibfnamefont{V.}~\bibnamefont{Sahni}} \bibnamefont{and}
  \bibinfo{author}{\bibfnamefont{A.~A.} \bibnamefont{Starobinsky}},
  \bibinfo{journal}{Int.J.Mod.Phys.} \textbf{\bibinfo{volume}{D9}},
  \bibinfo{pages}{373} (\bibinfo{year}{2000}), \eprint{astro-ph/9904398}.

\bibitem[{\citenamefont{Lidsey et~al.}(2002)\citenamefont{Lidsey, Matos, and
  Urena-Lopez}}]{Lidsey2002}
\bibinfo{author}{\bibfnamefont{J.~E.} \bibnamefont{Lidsey}},
  \bibinfo{author}{\bibfnamefont{T.}~\bibnamefont{Matos}}, \bibnamefont{and}
  \bibinfo{author}{\bibfnamefont{L.~A.} \bibnamefont{Urena-Lopez}},
  \bibinfo{journal}{Phys.Rev.} \textbf{\bibinfo{volume}{D66}},
  \bibinfo{pages}{023514} (\bibinfo{year}{2002}), \eprint{astro-ph/0111292}.

\bibitem[{\citenamefont{Pavluchenko}(2003)}]{Pavluchenko2003}
\bibinfo{author}{\bibfnamefont{S.~A.} \bibnamefont{Pavluchenko}},
  \bibinfo{journal}{Phys.Rev.} \textbf{\bibinfo{volume}{D67}},
  \bibinfo{pages}{103518} (\bibinfo{year}{2003}), \eprint{astro-ph/0304354}.

\bibitem[{\citenamefont{Urena-Lopez and Matos}(2000)}]{Urena-Lopez2000}
\bibinfo{author}{\bibfnamefont{L.~A.} \bibnamefont{Urena-Lopez}}
  \bibnamefont{and} \bibinfo{author}{\bibfnamefont{T.}~\bibnamefont{Matos}},
  \bibinfo{journal}{Phys.Rev.} \textbf{\bibinfo{volume}{D62}},
  \bibinfo{pages}{081302} (\bibinfo{year}{2000}), \eprint{astro-ph/0003364}.

\bibitem[{\citenamefont{Cardenas et~al.}(2003)\citenamefont{Cardenas, Gonzalez,
  Leiva, Martin, and Quiros}}]{Cardenas2003}
\bibinfo{author}{\bibfnamefont{R.}~\bibnamefont{Cardenas}},
  \bibinfo{author}{\bibfnamefont{T.}~\bibnamefont{Gonzalez}},
  \bibinfo{author}{\bibfnamefont{Y.}~\bibnamefont{Leiva}},
  \bibinfo{author}{\bibfnamefont{O.}~\bibnamefont{Martin}}, \bibnamefont{and}
  \bibinfo{author}{\bibfnamefont{I.}~\bibnamefont{Quiros}},
  \bibinfo{journal}{Phys.Rev.} \textbf{\bibinfo{volume}{D67}},
  \bibinfo{pages}{083501} (\bibinfo{year}{2003}), \eprint{astro-ph/0206315}.

\bibitem[{\citenamefont{Barreiro et~al.}(2000)\citenamefont{Barreiro, Copeland,
  and Nunes}}]{Barreiro2000}
\bibinfo{author}{\bibfnamefont{T.}~\bibnamefont{Barreiro}},
  \bibinfo{author}{\bibfnamefont{E.~J.} \bibnamefont{Copeland}},
  \bibnamefont{and} \bibinfo{author}{\bibfnamefont{N.}~\bibnamefont{Nunes}},
  \bibinfo{journal}{Phys.Rev.} \textbf{\bibinfo{volume}{D61}},
  \bibinfo{pages}{127301} (\bibinfo{year}{2000}), \eprint{astro-ph/9910214}.

\bibitem[{\citenamefont{Gonzalez et~al.}(2006)\citenamefont{Gonzalez, Leon, and
  Quiros}}]{Gonzalez2006}
\bibinfo{author}{\bibfnamefont{T.}~\bibnamefont{Gonzalez}},
  \bibinfo{author}{\bibfnamefont{G.}~\bibnamefont{Leon}}, \bibnamefont{and}
  \bibinfo{author}{\bibfnamefont{I.}~\bibnamefont{Quiros}},
  \bibinfo{journal}{Class.Quant.Grav.} \textbf{\bibinfo{volume}{23}},
  \bibinfo{pages}{3165} (\bibinfo{year}{2006}), \eprint{astro-ph/0702227}.

\bibitem[{\citenamefont{Gonzalez et~al.}(2007)\citenamefont{Gonzalez, Cardenas,
  Quiros, and Leyva}}]{Gonzalez2007}
\bibinfo{author}{\bibfnamefont{T.}~\bibnamefont{Gonzalez}},
  \bibinfo{author}{\bibfnamefont{R.}~\bibnamefont{Cardenas}},
  \bibinfo{author}{\bibfnamefont{I.}~\bibnamefont{Quiros}}, \bibnamefont{and}
  \bibinfo{author}{\bibfnamefont{Y.}~\bibnamefont{Leyva}},
  \bibinfo{journal}{Astrophys.Space Sci.} \textbf{\bibinfo{volume}{310}},
  \bibinfo{pages}{13} (\bibinfo{year}{2007}), \eprint{0707.2097}.

\bibitem[{\citenamefont{Leon et~al.}(2012)\citenamefont{Leon, Leyva, and
  Socorro}}]{Leon2012}
\bibinfo{author}{\bibfnamefont{G.}~\bibnamefont{Leon}},
  \bibinfo{author}{\bibfnamefont{Y.}~\bibnamefont{Leyva}}, \bibnamefont{and}
  \bibinfo{author}{\bibfnamefont{J.}~\bibnamefont{Socorro}}
  (\bibinfo{year}{2012}), \eprint{1208.0061}.

\bibitem[{\citenamefont{Lazkoz and Leon}(2006)}]{Lazkoz2006}
\bibinfo{author}{\bibfnamefont{R.}~\bibnamefont{Lazkoz}} \bibnamefont{and}
  \bibinfo{author}{\bibfnamefont{G.}~\bibnamefont{Leon}},
  \bibinfo{journal}{Phys.Lett.} \textbf{\bibinfo{volume}{B638}},
  \bibinfo{pages}{303} (\bibinfo{year}{2006}), \eprint{astro-ph/0602590}.

\bibitem[{\citenamefont{Lazkoz et~al.}(2007)\citenamefont{Lazkoz, Leon, and
  Quiros}}]{Lazkoz2007}
\bibinfo{author}{\bibfnamefont{R.}~\bibnamefont{Lazkoz}},
  \bibinfo{author}{\bibfnamefont{G.}~\bibnamefont{Leon}}, \bibnamefont{and}
  \bibinfo{author}{\bibfnamefont{I.}~\bibnamefont{Quiros}},
  \bibinfo{journal}{Phys.Lett.} \textbf{\bibinfo{volume}{B649}},
  \bibinfo{pages}{103} (\bibinfo{year}{2007}), \eprint{astro-ph/0701353}.

\bibitem[{\citenamefont{Leon}(2009)}]{Leon2009}
\bibinfo{author}{\bibfnamefont{G.}~\bibnamefont{Leon}},
  \bibinfo{journal}{Class.Quant.Grav.} \textbf{\bibinfo{volume}{26}},
  \bibinfo{pages}{035008} (\bibinfo{year}{2009}), \eprint{0812.1013}.

\bibitem[{\citenamefont{Leon et~al.}(2010)\citenamefont{Leon, Silveira, and
  Fadragas}}]{Leon2010}
\bibinfo{author}{\bibfnamefont{G.}~\bibnamefont{Leon}},
  \bibinfo{author}{\bibfnamefont{P.}~\bibnamefont{Silveira}}, \bibnamefont{and}
  \bibinfo{author}{\bibfnamefont{C.~R.} \bibnamefont{Fadragas}}
  (\bibinfo{year}{2010}), \eprint{1009.0689}.

\bibitem[{\citenamefont{Leon and Fadragas}(2012)}]{leon2011a}
\bibinfo{author}{\bibfnamefont{G.}~\bibnamefont{Leon}} \bibnamefont{and}
  \bibinfo{author}{\bibfnamefont{C.}~\bibnamefont{Fadragas}},
  \emph{\bibinfo{title}{Cosmological Dynamical Systems: And Their
  Applications}} (\bibinfo{publisher}{LAP Lambert Academic Publishing},
  \bibinfo{year}{2012}), ISBN \bibinfo{isbn}{9783847302339},
  \urlprefix\url{http://books.google.cl/books?id=dZm7pwAACAAJ}.

\bibitem[{\citenamefont{Maartens}(2000)}]{Maartens2000}
\bibinfo{author}{\bibfnamefont{R.}~\bibnamefont{Maartens}},
  \bibinfo{journal}{Phys.Rev.} \textbf{\bibinfo{volume}{D62}},
  \bibinfo{pages}{084023} (\bibinfo{year}{2000}), \eprint{hep-th/0004166}.

\bibitem[{\citenamefont{Brax and van~de Bruck}(2003)}]{Brax2003}
\bibinfo{author}{\bibfnamefont{P.}~\bibnamefont{Brax}} \bibnamefont{and}
  \bibinfo{author}{\bibfnamefont{C.}~\bibnamefont{van~de Bruck}},
  \bibinfo{journal}{Class.Quant.Grav.} \textbf{\bibinfo{volume}{20}},
  \bibinfo{pages}{R201} (\bibinfo{year}{2003}), \eprint{hep-th/0303095}.

\bibitem[{\citenamefont{Copeland et~al.}(1998)\citenamefont{Copeland, Liddle,
  and Wands}}]{Copeland1998}
\bibinfo{author}{\bibfnamefont{E.~J.} \bibnamefont{Copeland}},
  \bibinfo{author}{\bibfnamefont{A.~R.} \bibnamefont{Liddle}},
  \bibnamefont{and} \bibinfo{author}{\bibfnamefont{D.}~\bibnamefont{Wands}},
  \bibinfo{journal}{Phys.Rev.} \textbf{\bibinfo{volume}{D57}},
  \bibinfo{pages}{4686} (\bibinfo{year}{1998}), \eprint{gr-qc/9711068}.

\bibitem[{\citenamefont{Coley}(2003)}]{coley2003dynamical}
\bibinfo{author}{\bibfnamefont{A.~A.} \bibnamefont{Coley}},
  \emph{\bibinfo{title}{Dynamical Systems and Cosmology}}
  (\bibinfo{publisher}{Astrophysics and Space Science Library. Kluwer Academic
  Publishers}, \bibinfo{year}{2003}).

\bibitem[{\citenamefont{Dunsby et~al.}(2004)\citenamefont{Dunsby, Goheer,
  Bruni, and Coley}}]{Dunsby2004}
\bibinfo{author}{\bibfnamefont{P.}~\bibnamefont{Dunsby}},
  \bibinfo{author}{\bibfnamefont{N.}~\bibnamefont{Goheer}},
  \bibinfo{author}{\bibfnamefont{M.}~\bibnamefont{Bruni}}, \bibnamefont{and}
  \bibinfo{author}{\bibfnamefont{A.}~\bibnamefont{Coley}},
  \bibinfo{journal}{Phys.Rev.} \textbf{\bibinfo{volume}{D69}},
  \bibinfo{pages}{101303} (\bibinfo{year}{2004}), \eprint{hep-th/0312174}.

\bibitem[{\citenamefont{Goheer et~al.}(2004)\citenamefont{Goheer, Dunsby,
  Coley, and Bruni}}]{Goheer2004}
\bibinfo{author}{\bibfnamefont{N.}~\bibnamefont{Goheer}},
  \bibinfo{author}{\bibfnamefont{P.~K.} \bibnamefont{Dunsby}},
  \bibinfo{author}{\bibfnamefont{A.}~\bibnamefont{Coley}}, \bibnamefont{and}
  \bibinfo{author}{\bibfnamefont{M.}~\bibnamefont{Bruni}},
  \bibinfo{journal}{Phys.Rev.} \textbf{\bibinfo{volume}{D70}},
  \bibinfo{pages}{123517} (\bibinfo{year}{2004}), \eprint{hep-th/0408092}.

\bibitem[{\citenamefont{Leach et~al.}(2006)\citenamefont{Leach, Carloni, and
  Dunsby}}]{Leach2006}
\bibinfo{author}{\bibfnamefont{J.~A.} \bibnamefont{Leach}},
  \bibinfo{author}{\bibfnamefont{S.}~\bibnamefont{Carloni}}, \bibnamefont{and}
  \bibinfo{author}{\bibfnamefont{P.~K.} \bibnamefont{Dunsby}},
  \bibinfo{journal}{Class.Quant.Grav.} \textbf{\bibinfo{volume}{23}},
  \bibinfo{pages}{4915} (\bibinfo{year}{2006}), \eprint{gr-qc/0603012}.

\bibitem[{\citenamefont{Goheer et~al.}(2007)\citenamefont{Goheer, Leach, and
  Dunsby}}]{Goheer2007}
\bibinfo{author}{\bibfnamefont{N.}~\bibnamefont{Goheer}},
  \bibinfo{author}{\bibfnamefont{J.~A.} \bibnamefont{Leach}}, \bibnamefont{and}
  \bibinfo{author}{\bibfnamefont{P.~K.} \bibnamefont{Dunsby}},
  \bibinfo{journal}{Class.Quant.Grav.} \textbf{\bibinfo{volume}{24}},
  \bibinfo{pages}{5689} (\bibinfo{year}{2007}), \eprint{0710.0814}.

\bibitem[{\citenamefont{Goheer et~al.}(2008)\citenamefont{Goheer, Leach, and
  Dunsby}}]{Goheer2008}
\bibinfo{author}{\bibfnamefont{N.}~\bibnamefont{Goheer}},
  \bibinfo{author}{\bibfnamefont{J.~A.} \bibnamefont{Leach}}, \bibnamefont{and}
  \bibinfo{author}{\bibfnamefont{P.~K.} \bibnamefont{Dunsby}},
  \bibinfo{journal}{Class.Quant.Grav.} \textbf{\bibinfo{volume}{25}},
  \bibinfo{pages}{035013} (\bibinfo{year}{2008}), \eprint{0710.0819}.

\bibitem[{\citenamefont{Leon and Saridakis}(2011)}]{Leon2011}
\bibinfo{author}{\bibfnamefont{G.}~\bibnamefont{Leon}} \bibnamefont{and}
  \bibinfo{author}{\bibfnamefont{E.~N.} \bibnamefont{Saridakis}},
  \bibinfo{journal}{Class.Quant.Grav.} \textbf{\bibinfo{volume}{28}},
  \bibinfo{pages}{065008} (\bibinfo{year}{2011}), \eprint{1007.3956}.

\bibitem[{\citenamefont{Foster}(1998)}]{Foster1998}
\bibinfo{author}{\bibfnamefont{S.}~\bibnamefont{Foster}},
  \bibinfo{journal}{Class.Quant.Grav.} \textbf{\bibinfo{volume}{15}},
  \bibinfo{pages}{3485} (\bibinfo{year}{1998}), \eprint{gr-qc/9806098}.

\bibitem[{\citenamefont{Sahni et~al.}(2005)\citenamefont{Sahni, Shtanov, and
  Viznyuk}}]{Sahni2005}
\bibinfo{author}{\bibfnamefont{V.}~\bibnamefont{Sahni}},
  \bibinfo{author}{\bibfnamefont{Y.}~\bibnamefont{Shtanov}}, \bibnamefont{and}
  \bibinfo{author}{\bibfnamefont{A.}~\bibnamefont{Viznyuk}},
  \bibinfo{journal}{JCAP} \textbf{\bibinfo{volume}{0512}}, \bibinfo{pages}{005}
  (\bibinfo{year}{2005}), \eprint{astro-ph/0505004}.

\bibitem[{\citenamefont{Astashenok
  et~al.}(2012{\natexlab{a}})\citenamefont{Astashenok, Nojiri, Odintsov, and
  Scherrer}}]{Astashenok2012b}
\bibinfo{author}{\bibfnamefont{A.~V.} \bibnamefont{Astashenok}},
  \bibinfo{author}{\bibfnamefont{S.}~\bibnamefont{Nojiri}},
  \bibinfo{author}{\bibfnamefont{S.~D.} \bibnamefont{Odintsov}},
  \bibnamefont{and} \bibinfo{author}{\bibfnamefont{R.~J.}
  \bibnamefont{Scherrer}}, \bibinfo{journal}{Phys.Lett.}
  \textbf{\bibinfo{volume}{B713}}, \bibinfo{pages}{145}
  (\bibinfo{year}{2012}{\natexlab{a}}), \eprint{1203.1976}.

\bibitem[{\citenamefont{Astashenok et~al.}(2013)\citenamefont{Astashenok,
  Elizalde, de~Haro, Odintsov, and Yurov}}]{Astashenok2013}
\bibinfo{author}{\bibfnamefont{A.~V.} \bibnamefont{Astashenok}},
  \bibinfo{author}{\bibfnamefont{E.}~\bibnamefont{Elizalde}},
  \bibinfo{author}{\bibfnamefont{J.}~\bibnamefont{de~Haro}},
  \bibinfo{author}{\bibfnamefont{S.~D.} \bibnamefont{Odintsov}},
  \bibnamefont{and} \bibinfo{author}{\bibfnamefont{A.~V.} \bibnamefont{Yurov}},
  \bibinfo{journal}{Astrophys.Space Sci.} \textbf{\bibinfo{volume}{347}},
  \bibinfo{pages}{1} (\bibinfo{year}{2013}), \eprint{1301.6344}.

\bibitem[{\citenamefont{Astashenok and Odintsov}(2013)}]{Astashenok2013a}
\bibinfo{author}{\bibfnamefont{A.~V.} \bibnamefont{Astashenok}}
  \bibnamefont{and} \bibinfo{author}{\bibfnamefont{S.~D.}
  \bibnamefont{Odintsov}}, \bibinfo{journal}{Phys.Lett.}
  \textbf{\bibinfo{volume}{B718}}, \bibinfo{pages}{1194}
  (\bibinfo{year}{2013}), \eprint{1211.1888}.

\bibitem[{\citenamefont{Amanullah et~al.}(2010)\citenamefont{Amanullah, Lidman,
  Rubin, Aldering, Astier et~al.}}]{Amanullah2010}
\bibinfo{author}{\bibfnamefont{R.}~\bibnamefont{Amanullah}},
  \bibinfo{author}{\bibfnamefont{C.}~\bibnamefont{Lidman}},
  \bibinfo{author}{\bibfnamefont{D.}~\bibnamefont{Rubin}},
  \bibinfo{author}{\bibfnamefont{G.}~\bibnamefont{Aldering}},
  \bibinfo{author}{\bibfnamefont{P.}~\bibnamefont{Astier}},
  \bibnamefont{et~al.}, \bibinfo{journal}{Astrophys.J.}
  \textbf{\bibinfo{volume}{716}}, \bibinfo{pages}{712} (\bibinfo{year}{2010}),
  \eprint{1004.1711}.

\bibitem[{\citenamefont{Blake et~al.}(2011)\citenamefont{Blake, Kazin, Beutler,
  Davis, Parkinson et~al.}}]{Blake2011a}
\bibinfo{author}{\bibfnamefont{C.}~\bibnamefont{Blake}},
  \bibinfo{author}{\bibfnamefont{E.}~\bibnamefont{Kazin}},
  \bibinfo{author}{\bibfnamefont{F.}~\bibnamefont{Beutler}},
  \bibinfo{author}{\bibfnamefont{T.}~\bibnamefont{Davis}},
  \bibinfo{author}{\bibfnamefont{D.}~\bibnamefont{Parkinson}},
  \bibnamefont{et~al.}, \bibinfo{journal}{Mon.Not.Roy.Astron.Soc.}
  \textbf{\bibinfo{volume}{418}}, \bibinfo{pages}{1707} (\bibinfo{year}{2011}),
  \eprint{1108.2635}.

\bibitem[{\citenamefont{Stern et~al.}(2010)\citenamefont{Stern, Jimenez, Verde,
  Kamionkowski, and Stanford}}]{Stern2010}
\bibinfo{author}{\bibfnamefont{D.}~\bibnamefont{Stern}},
  \bibinfo{author}{\bibfnamefont{R.}~\bibnamefont{Jimenez}},
  \bibinfo{author}{\bibfnamefont{L.}~\bibnamefont{Verde}},
  \bibinfo{author}{\bibfnamefont{M.}~\bibnamefont{Kamionkowski}},
  \bibnamefont{and} \bibinfo{author}{\bibfnamefont{S.~A.}
  \bibnamefont{Stanford}}, \bibinfo{journal}{JCAP}
  \textbf{\bibinfo{volume}{1002}}, \bibinfo{pages}{008} (\bibinfo{year}{2010}),
  \eprint{0907.3149}.

\bibitem[{\citenamefont{Christopherson}(2010)}]{Christopherson2010}
\bibinfo{author}{\bibfnamefont{A.~J.} \bibnamefont{Christopherson}},
  \bibinfo{journal}{Phys.Rev.} \textbf{\bibinfo{volume}{D82}},
  \bibinfo{pages}{083515} (\bibinfo{year}{2010}), \eprint{1008.0811}.

\bibitem[{\citenamefont{Holanda et~al.}(2013)\citenamefont{Holanda, Silva, and
  Dahia}}]{Holanda2013}
\bibinfo{author}{\bibfnamefont{R.}~\bibnamefont{Holanda}},
  \bibinfo{author}{\bibfnamefont{J.}~\bibnamefont{Silva}}, \bibnamefont{and}
  \bibinfo{author}{\bibfnamefont{F.}~\bibnamefont{Dahia}}
  (\bibinfo{year}{2013}), \eprint{1304.4746}.

\bibitem[{\citenamefont{Garcia-Aspeitia}(2013)}]{Garcia-Aspeitia2013}
\bibinfo{author}{\bibfnamefont{M.~A.} \bibnamefont{Garcia-Aspeitia}}
  (\bibinfo{year}{2013}), \eprint{1306.1283}.

\bibitem[{\citenamefont{Wald}(1983)}]{Wald}
\bibinfo{author}{\bibfnamefont{R.~M.} \bibnamefont{Wald}},
  \bibinfo{journal}{Phys.Rev.} \textbf{\bibinfo{volume}{D28}},
  \bibinfo{pages}{2118} (\bibinfo{year}{1983}).

\bibitem[{\citenamefont{Nojiri et~al.}(2005)\citenamefont{Nojiri, Odintsov, and
  Tsujikawa}}]{Nojiri2005}
\bibinfo{author}{\bibfnamefont{S.}~\bibnamefont{Nojiri}},
  \bibinfo{author}{\bibfnamefont{S.~D.} \bibnamefont{Odintsov}},
  \bibnamefont{and}
  \bibinfo{author}{\bibfnamefont{S.}~\bibnamefont{Tsujikawa}},
  \bibinfo{journal}{Phys.Rev.} \textbf{\bibinfo{volume}{D71}},
  \bibinfo{pages}{063004} (\bibinfo{year}{2005}), \eprint{hep-th/0501025}.

\bibitem[{\citenamefont{Nojiri and Odintsov}(2005)}]{Nojiri2005a}
\bibinfo{author}{\bibfnamefont{S.}~\bibnamefont{Nojiri}} \bibnamefont{and}
  \bibinfo{author}{\bibfnamefont{S.~D.} \bibnamefont{Odintsov}},
  \bibinfo{journal}{Phys.Lett.} \textbf{\bibinfo{volume}{B631}},
  \bibinfo{pages}{1} (\bibinfo{year}{2005}), \eprint{hep-th/0508049}.

\bibitem[{\citenamefont{Bamba et~al.}(2012)\citenamefont{Bamba, Capozziello,
  Nojiri, and Odintsov}}]{Bamba2012g}
\bibinfo{author}{\bibfnamefont{K.}~\bibnamefont{Bamba}},
  \bibinfo{author}{\bibfnamefont{S.}~\bibnamefont{Capozziello}},
  \bibinfo{author}{\bibfnamefont{S.}~\bibnamefont{Nojiri}}, \bibnamefont{and}
  \bibinfo{author}{\bibfnamefont{S.~D.} \bibnamefont{Odintsov}},
  \bibinfo{journal}{Astrophys.Space Sci.} \textbf{\bibinfo{volume}{342}},
  \bibinfo{pages}{155} (\bibinfo{year}{2012}), \eprint{1205.3421}.

\bibitem[{\citenamefont{Astashenok
  et~al.}(2012{\natexlab{b}})\citenamefont{Astashenok, Elizalde, Odintsov, and
  Yurov}}]{Astashenok2012}
\bibinfo{author}{\bibfnamefont{A.~V.} \bibnamefont{Astashenok}},
  \bibinfo{author}{\bibfnamefont{E.}~\bibnamefont{Elizalde}},
  \bibinfo{author}{\bibfnamefont{S.~D.} \bibnamefont{Odintsov}},
  \bibnamefont{and} \bibinfo{author}{\bibfnamefont{A.~V.} \bibnamefont{Yurov}},
  \bibinfo{journal}{Eur.Phys.J.} \textbf{\bibinfo{volume}{C72}},
  \bibinfo{pages}{2260} (\bibinfo{year}{2012}{\natexlab{b}}),
  \eprint{1206.2192}.

\bibitem[{\citenamefont{Barrow}(2004)}]{Barrow2004}
\bibinfo{author}{\bibfnamefont{J.~D.} \bibnamefont{Barrow}},
  \bibinfo{journal}{Class.Quant.Grav.} \textbf{\bibinfo{volume}{21}},
  \bibinfo{pages}{L79} (\bibinfo{year}{2004}), \eprint{gr-qc/0403084}.

\bibitem[{\citenamefont{Nojiri and Odintsov}(2004{\natexlab{a}})}]{Nojiri2004}
\bibinfo{author}{\bibfnamefont{S.}~\bibnamefont{Nojiri}} \bibnamefont{and}
  \bibinfo{author}{\bibfnamefont{S.~D.} \bibnamefont{Odintsov}},
  \bibinfo{journal}{Phys.Lett.} \textbf{\bibinfo{volume}{B595}},
  \bibinfo{pages}{1} (\bibinfo{year}{2004}{\natexlab{a}}),
  \eprint{hep-th/0405078}.

\bibitem[{\citenamefont{Nojiri and Odintsov}(2004{\natexlab{b}})}]{Nojiri2004a}
\bibinfo{author}{\bibfnamefont{S.}~\bibnamefont{Nojiri}} \bibnamefont{and}
  \bibinfo{author}{\bibfnamefont{S.~D.} \bibnamefont{Odintsov}},
  \bibinfo{journal}{Phys.Rev.} \textbf{\bibinfo{volume}{D70}},
  \bibinfo{pages}{103522} (\bibinfo{year}{2004}{\natexlab{b}}),
  \eprint{hep-th/0408170}.

\bibitem[{\citenamefont{Allen and Wands}(2004)}]{Allen2004}
\bibinfo{author}{\bibfnamefont{L.~E.} \bibnamefont{Allen}} \bibnamefont{and}
  \bibinfo{author}{\bibfnamefont{D.}~\bibnamefont{Wands}},
  \bibinfo{journal}{Phys.Rev.} \textbf{\bibinfo{volume}{D70}},
  \bibinfo{pages}{063515} (\bibinfo{year}{2004}), \eprint{astro-ph/0404441}.

\bibitem[{\citenamefont{Biswas et~al.}(2006)\citenamefont{Biswas, Mazumdar, and
  Siegel}}]{Biswas2006}
\bibinfo{author}{\bibfnamefont{T.}~\bibnamefont{Biswas}},
  \bibinfo{author}{\bibfnamefont{A.}~\bibnamefont{Mazumdar}}, \bibnamefont{and}
  \bibinfo{author}{\bibfnamefont{W.}~\bibnamefont{Siegel}},
  \bibinfo{journal}{JCAP} \textbf{\bibinfo{volume}{0603}}, \bibinfo{pages}{009}
  (\bibinfo{year}{2006}), \eprint{hep-th/0508194}.

\bibitem[{\citenamefont{Novello and Bergliaffa}(2008)}]{Novello2008}
\bibinfo{author}{\bibfnamefont{M.}~\bibnamefont{Novello}} \bibnamefont{and}
  \bibinfo{author}{\bibfnamefont{S.~P.} \bibnamefont{Bergliaffa}},
  \bibinfo{journal}{Phys.Rept.} \textbf{\bibinfo{volume}{463}},
  \bibinfo{pages}{127} (\bibinfo{year}{2008}), \eprint{0802.1634}.

\bibitem[{\citenamefont{Cai and Saridakis}(2009)}]{Cai2009}
\bibinfo{author}{\bibfnamefont{Y.-F.} \bibnamefont{Cai}} \bibnamefont{and}
  \bibinfo{author}{\bibfnamefont{E.~N.} \bibnamefont{Saridakis}},
  \bibinfo{journal}{JCAP} \textbf{\bibinfo{volume}{0910}}, \bibinfo{pages}{020}
  (\bibinfo{year}{2009}), \eprint{0906.1789}.

\bibitem[{\citenamefont{Cai et~al.}(2011)\citenamefont{Cai, Chen, Dent, Dutta,
  and Saridakis}}]{Cai2011}
\bibinfo{author}{\bibfnamefont{Y.-F.} \bibnamefont{Cai}},
  \bibinfo{author}{\bibfnamefont{S.-H.} \bibnamefont{Chen}},
  \bibinfo{author}{\bibfnamefont{J.~B.} \bibnamefont{Dent}},
  \bibinfo{author}{\bibfnamefont{S.}~\bibnamefont{Dutta}}, \bibnamefont{and}
  \bibinfo{author}{\bibfnamefont{E.~N.} \bibnamefont{Saridakis}},
  \bibinfo{journal}{Class.Quant.Grav.} \textbf{\bibinfo{volume}{28}},
  \bibinfo{pages}{215011} (\bibinfo{year}{2011}), \eprint{1104.4349}.

\bibitem[{\citenamefont{Saridakis}(2009)}]{Saridakis2009}
\bibinfo{author}{\bibfnamefont{E.}~\bibnamefont{Saridakis}},
  \bibinfo{journal}{Nucl.Phys.} \textbf{\bibinfo{volume}{B808}},
  \bibinfo{pages}{224} (\bibinfo{year}{2009}), \eprint{0710.5269}.

\bibitem[{\citenamefont{Carloni et~al.}(2006)\citenamefont{Carloni, Dunsby, and
  Solomons}}]{Carloni2006}
\bibinfo{author}{\bibfnamefont{S.}~\bibnamefont{Carloni}},
  \bibinfo{author}{\bibfnamefont{P.~K.} \bibnamefont{Dunsby}},
  \bibnamefont{and} \bibinfo{author}{\bibfnamefont{D.~M.}
  \bibnamefont{Solomons}}, \bibinfo{journal}{Class.Quant.Grav.}
  \textbf{\bibinfo{volume}{23}}, \bibinfo{pages}{1913} (\bibinfo{year}{2006}),
  \eprint{gr-qc/0510130}.

\bibitem[{\citenamefont{Barragan and Olmo}(2010)}]{Barragan2010}
\bibinfo{author}{\bibfnamefont{C.}~\bibnamefont{Barragan}} \bibnamefont{and}
  \bibinfo{author}{\bibfnamefont{G.~J.} \bibnamefont{Olmo}},
  \bibinfo{journal}{Phys.Rev.} \textbf{\bibinfo{volume}{D82}},
  \bibinfo{pages}{084015} (\bibinfo{year}{2010}), \eprint{1005.4136}.

\bibitem[{\citenamefont{Khoury et~al.}(2001)\citenamefont{Khoury, Ovrut,
  Steinhardt, and Turok}}]{Khoury2001}
\bibinfo{author}{\bibfnamefont{J.}~\bibnamefont{Khoury}},
  \bibinfo{author}{\bibfnamefont{B.~A.} \bibnamefont{Ovrut}},
  \bibinfo{author}{\bibfnamefont{P.~J.} \bibnamefont{Steinhardt}},
  \bibnamefont{and} \bibinfo{author}{\bibfnamefont{N.}~\bibnamefont{Turok}},
  \bibinfo{journal}{Phys.Rev.} \textbf{\bibinfo{volume}{D64}},
  \bibinfo{pages}{123522} (\bibinfo{year}{2001}), \eprint{hep-th/0103239}.

\bibitem[{\citenamefont{Shtanov and Sahni}(2003)}]{Shtanov2003}
\bibinfo{author}{\bibfnamefont{Y.}~\bibnamefont{Shtanov}} \bibnamefont{and}
  \bibinfo{author}{\bibfnamefont{V.}~\bibnamefont{Sahni}},
  \bibinfo{journal}{Phys.Lett.} \textbf{\bibinfo{volume}{B557}},
  \bibinfo{pages}{1} (\bibinfo{year}{2003}), \eprint{gr-qc/0208047}.

\bibitem[{\citenamefont{Mukherji and Peloso}(2002)}]{Mukherji2002}
\bibinfo{author}{\bibfnamefont{S.}~\bibnamefont{Mukherji}} \bibnamefont{and}
  \bibinfo{author}{\bibfnamefont{M.}~\bibnamefont{Peloso}},
  \bibinfo{journal}{Phys.Lett.} \textbf{\bibinfo{volume}{B547}},
  \bibinfo{pages}{297} (\bibinfo{year}{2002}), \eprint{hep-th/0205180}.

\bibitem[{\citenamefont{Brandenberger et~al.}(2007)\citenamefont{Brandenberger,
  Firouzjahi, and Saremi}}]{Brandenberger2007}
\bibinfo{author}{\bibfnamefont{R.}~\bibnamefont{Brandenberger}},
  \bibinfo{author}{\bibfnamefont{H.}~\bibnamefont{Firouzjahi}},
  \bibnamefont{and} \bibinfo{author}{\bibfnamefont{O.}~\bibnamefont{Saremi}},
  \bibinfo{journal}{JCAP} \textbf{\bibinfo{volume}{0711}}, \bibinfo{pages}{028}
  (\bibinfo{year}{2007}), \eprint{0707.4181}.

\bibitem[{\citenamefont{Dominguez et~al.}(2013)\citenamefont{Dominguez, Finke,
  Prada, Primack, Kitaura et~al.}}]{Dominguez2013}
\bibinfo{author}{\bibfnamefont{A.}~\bibnamefont{Dominguez}},
  \bibinfo{author}{\bibfnamefont{J.}~\bibnamefont{Finke}},
  \bibinfo{author}{\bibfnamefont{F.}~\bibnamefont{Prada}},
  \bibinfo{author}{\bibfnamefont{J.}~\bibnamefont{Primack}},
  \bibinfo{author}{\bibfnamefont{F.}~\bibnamefont{Kitaura}},
  \bibnamefont{et~al.}, \bibinfo{journal}{Astrophys.J.}
  \textbf{\bibinfo{volume}{770}}, \bibinfo{pages}{77} (\bibinfo{year}{2013}),
  \eprint{1305.2162}.

\bibitem[{\citenamefont{Dominguez and Prada}(2013)}]{Dominguez2013a}
\bibinfo{author}{\bibfnamefont{A.}~\bibnamefont{Dominguez}} \bibnamefont{and}
  \bibinfo{author}{\bibfnamefont{F.}~\bibnamefont{Prada}}
  (\bibinfo{year}{2013}), \eprint{1305.2163}.

\bibitem[{\citenamefont{Suyu et~al.}(2012)\citenamefont{Suyu, Treu, Blandford,
  Freedman, Hilbert et~al.}}]{Suyu2012}
\bibinfo{author}{\bibfnamefont{S.}~\bibnamefont{Suyu}},
  \bibinfo{author}{\bibfnamefont{T.}~\bibnamefont{Treu}},
  \bibinfo{author}{\bibfnamefont{R.}~\bibnamefont{Blandford}},
  \bibinfo{author}{\bibfnamefont{W.}~\bibnamefont{Freedman}},
  \bibinfo{author}{\bibfnamefont{S.}~\bibnamefont{Hilbert}},
  \bibnamefont{et~al.} (\bibinfo{year}{2012}), \eprint{1202.4459}.

\bibitem[{\citenamefont{Weinberg et~al.}(2013)\citenamefont{Weinberg,
  Mortonson, Eisenstein, Hirata, Riess et~al.}}]{Weinberg2013}
\bibinfo{author}{\bibfnamefont{D.~H.} \bibnamefont{Weinberg}},
  \bibinfo{author}{\bibfnamefont{M.~J.} \bibnamefont{Mortonson}},
  \bibinfo{author}{\bibfnamefont{D.~J.} \bibnamefont{Eisenstein}},
  \bibinfo{author}{\bibfnamefont{C.}~\bibnamefont{Hirata}},
  \bibinfo{author}{\bibfnamefont{A.~G.} \bibnamefont{Riess}},
  \bibnamefont{et~al.}, \bibinfo{journal}{Phys.Rept.}
  \textbf{\bibinfo{volume}{530}}, \bibinfo{pages}{87} (\bibinfo{year}{2013}),
  \eprint{1201.2434}.

\bibitem[{\citenamefont{Freedman and Madore}(2010)}]{Freedman2010}
\bibinfo{author}{\bibfnamefont{W.~L.} \bibnamefont{Freedman}} \bibnamefont{and}
  \bibinfo{author}{\bibfnamefont{B.~F.} \bibnamefont{Madore}},
  \bibinfo{journal}{Ann.Rev.Astron.Astrophys.} \textbf{\bibinfo{volume}{48}},
  \bibinfo{pages}{673} (\bibinfo{year}{2010}), \eprint{1004.1856}.

\end{thebibliography}

\end{document}